%% file: ms.tex
\documentclass[twocolumn,floatfix,times,twocolappendix]{aastex63}

\usepackage{graphicx}
\usepackage{longtable} \LTcapwidth=\textwidth
\usepackage{multirow}
\usepackage{subfigure}
\usepackage{ulem}
\usepackage{soul}
\usepackage{lipsum, babel}
\usepackage{enumitem}
\usepackage{amsmath}
\usepackage{euscript}

\newcommand{\Hii}{H$_2$}
\newcommand{\DEG}{$^{\circ}$}
\newcommand{\Tex}{$T_\mathrm{ex}$}
\newcommand{\Ntot}{$N$}

\newcommand{\vLSR}{$v_\mathrm{LSR}$}

\newcommand{\deltav}{$\delta v$}
\newcommand{\Spitzer}{\textit{Spitzer}}
\newcommand{\Herschel}{\textit{Herschel}}
\newcommand{\frest}{$f_\mathrm{rest}$}
\newcommand{\Aij}{$A_\mathrm{ij}$}
\newcommand{\Eu}{$E_\mathrm{up}$}

\newcommand{\gu}{$g_\mathrm{up}$}
\newcommand{\CCIso}{$^{12}$C/$^{13}$C}
\newcommand{\OOIso}{$^{16}$O/$^{18}$O}

\newcommand{\CCIsoOrion}{50}

\newcommand{\Lbol}{$L_\mathrm{bol}$}
\newcommand{\Tbol}{$T_\mathrm{bol}$}

\newcommand{\tstar}{$t_\mathrm{\star}$}
\newcommand{\Mstar}{$M_\mathrm{\star}$}
\newcommand{\Rstar}{$R_\mathrm{\star}$}
\newcommand{\Tstar}{$T_\mathrm{\star}$}
\newcommand{\Mdisk}{$M_\mathrm{disk}$}
\newcommand{\RMAXD}{$R^\mathrm{outer}_\mathrm{disk}$}
\newcommand{\RMAXE}{$R^\mathrm{outer}_\mathrm{env}$}
\newcommand{\rhoamb}{$\rho_\mathrm{amb}$}
\newcommand{\rhocav}{$\rho_\mathrm{cav}$}
\newcommand{\Msun}{$M_\odot$}
\newcommand{\Lsun}{$L_\odot$}
\newcommand{\Rsun}{$R_\odot$}
\newcommand{\MdotD}{$\dot{M}_\mathrm{disk}$}
\newcommand{\MdotE}{$\dot{M}_\mathrm{env}$}
\newcommand{\RMIND}{$R^\mathrm{inner}_\mathrm{disk}$}
\newcommand{\ZFAC}{$z^\mathrm{scale}_\mathrm{disk}$}
\newcommand{\incl}{$\cos\varphi$}

\newcommand{\ThetaIV}{$\Theta_\mathrm{400}$}

\setstcolor{red}

\accepted{\today}
\submitjournal{ApJ}

\shorttitle{COMs in Orion Protostars}
\shortauthors{Hsu et al.}
\graphicspath{{./}{figures/}}

\begin{document}

\title{ALMA Survey of Orion Planck Galactic Cold Clumps (ALMASOP):\\ 
A Hot Corino Survey toward Protostellar Cores in the Orion Cloud}

\input{AUTHOR.tex}

\begin{abstract}
The presence of complex organic molecules (COMs) in the interstellar medium (ISM) is of great interest since it may link to the origin and prevalence of life in the universe.
Aiming to investigate the occurrence of COMs and their possible origins, we conducted a chemical census toward a sample of protostellar cores as part of the ALMA Survey of Orion Planck Galactic Cold Clumps (ALMASOP) project.
We report the detection of 11 hot corino sources, which exhibit compact emissions from warm and abundant COMs, among 56 Class 0/I protostellar cores.
All the hot corino sources discovered are likely Class 0 and their sizes of the warm region ($>$ 100 K)  are comparable to 100 au.
The luminosity of the hot corino sources exhibits positive correlations with the total number of methanol and the extent of its emissions.
Such correlations are consistent with the thermal desorption picture for the presence of hot corino and suggest that the lower luminosity (Class 0) sources likely have a smaller region with COMs emissions. 
With the same sample selection method and detection criteria being applied, the detection rates of the warm methanol in the Orion cloud (15/37) and the Perseus cloud (28/50) are statistically similar when the cloud distances and the limited sample size are considered.
Observing the same set of COM transitions will bring a more informative comparison between the cloud properties.
\end{abstract}


\input{sec_Intro.tex}

\input{sec_Obs.tex}
\input{sec_XCLASS.tex}
\input{sec_SED.tex}
\section{Discussion}
\label{sec:Disc}
\input{disc_class}
\input{disc_rate}
\input{disc_COMs}
\input{sec_Conclusion.tex}


\acknowledgments
This paper makes use of the following ALMA data: ADS/JAO.ALMA\#2018.1.00302.S. ALMA is a partnership of ESO (representing its member states), NSF (USA) and NINS (Japan), together with NRC (Canada), MOST and ASIAA (Taiwan), and KASI (Republic of Korea), in cooperation with the Republic of Chile. The Joint ALMA Observatory is operated by ESO, AUI/NRAO and NAOJ.
SYH and SYL acknowledge support from the Ministry of Science and Technology (MoST) with grants 110-2112-M-001-056-.
Tie Liu acknowledges the supports by National Natural Science Foundation of China (NSFC) through grants No.12073061 and No.12122307, the international partnership program of Chinese Academy of Sciences through grant No.114231KYSB20200009, and Shanghai Pujiang Program 20PJ1415500."
N.H. acknowledges a grant from the Ministry of Science and Technology (MoST) of Taiwan (MoST109-2112-M-001-023- and MoST 110-2112-M-001-048-).
Y.-L. Yang acknowledges the supports from the Virginia Initiative of Cosmic Origins (VICO) Postdoctoral Fellowship
Doug Johnstone is supported by the National Research Council of Canada and by an NSERC Discovery Grant. 
L.B. gratefully acknowledges support from ANID BASAL project FB210003. 
This research has made use of the Spanish Virtual Observatory (http://svo.cab.inta-csic.es) supported from the Spanish MICINN/FEDER through grant AyA2017-84089.
S.L. Qin is supported by  the National Natural Science Foundation of China under grant No. 12033005.
C.W.L. is supported by the Basic Science Research Program through the National Research Foundation of Korea (NRF) funded by the Ministry of Education, Science and Technology (NRF- 2019R1A2C1010851)
P.S. was partially supported by a Grant-in-Aid for Scientific Research (KAKENHI Number 18H01259) of the Japan Society for the Promotion of Science (JSPS).
This research made use of Astropy,\footnote{http://www.astropy.org} a community-developed core Python package for Astronomy \citep{astropy:2013, astropy:2018}. 

%



\software{
astropy \citep{astropy:2013, astropy:2018},
CASA \citep{2007McMullin_CASA},
HO--CHUNK \citep{2003Whitney_SED}
SED Fitter \citep{2006Robitaille_sedfitter, 2007Robitaille_sedfitter}
XCLASS \citep{2017Moller_XCLASS}
}

\clearpage
\appendix
\section{Source Overview \label{appx:source}}
\restartappendixnumbering
\input{appx_source.tex}

\clearpage
\section{Molecular Analysis \label{appx:xclass}}
\restartappendixnumbering
\input{appx_xclass.tex}

\clearpage
\section{SED Analysis \label{appx:sed}}
\restartappendixnumbering
\input{appx_sed.tex}

\bibliography{REFERENCE.bib}{}
\bibliographystyle{aasjournal}




\end{document}

%% file: AUTHOR.tex

\author[0000-0002-1369-1563]{Shih-Ying Hsu}
\email{seansyhsu@gmail.com}
\affiliation{National Taiwan University (NTU), No. 1, Section 4, Roosevelt Rd, Taipei 10617, Taiwan (R.O.C.)}
\affiliation{Institute of Astronomy and Astrophysics, Academia Sinica, No.1, Sec. 4, Roosevelt Rd, Taipei 10617, Taiwan (R.O.C.)}

\author[0000-0012-3245-1234]{Sheng-Yuan Liu}
\email{syliu@asiaa.sinica.edu.tw}
\affiliation{Institute of Astronomy and Astrophysics, Academia Sinica, No.1, Sec. 4, Roosevelt Rd, Taipei 10617, Taiwan (R.O.C.)}

\author[0000-0002-5286-2564]{Tie Liu}
\affiliation{Key Laboratory for Research in Galaxies and Cosmology, Shanghai Astronomical Observatory, Chinese Academy of Sciences, 80 Nandan Road, Shanghai 200030, People’s Republic of China}

\author[0000-0002-4393-3463]{Dipen Sahu}
\affiliation{Institute of Astronomy and Astrophysics, Academia Sinica, No.1, Sec. 4, Roosevelt Rd, Taipei 10617, Taiwan (R.O.C.)}

\author[0000-0002-3024-5864]{Chin-Fei Lee}
\affiliation{Institute of Astronomy and Astrophysics, Academia Sinica, No.1, Sec. 4, Roosevelt Rd, Taipei 10617, Taiwan (R.O.C.)}

\author[0000-0002-8149-8546]{Kenichi Tatematsu}
\affiliation{Nobeyama Radio Observatory, National Astronomical Observatory of Japan, National Institutes of Natural Sciences, 462-2 Nobeyama, Minamimaki, Minamisaku, Nagano 384-1305, Japan}
\affiliation{Department of Astronomical Science, The Graduate University for Advanced Studies, SOKENDAI,
2-21-1 Osawa, Mitaka, Tokyo 181-8588, Japan}

\author[0000-0003-2011-8172]{Kee-Tae Kim}
\affiliation{Korea Astronomy and Space Science Institute (KASI), 776 Daedeokdae-ro, Yuseong-gu, Daejeon 34055, Republic of Korea}
\affiliation{University of Science and Technology, Korea (UST), 217 Gajeong-ro, Yuseong-gu, Daejeon 34113, Republic of Korea}

\author[0000-0001-9304-7884]{Naomi Hirano}
\affiliation{Institute of Astronomy and Astrophysics, Academia Sinica, No.1, Sec. 4, Roosevelt Rd, Taipei 10617, Taiwan (R.O.C.)}

\author[0000-0001-8227-2816]{Yao-Lun Yang}
\affiliation{Department of Astronomy, University of Virginia, Charlottesville, VA 22904, USA}

\author[0000-0002-6773-459X]{Doug Johnstone}
\affiliation{NRC Herzberg Astronomy and Astrophysics, 5071 West Saanich Rd, Victoria, BC, V9E 2E7, Canada}
\affiliation{Department of Physics and Astronomy, University of Victoria, Victoria, BC, V8P 5C2, Canada}

\author[0000-0003-3343-9645]{Hongli Liu}
\affiliation{Department of Astronomy, Yunnan University, Kunming 650091, People’s Republic of China}

\author[0000-0002-5809-4834]{Mika Juvela}
\affiliation{Department of Physics, P.O.Box 64, FI-00014, University of Helsinki, Finland}

\author[0000-0002-9574-8454]{Leonardo Bronfman}
\affiliation{Departamento de Astronom\'{i}a, Universidad de Chile, Casilla 36-D, Santiago, Chile}

\author[0000-0002-9774-1846]{Huei-Ru Vivien Chen}
\affiliation{Department of Physics and Institute of Astronomy, National Tsing Hua University, Hsinchu, 30013, Taiwan}

\author[0000-0002-2338-4583]{Somnath Dutta}
\affiliation{Institute of Astronomy and Astrophysics, Academia Sinica, No.1, Sec. 4, Roosevelt Rd, Taipei 10617, Taiwan (R.O.C.)}

\author[0000-0002-5881-3229]{David J. Eden}
\affiliation{Astrophysics Research Institute, Liverpool John Moores University, iC2, Liverpool Science Park, 146 Brownlow Hill, Liverpool, L3 5RF, UK.}

\author[0000-0003-2069-1403]{Kai-Syun Jhan}
\affiliation{National Taiwan University (NTU), No. 1, Section 4, Roosevelt Rd, Taipei 10617, Taiwan (R.O.C.)}
\affiliation{Institute of Astronomy and Astrophysics, Academia Sinica, No.1, Sec. 4, Roosevelt Rd, Taipei 10617, Taiwan (R.O.C.)}

\author[0000-0002-4336-0730]{Yi-Jehng Kuan}
\affiliation{Department of Earth Sciences, National Taiwan Normal University, Taipei, Taiwan (R.O.C.)}
\affiliation{Institute of Astronomy and Astrophysics, Academia Sinica, No.1, Sec. 4, Roosevelt Rd, Taipei 10617, Taiwan (R.O.C.)}

\author[0000-0002-3179-6334]{Chang Won Lee}
\affiliation{Korea Astronomy and Space Science Institute (KASI), 776 Daedeokdae-ro, Yuseong-gu, Daejeon 34055, Republic of Korea}
\affiliation{University of Science and Technology, Korea (UST), 217 Gajeong-ro, Yuseong-gu, Daejeon 34113, Republic of Korea}

\author[0000-0003-3119-2087]{Jeong-Eun Lee}
\affiliation{School of Space Research, Kyung Hee University, 1732, Deogyeong-Daero, Giheung-gu Yongin-shi, Gyunggi-do 17104, Korea}

\author[0000-0003-1275-5251]{Shanghuo Li}
\affiliation{Korea Astronomy and Space Science Institute, 776 Daedeokdae-ro, Yuseong-gu, Daejeon 34055, Republic of Korea.}

\author[0000-0002-1624-6545]{Chun-Fan Liu}
\affiliation{Institute of Astronomy and Astrophysics, Academia Sinica, No.1, Sec. 4, Roosevelt Rd, Taipei 10617, Taiwan (R.O.C.)}

\author[0000-0003-2302-0613]{Sheng-Li Qin}
\affiliation{Department of Astronomy, Yunnan University, and Key Laboratory of Astroparticle Physics of Yunnan Province, Kunming, 650091, People's Republic of China}

\author[0000-0002-7125-7685]{Patricio Sanhueza}
\affiliation{National Astronomical Observatory of Japan, National Institutes of Natural Sciences, 2-21-1 Osawa, Mitaka, Tokyo 181-8588, Japan}
\affiliation{Department of Astronomical Science, SOKENDAI (The Graduate University for Advanced Studies), 2-21-1 Osawa, Mitaka, Tokyo 181-8588, Japan}

\author[0000-0001-8385-9838]{Hsien Shang}
\affiliation{Institute of Astronomy and Astrophysics, Academia Sinica, No.1, Sec. 4, Roosevelt Rd, Taipei 10617, Taiwan (R.O.C.)}

\author[0000-0002-6386-2906]{Archana Soam}
\affiliation{SOFIA Science Center, USRA, NASA Ames Research Center, M.S.-12, N232, Moffett Field, CA 94035, USA}

\author[0000-0003-1665-6402]{Alessio Traficante} 
\affiliation{INAF-IAPS, via Fosso del Cavaliere, 100. 00133, Rome, IT}

\author[0000-0003-0356-818X]{Jianjun Zhou}
\affiliation{Xinjiang Astronomical Observatory, Chinese Academy of Sciences 150 Science 1-Street, 830011 Urumqi, People's Republic of China}

%% file: sec_Intro.tex
\section{Introduction}
\label{sec:Intro}



The presence of complex organic molecules, those organic molecules consisting of six or more atoms, in the interstellar medium (ISM) is particularly of great interest since it may link to the origin and prevalence of organic matter and even life in the Universe.
Saturated complex organic molecules (also often referred as complex organic molecules, COMs, interstellar COMs, or iCOMs) have been discovered in various star--forming environments, including prestellar cores \citep[e.g., L1689B][]{2012Bacmann_COM_prestellar}, protostellar cores \citep[e.g., IRAS 16293--2422 A and B][]{2004Bottinelli_IRAS16293A_COM, 2004Kuan_IRAS16293B_COM, 2005Huang_IRAS16293B_COM}, outflows \citep[e.g., L1157][]{2008Arce_L1157_COMs_outflow}, and most recently protostellar disks \citep[e.g., HH-212 and V883 Ori3][]{2017Lee_HH212, 2019Lee_V883Ori_COMs_disk}.
In particular, hot corino sources are identified by localized zones, which surround low- or intermediate-mass forming stars and harbor warm ($\sim$ 100 K), abundant (relative column density with respect to molecular hydrogen $X>10^{-8}$), and compact ($\sim$ 100 au) COM emissions \citep{2003Cazaux_IRAS16293-2422}.
Since its first identification in 2004 \citep{2004Ceccarelli_HotCorino}, to date only about a couple tens of hot corino sources have been discovered, including, for example, IRAS 16293--2422 B \citep{2004Kuan_IRAS16293B_COM, 2005Huang_IRAS16293B_COM}, B335 \citep{2016Imai_B335}, HH--212 \citep{2016Codella_HH212, 2017Lee_HH212, 2019Lee_HH212}, NGC 1333 IRAS 4A1 \citep{2019Sahu_IRAS4A1}, and BHR--71 IRS1 \citep{2020Yang_BHR-71-IRS1}.

Recently, there have emerged studies investigating the statistical nature of COM emissions towards selected samples of protostellar cores.
\citet{2020Belloche_COM_CALYPSO} observed 26 solar--type star--forming regions, including 22 Class 0 and four Class I protostars, under the project ``Continuum And Lines in Young ProtoStellar Objects (CALYPSO)" conducted with Plateau de Bure Interferometer (PdBI).
Based on the chemical differentiation in multiple systems, they raised the question of whether the hot corino phase is a common stage of star formation.
\citet{2020vanGelder_COMs} investigated COM emission on the scale of $\sim$ 100 au toward seven Class 0 protostellar cores, including four sources in the Perseus Barnard 1 cloud and three sources in the Serpens Main region.
They found three of these seven sources to be COM--rich and among these three sources similar abundance ratios of COMs, including deuterated methanol (CH$_2$DOH), methyl formate (HCOOCH$_3$), and dimethyl ether (CH$_3$OCH$_3$), with respect to CH$_3$OH.
Since such comparable abundance ratios have also been reported toward two other COM--rich sources in different clouds: IRAS 16293--2422 B in the Ophiuchus cloud and HH 212 in the Orion cloud, \citet{2020vanGelder_COMs} suggested that the abundances of most O--bearing COMs are similar among different star--forming regions at their prestellar stages.
For a few COMs having varying abundance ratios with respect to CH$_3$OH, such as acetaldehyde (CH$_3$CHO) and ethanol (C$_2$H$_5$OH), \citet{2020vanGelder_COMs} on the other hand suggested that they were affected by the local environment of the parent cloud.
Toward the same sample of \citet{2020vanGelder_COMs}, the similar abundances of N-bearing molecules with respect to isocyanic acid (HNCO) suggest a shared chemical history \citep{2021Nazari_COMs}. 
\citet{2021Yang_PEACHES} carried out a survey of COMs toward 50 protostars in the Perseus cloud. The sample was selected to have the following properties: (1) the central source is Class 0/I,  (2) the bolometric luminosity $L_\mathrm{bol} > 1 L_\sun$ (except for B1--5), and (3) the envelope mass $M_\mathrm{env} > 1 M_\sun$ to ensure the association of a substantial amount of molecular gas.
They reported 28 detections of CH$_3$OH emission among 50 embedded Class 0/I protostars, and also suggested a possible chemical relation between methanol (CH$_3$OH) and methyl cyanide (CH$_3$CN).

To investigate the occurrence of COM emissions and their possible origins, we conducted a chemical census toward a sample of protostellar cores as part of the ALMA Survey of Orion Planck Galactic Cold Clumps (ALMASOP) project.
Starting with a sample of Planck Galactic Cold Clumps \citep[PGCCs, ][]{2016Planck_PGCC} in the Orion Molecular Complex, which are cold and dense and likely star--forming dust condensations, we made a series of observations and studies \citep{2016Tatematsu_PGCC_N2Dp, 2018Liu_TOP-SCOPE, 2018Yi_PGCC_Orion, 2019Eden_SCOPE, 2020Tatematsu_ALMASOP, 2020Kim_ALMASOP_Nobeyama}.
Based on their 850 \micron~continuum and N$_2$D$^+$ emission, we selected 72 clumps as the ALMASOP observation targets.
Some of these 72 targets harbor more than one core which may be starless, prestellar, or protostellar \citep{2020Dutta_ALMASOP}.
See \citet{Hsu2020_ALMASOP}, \citet{2020Dutta_ALMASOP}, and \citet{2021Sahu_ALMASOP} for more details about the project and the sample selection.

\citet{Hsu2020_ALMASOP} reported the identification of four hot corino sources, which are G211.47--19.27S (HOPS--288), G208.68--19.20N1 (HOPS--87) , G210.49-19.79W--A (HOPS--168), and G192.12--11.10, based on the observational data obtained from the ACA (Atacama Compact Array), a 7--m array of ALMA (Atacama Large Millimeter/submillimeter Array).
These four hot corinos harbor emission from CH$_3$OH and other oxygen--bearing COMs (in G211.47--19.27S and G208.68--19.20N1) as well as NH$_2$CHO (in G211.47--19.27S), which is of prebiotic interests.
Although the spatial resolution achieved with the ACA was not sufficient to resolve most of the molecular emission distributions, the large line widths (4--9 km s$^{-1}$) and high rotational temperatures ($>$ 100 K) of the detected COMs suggest that they likely reside in the hotter and innermost region immediately surrounding the protostars.
The occurrence of hot corino sources was about 8 \%, based on the detection of warm ($>100$ K) methanol among the 48 Class 0/I protostellar cores identified in the ACA--only data.

In this paper, Sect. \ref{sec:Obs} introduces the observation of the ALMASOP project.
We report the detection of 11 hot corino sources among 56 Class 0/I protostellar cores from the combined data of ALMASOP in Sect. \ref{sec:XCLASS}.
Sect. \ref{sec:SED} presents our spectral energy distribution modeling, based on which we infer the thermal structure and the extent of the warm region ($>100$ K) in these sources.
In Sect. \ref{sec:disc_class}, we discuss the spectral energy distribution ranges and the classifications (i.e., Class 0/I) of the detected hot corinos.
We show the correlation between the luminosity and the detection of hot corino (Sect. \ref{sec:disc_rate_luminosity}) and demonstrate how sensitivity impacts the detection rate of hot corinos
(Sect. \ref{sec:disc_rate_sensitivity}).
We also compare the detection rate of warm methanol, an indicator of the occurrence of hot corino, in our survey toward the Orion cloud and that toward the Perseus cloud (Sect. \ref{sec:disc_rate_PEACEHS}).
We discuss the chemical relations between the identified organic molecules based on their column densities (Sect. \ref{sec:disc_mol}), and finally summarize our findings in Sect. \ref{sec:Conclusions}.

%% file: sec_Obs.tex
\section{Observations}
\label{sec:Obs}

The observations toward the 72 targets of the ALMASOP project were carried out with both the 12-m array (in two configurations: C43--5 and C43--2 represented respectively by TM1 and TM2) and the 7--m array (ACA) of ALMA in the ALMA Cycle 6 operation (\#2018.1.003.2.S, PI: Tie Liu). 
Following the successful identifications of the hot corino sources with the ACA data in \citet{Hsu2020_ALMASOP}, we further combined the ALMA and ACA data for gaining a better angular resolution.
The imaging was carried out using \texttt{tclean} in Common Astronomy Software Applications \citep[CASA, ][]{2007McMullin_CASA} 5.6 with the following parameters: a robust value of 2.0, the thresholds for the spectral cube and the continuum mfs image of 20 mJy and 60 $\mu$Jy, respectively.
The resulting typical angular resolution is 0\farcs{35}, which corresponds to about 140 au for a distance of $\sim400$ pc.
The sensitivity of the image reaches $\sim$ 2.8 mJy beam$^{-1}$ and $\sim$ 12 $\mu$Jy beam$^{-1}$ for each channel (1.129 MHz) and the full--band (7.5 GHz) continuum, respectively.

%% file: sec_XCLASS.tex
\section{Molecule Detection \label{sec:XCLASS}}
\input{tab_srcCoord.tex}

\subsection{Data and Tools}
To have a quick search of the presence of COMs, we first extract the spectra at the continuum peak pixel for the protostellar cores from the ALMASOP sample. 
We select the sources where at least two methanol transitions are detected (SNR $>$ 5) and define their emission extent based on the 2D Gaussian fitting of their CH$_3$OH moment--0 images.
The CH$_3$OH transition we use is at $J=4-3$; $K_a=2-1$; $K_c=3-2$; \Eu~$=46$ K; and \frest~$ = 218440$ MHz (hereafter CH$_3$OH--46K transition), which is the strongest methanol transition in our spectral windows.
We then extract the full spectra within the same extent toward each source.
Following the similar procedures in \citet{Hsu2020_ALMASOP}, we use XCLASS \citep[eXtended CASA Line Analysis Software Suite, ][]{2017Moller_XCLASS} to search for the molecular carrier candidates with its \texttt{LineIdentification} function, and to simulate synthetic spectra with input molecular parameters including the excitation temperature (\Tex), total column density (\Ntot), line width in velocity (\deltav), and local-standard--of--rest velocity (\vLSR). 
Each (gaseous) molecular component is assumed to be in its individual local thermodynamic equilibrium (LTE).
We further use MAGIX (Modeling and Analysis Generic Interface for eXternal numerical codes), a package of XCLASS, to optimize the above parameters.
The optimization algorithms in MAGIX we use are ``Bees algorithms'' and ``Levenberg–Marquardt (LM) algorithms''.
The former is a swarm algorithm providing a good overview for all parameter combinations within given ranges.
The latter is also known as the ``damped least-squares non-linear method'' for finding local minima in non-linear least-squares problems.
To have a first estimation, we make individual fittings for each molecular species with the Bees algorithm, and the range of \Tex~is limited to be (0, 400).
We then run overall fittings for all species simultaneously with Bees algorithm and LM algorithm for each source to take into account blending of lines.
We assume the emission size of all species to be the (2D Gaussian) extent of the CH$_3$OH--46K transition deconvolved by the beam size.

For molecules detected with only one transition, we assume their \Tex\ to be 100~K, the typical temperature of a hot corino.
The estimate of column density (\Ntot) in these cases hence may not be less precise.

\subsection{Results}

\begin{figure}
\epsscale{1.1}
\plotone{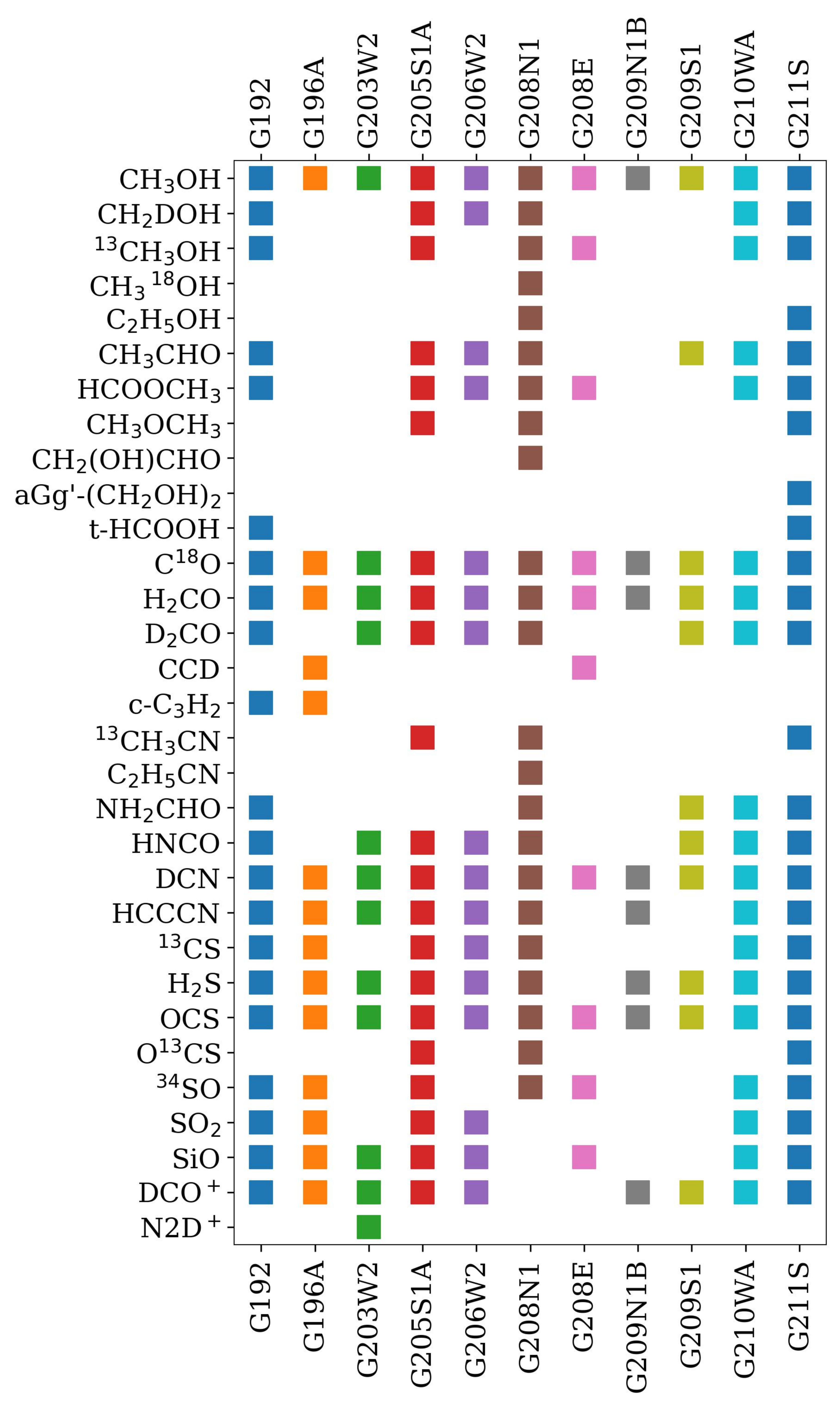}
\caption{\label{fig:scatter_mol_exist} 
Summary of the molecular detection in each source.
}
\end{figure}
The combined data (TM1+TM2+ACA) of ALMASOP, with a higher angular resolution, resolve more protostellar cores compared to the ACA data.
Consequently, \citet{2020Dutta_ALMASOP} reported the identification of 56 Class 0/I protostellar objects.
Among the 56 protostellar cores, we find 11 sources with at least two CH$_3$OH transitions including at least one transition with upper energy \Eu\ $>100$ K getting detected at a level above 5$\sigma$. 
They are G192, G196A, G203W2, G205S1A, G206W2, G208N1, G208E, G209N1B, G209S1, G210WA and G211S. 
See Table \ref{tab:srcCoord} for their corresponding full names in the ALMASOP observation.
Table \ref{tab:srcCoord} lists the parent clouds, coordinates, bolometric luminosity and temperature, and classifications of the 11 sources.
Seven of these 11 sources were observed by the Herschel Orion Protostar Survey (HOPS) and the correspondence is listed in Table \ref{tab:srcCoord}.
See Appendix \ref{appx:source} for more details about the 11 sources.
Tables \ref{tab:molfit_G192} -- \ref{tab:molfit_G211S} display the molecular parameters (i.e. \Tex, \Ntot, \deltav, and \vLSR) and their errors estimated by the Markov chain Monte Carlo (MCMC) method in XCLASS of these 11 sources.
Fig. set \ref{fig:spec_xclass} shows the spectra and the identified transitions for each source (see Fig. \ref{fig:spec_xclass} for an example).
The emission region of the detected COMs, such as methanol and methyl formate, based on their integrated intensity (moment--0) images, are mostly unresolved or marginally resolved, as demonstrated in Fig. \ref{fig:mom0_CH3OH} and Fig. \ref{fig:mom0_G211S}, except for a few low--excitation transitions.

Based on the presence of compact (a few hundred au), warm (\Tex~$> 100$ K) and abundant ($X>10^{-8}$) CH$_3$OH, we consider these 11 cores to be hot corinos. 
We detect toward them in total 36 molecules, including simple diatomic species such as carbon monoxide isotopologues (CO and C$^{18}$O), simple organics such as formaldehyde isotopologues (H$_2$CO and D$_2$CO), and COMs such as methanol isotopologues (CH$_3$OH, CH$_2$DOH, $^{13}$CH$_3$OH, and CH$_3\,^{18}$OH), acetaldehyde isotopologues (CH$_3$CHO and CH$_3$CHD), methyl formate (HCOOCH$_3$), ethanol (C$_2$H$_5$OH),  $^{13}$C substituted methyl cyanide ($^{13}$CH$_3$CN), formamide (NH$_2$CHO), etc.
See Table \ref{tab:trans_all} for a complete list of the molecular species and their detected transitions.

Fig. \ref{fig:scatter_mol_exist} summarizes the occurrence of the identified molecules in the 11 sources.
There are three sources (i.e., G203W2, G209N1B and G196A) harboring only one COM, which is CH$_3$OH.
Their weaker continuum emission, corresponding to lower \Hii~column densities among the sample, may have prevented the detection of other COMs toward them, should other COMs bear similar fractional abundances as in the stronger sources.


Among these 11 sources, four of them were previously reported as hot corinos in \citet{Hsu2020_ALMASOP}, namely G192, G208N1, G210WA, and G211S.
For the methanol detected in the combined data (TM1+TM2+ACA) of these four sources, their \Tex~is higher, \Ntot~is larger, and \deltav~is larger than for those detected in the ACA--only data.
The comparison suggests that the combined data enables better detection of methanol from more compact, denser, hotter, and perhaps inner or more turbulent regions.

%% file: tab_srcCoord.tex
\begin{deluxetable*}{llrrrrrcl}[htb]
\tablecaption{\label{tab:srcCoord}
Source Information}
\tablewidth{2pt}
\tablehead{\colhead{Name} & \colhead{Short Name} & \colhead{Cloud} & \colhead{$\mathrm{\alpha_{J2000}}$} & \colhead{$\mathrm{\delta_{J2000}}$} & \colhead{\Lbol~(\Lsun)} & \colhead{\Tbol~(K)} & \colhead{Class} &  \colhead{HOPS Index}  } 
\startdata
G192.12--11.10 & G192 & $\lambda$ Orionis & 05h32m19.4 & +12d49m40.92 & 9.5$\pm$4.0 & 44$\pm$15 & 0 & \nodata \\
G196.92--10.37--A & G196A & $\lambda$ Orionis & 05h44m29.2 & +09d08m52.18 & \nodata & \nodata & (0) & \nodata \\
G203.21--11.20W2 & G203W2 & Orion B & 05h53m39.5 & +03d22m23.89 & 0.5$\pm$0.3 & 15$\pm$5 & (0) & \nodata \\
G205.46--14.56S1--A & G205S1A & Orion B & 05h46m07.2 & -00d13m30.24 & 22.0$\pm$8.0 & 44$\pm$19 & 0 & HOPS--358 \\
G206.93--16.61W2 & G206W2 & Orion B & 05h41m24.9 & -02d18m06.75 & 6.3$\pm$3.0 & 31$\pm$10 & 0 & HOPS--399 \\
G208.68--19.20N1 & G208N1 & Orion A & 05h35m23.4 & -05d01m30.60 & 36.7$\pm$14.5 & 38$\pm$13 & 0 & HOPS--87 \\
G208.89--20.04E & G208E & Orion A & 05h32m48.1 & -05d34m41.46 & 2.2$\pm$1.0 & 108$\pm$25 & (I) & \nodata \\
G209.55--19.68N1--B & G209N1B & Orion A & 05h35m08.6 & -05d55m54.67 & 9.0$\pm$3.7 & 47$\pm$13 & 0 & HOPS--12 \\
G209.55--19.68S1 & G209S1 & Orion A & 05h35m13.4 & -05d57m57.89 & 9.1$\pm$3.6 & 50$\pm$15 & 0 & HOPS--11 \\
G210.49--19.79W--A & G210WA & Orion A & 05h36m18.9 & -06d45m23.55 & 60.0$\pm$24.0 & 51$\pm$20 & 0 & HOPS--168 \\
G211.47--19.27S & G211S & Orion A & 05h39m56.0 & -07d30m27.62 & 180.0$\pm$70.0 & 49$\pm$21 & 0 & HOPS--288 \\
\enddata
\tablecomments{
$\mathrm{\alpha_{J2000}}$ and $\mathrm{\delta_{J2000}}$ are the right ascension and declination, respectively, of the peak position in our combined 1.3~mm continuum observations.
\Tbol~and \Lbol are bolometric temperature and bolometric luminosity adapted from \citet{2020Dutta_ALMASOP}.
The classes of the sources without photometric data at 70 $\sim$ 100 \micron~in \citet{2020Dutta_ALMASOP} are marked in parentheses.
For the four sources which are not HOPS objects: G192, G196A, and G203W2 are not in the spatial coverage of the HOPS observations; and G208E does not have 24 \micron~photometric data \citep{2012Megeath_MGM2012, 2012Fischer_HOPS}.
}
\tablerefs{
\Lbol, \Tbol, and Class: \citet{2020Dutta_ALMASOP}; 
HOPS:~\cite{2016Furlan_HOPS_SED} 
}
\end{deluxetable*}

%% file: sec_SED.tex
\section{Spectral Energy Distribution (SED) Modeling}
\label{sec:SED}
Aiming to infer the physical structure and the potential origin of COM emissions in our hot corino sources, we carried out spectral energy distribution (SED) modeling and obtain the YSO models which may best describe the sources.
At the protostellar stage, the short-wavelength radiation emitted by the central protostar and the accretion shock are absorbed and re-emitted by the dust in the disk and envelope with its SED peaking at $\sim$ 100 \micron~ \citep{2016Furlan_HOPS_SED}.

\subsection{Data \label{sec:sed_data}}

The archival SED data of the 11 sources collected by \citet{2020Dutta_ALMASOP} cover from 3.6 \micron~to 870 \micron.
These data were extracted from the UKIRT Infrared Deep Sky Survey \citep[UKIDSS, ][]{2007Lawrence_SED_UKIDSS}, the Spitzer Space Telescope Survey of Orion A-B \citep{2012Megeath_MGM2012}, the Wide-field Infrared Survey Explorer \citep[WISE, ][]{2010Wright_SED_WISE}, the AKARI/IRC All--Sky Survey Point Source Catalogue, the AKARI/FIS All--Sky Survey Bright Source Catalogue \citep[AKARIPSC and AKARIBSC, ][]{2010Ishihara_SED_AKARI_IRC, 2010Yamamura_SED_AKARI_FIS}, the Herschel Orion Protostellar Survey \citep[HOPS, ][]{2013Stutz_HOPS_APEX, 2015Tobin_SED_HOPS}, the Atacama Pathfinder Experiment \citep[APEX, ][]{2013Stutz_HOPS_APEX}, and the 850 \micron\ JCMT/SCUBA2 observations by \citet{2018Yi_PGCC_Orion}.
All the flux data can be found in Table 6 of \citet{2020Dutta_ALMASOP} and shown in Fig. \ref{fig:sed_best}.
See Table \ref{tab:sed_info} for more information of the archival SED data points including their representative wavelengths and apertures.


On top of these SED data points, we append two data points at 1.3~mm based on the continuum images made with the ALMASOP ACA--only data and the combined data (TM1+TM2+ACA).
We extracted the flux density at the emission peak on the 1.3~mm continuum images within an aperture which is the same as the beam size (Table \ref{tab:continuum}).

\subsection{Tools and Methods}
\label{sec:SED_Tool_Method}
We employed the SED Fitter \footnote{https://sedfitter.readthedocs.io/en/stable/}\citep{2007Robitaille_sedfitter} for deriving the YSO models best matching with the observed SED data points.
In the SED Fitter, users have to provide a grid of SED models, a range of foreground extinction ($A_V$), a range of source distance ($D$), a list of observed SED data points (flux and the corresponding uncertainty), and the filter response and the aperture of the observation for each SED data point.
The SED Fitter will evaluate the $\chi^2$ from the observed flux and the modeled flux under a combination of a SED grid model, $A_V$, and $D$, and list the best combinations of parameters with the minimum $\chi^2$ values.
In our study, the range of $A_V$ was set to [0, 250] mag and the distance $D$ was fixed to be 398~pc, 404~pc, and 404~pc, respectively, for sources in the Orion A, Orion B, and $\lambda$ Orionis clouds \citep{2018Kounkel_Orion_distance}.

We adopted the model grid published by \citet{2006Robitaille_sedfitter} (hereafter ``R06 grid''), which contains the SEDs of 200,000 axis-symmetric 2-D YSO models covering a wide range of stellar masses (from 0.1 to 50 \Msun) and evolutionary ages (from $10^3$ to $10^7$ years).
The SED of each YSO model was produced by the HO-CHUNK package\footnote{https://gemelli.colorado.edu/~bwhitney/codes/codes.html} \citep{2003Whitney_SED}, a Monte Carlo code simulating the radiation transfer of YSO models.
The physical structural setup of the YSO model in the HO-CHUNK package includes a central star, an accretion disk, an envelope, and an optional outflow cavity, and each of them has its corresponding parameters.

In the R06 grid there are 14 model parameters, including stellar mass (\Mstar), stellar radius (\Rstar), stellar temperature (\Tstar), envelope accretion rate (\MdotE), envelope outer radius (\RMAXE), cavity density (\rhocav), cavity opening angle ($\theta$), disk mass (\Mdisk), disk outer radius (\RMAXD), inner radius of disk and envelope (\RMIND), disk accretion rate (\MdotD), disk scale height factor (\ZFAC), disk flaring power component ($B$), and ambient density (\rhoamb).
\citet{2006Robitaille_sedfitter} further introduced an additional parameter into the YSO model, the stellar age (\tstar).
While some of these parameters (such as stellar age) are uniformly sampled within their given ranges, there are also parameters which correlate with each other (such as stellar mass and stellar radius).
For each set of these parameters, the R06 grid model SEDs are evaluated at 10 different inclination values (\incl) equally ranging from 0.05 to 0.95 and 50 different apertures ranging from 100 to 100,000 au.
See \citet{2003Whitney_SED}, \citet{2006Robitaille_sedfitter}, \citet{2007Robitaille_sedfitter}, and the instruction file of the HO-CHUNK for more information.

In addition to the photometric data points,
the continuum and molecular line images obtained with ALMASOP readily yield constraints to the acceptable model.
Here we briefly introduce these constraints.
First, the disk size (disk outer radius) should be less than 140 au due to the fact that there were no clear disk signatures in the C$^{18}$O $J=2-1$ moment--0 images for all the hot corino sources (except G192). 
Second, all the sources (except G209N1B) have a clear bipolar outflow in their CO $J=2-1$ moment--0 images.
The effective opening angle ($\Theta$) of a cavity can be described by the true opening angle ($\theta$) and the inclination angle ($\varphi$, where $\varphi = 0$ corresponds to an edge-on viewing down the longitudinal axis):
\begin{equation}
\tan\Theta = \frac{\tan\theta}{\cos(90^\circ-\varphi)}
\end{equation}
The observed effective opening angle of the outflow ($\Theta^\mathrm{obs}$) therefore helps constraining the cavity opening angle and the inclination of the YSO model (See Fig. \ref{fig:mom0_srcs_cavity} and Table \ref{tab:sed_cavity}).
Appendix~\ref{appx:sed} elaborates in more detail the methods. 
We applied the SED fitting analysis only to eight sources, including G192, G205S1A, G206W2, G208N1, G209N1B, 209S1, G210WA, and G211S, as the other three targets lack photometric data at [70 --- 100] \micron\ band, the critical wavelengths for constraining the total luminosity.


\subsection{Results}
We present in Table \ref{tab:sed_fit_top5} the five best-fit YSO models exported by the SED Fitter for each source and examine the common features among these best models for each source:

\begin{itemize}
    \item \textit{Central Protostar}: 
    Despite the R06 grid having a wide range of stellar age ($10^3$ to $10^7$ years), the best-fit $t_\star$ of the sources are all less or comparable to $10^5$ years.
    This general youth is expected since all the sources are at their early evolutionary stages based on our original sample selection criteria.
    The other three stellar parameters, namely $T_\star$, $R_\star$, and $M_\star$, are generally similar among the five best-fit models for each source.
    In the R06 grid, the luminosity is primarily contributed by the stellar radiation (rather than by the disk accretion), which is likely to constrain "uniquely" the stellar parameters.

    \item \textit{Disk}: 
    Although the disk sizes (disk outer radius or \RMAXD) are more diverse than the stellar parameters between the five best-fit models for each source (e.g., 11 au $<$\RMAXD$<$ 127 au for G210WA), all of the disk sizes are below or comparable to the given maximum outer disk radius, 140 au, which is constrained by the lack of velocity gradients in the C$^{18}$O $J=2-1$ images in most sources, and are significantly smaller than the sizes deduced by \citet{2016Furlan_HOPS_SED} (except for G210WA).
    The ALMASOP photometric measurements of the 1.3~mm continuum, which traces cool dust at hundred-au scale, likely have helped in constraining the disk sizes.
    
    
    \item \textit{Envelope and cavity}: 
    In spite of the potential large degeneracy in the cavity opening angle in combination with the wide range of the inclination, the inferred cavity opening angle appears generally constrained for each source (e.g., $2.51^\circ<\theta<8.77^\circ$ in G206W2 and $9^\circ<\theta<11^\circ$ in G209S1).

\end{itemize}

\subsection{Warm Region Sizes in the YSO Models}
We further reconstruct the physical model using the 2008 version of the HO-CHUNK code with the best-fit parameters obtained from the SED Fitter despite a few limitations noted below.
These limitations are: (1) the version of the HO-CHUNK code for making the R06 grid was released in 2006 but is no longer available; (2) part of the HO-CHUNK R06 model grid parameters, such as the magnetosphere co-rotation radius, are not advertised, and (3) some input models (e.g. the grain dust model and the stellar photosphere model) used by HO-CHUNK are not directly provided.
More details about these limitations and how we mitigate them are given in Appendix~\ref{appx:sed}.

\begin{figure}[htb!]
    \gridline{\fig{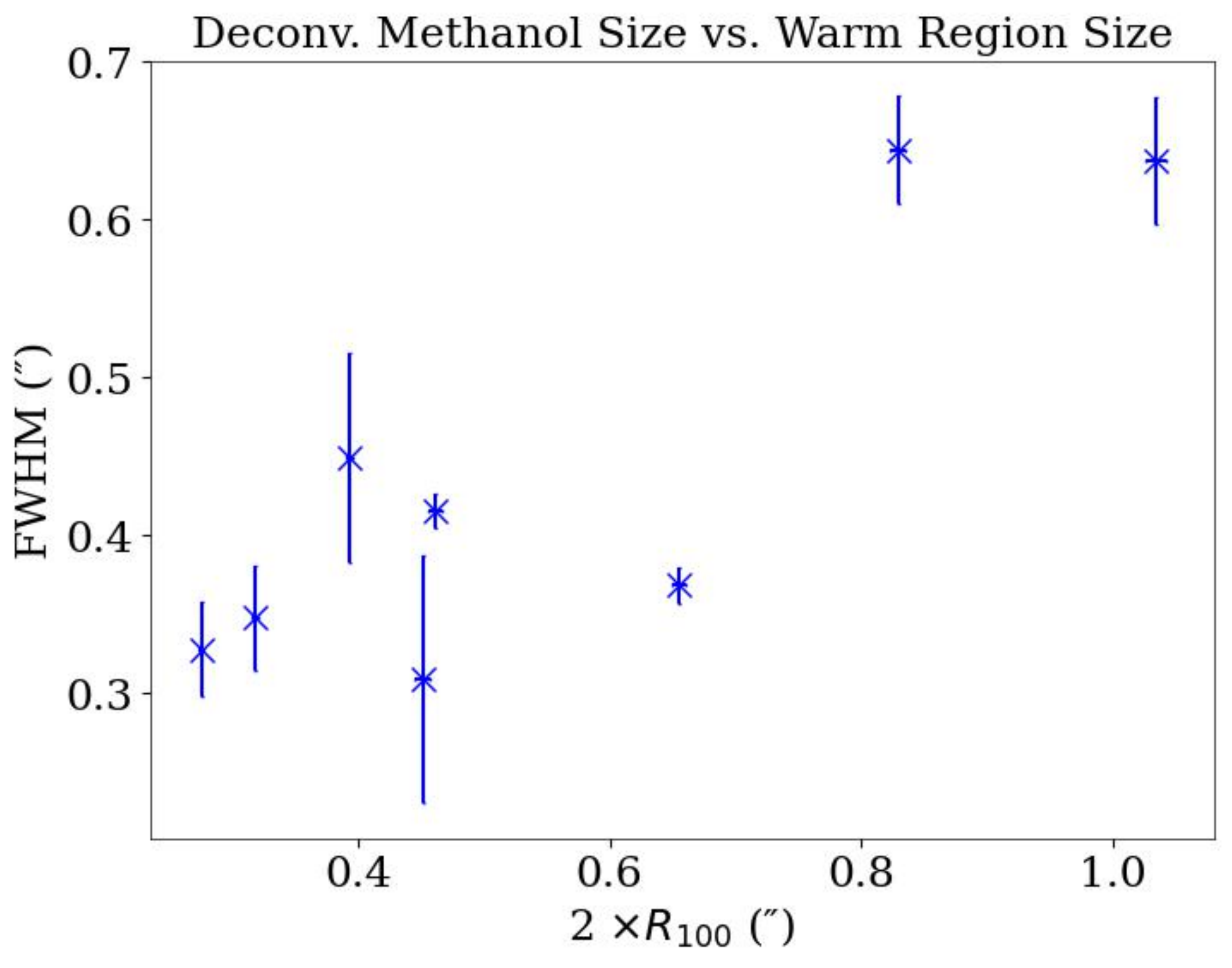}{0.45\textwidth}{}}
\caption{\label{fig:FWHM_2R100} 
The deconvolved size of methanol versus the diameter of the 100--K boundary.
The methanol transition is the CH$_3$OH--46 transition, which is the strongest transition in this study.}
\end{figure}

While the inferred physical (thermal) structures of the sources are model dependent and the real situation may be far more complicated than what has been assumed in the SED models, the results from our analysis are nevertheless indicative.
Fig. \ref{fig:FWHM_2R100} plots the deconvolved source size of the CH$_3$OH--46 transition versus the warm region size ($>$ 100 K, the typical temperature of a hot corino) inferred from the SED fitting.
{There exists a positive correlation between the two quantities with a ratio on the order of unity, indicating that the YSO model exported by the SED fitting is reflecting a coherent trend seen in the real observations.}
In addition, the sizes of the warm region are larger than the inferred disk sizes, hinting that the warm regions, where COM emissions originate, extend to the inner envelopes.


%% file: disc_class.tex
\subsection{SED and Classification of Hot Corino}
\label{sec:disc_class}

\begin{figure}[htb!]
    \gridline{\fig{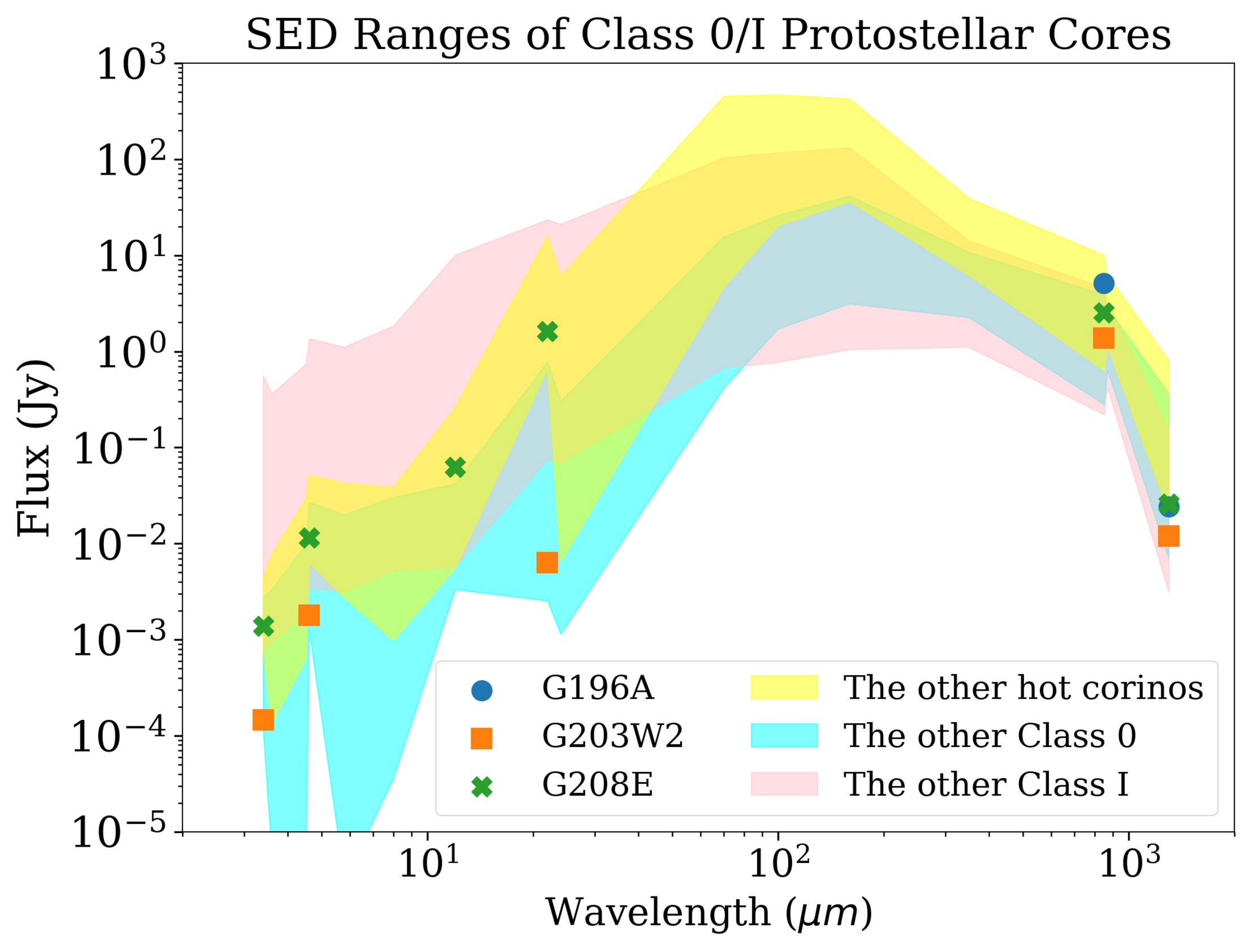}{0.45\textwidth}{}}
\caption{\label{fig:SEDRange} SED ranges of the Class 0/I protostellar cores in ALMASOP.
The pink and blue shaded areas represent the Class I and Class 0 protostellar cores which are not hot corinos, respectively.
The yellow shaded area represents the hot corinos having [70 --- 160] \micron~photometric data.
The scatters show the photometric data points of the three hot corinos lacking [70 --- 160] \micron~photometric data.
The shaded areas are made by the interpolation based on the 15 wavelengths, where most of the sources have photometric data at.
The photometric data are adopted from WISE, \Spitzer~IRAC, \Spitzer~MIPS, \Herschel~PACS, APEX, JCMT/SCUBA--2, and ALMASOP.
Sources with data at less than 10 of the 15 wavelengths are not used.
See Table \ref{tab:sed_info} for the references.
}
\end{figure}

We report toward our ALMASOP sample the detection of 11 hot corinos, among which 10 were classified as Class~0 and one was classified as Class~I by \citet{2020Dutta_ALMASOP} based on their bolometric temperatures. 
G208E, the only Class~I hot corino, appears special, as there have been so far only a very limited number  of Class~I hot corinos reported, namely SVS 13A \citep{2019Bianchi_SVS13A}, Ser--emb 17 \citep{2019Bergner_Ser-emb_COM}, and L1551 IRS5 \citep{2020Bianchi_L1551-IRS5}. 
We note, however, that the classification of G208E may not be secure.
Fig. \ref{fig:SEDRange} displays the range of SEDs for the Class~0 sources with hot corino detection, without hot corino detection, and the Class~I sources in yellow, blue, and pink, respectively.
The fact that G208E, marked by green asterisks, lacks the photometric data at [70 --- 160] \micron~band where the SED of a (Class~0) protostellar core typically peaks, makes its \Tbol\ uncertain.
In fact, the photometric fluxes of G208E in the detected bands are all well within the range spanned by the other Class~0 hot corino sources.
The tentative steep slope at near--IR (12 and 22 \micron) also makes G208E more similar to a Class~0 source. 
As a result, there may not be any Class I hot corino among the ALMASOP sample.
As discussed in Sect. \ref{sec:disc_rate}, the COM emission likely originates from the warm ($> 100$ K) inner envelopes,  the lack of COM emission in Class~I YSOs may be linked to their tenuous envelope remaining around the central YSOs. 
Future (SED) modeling of the Class I source in our sample may help verify this speculation.

There are two other hot corino sources, G196A and G203W2, with uncertain bolometric luminosity and bolometric temperature due to their lack of [70 $\sim$ 160]~\micron\ photometric data points.
\citet{2020Dutta_ALMASOP} previously has reported that both G196A and G203W2 are Class~0 protostellar cores. 
Since G196A has only one photometric data point in addition to the ALMASOP band, \citet{2020Dutta_ALMASOP} did not report its \Tbol\ and \Lbol .
G203W2 was reported with a low bolometric temperature \citep[\Tbol~$=15$ K,][]{2020Dutta_ALMASOP}.
However, its SED curve is close to the lower boundary of the hot corino range and there appears no strong evidence for an exceptionally low \Tbol\ (Fig. \ref{fig:SEDRange}). 

The lack of photometric data points at [70 --- 160] \micron\ also restricts the derivation of \Lbol\ in addition to \Tbol for G203W2 and G208E.
As shown in Fig. \ref{fig:SEDRange}, the SED of Class 0/I protostellar cores typically peak at [70 --- 160] \micron\ band.
The flux at [70 --- 160] \micron\ band can be orders of magnitude stronger than the other bands.
Therefore, the reported \Lbol\ values of these sources (G203W2 and G208E) may be severely underestimated and represent the lower limits.

%% file: disc_rate.tex
\subsection{Detection Rate of Hot Corinos}
\label{sec:disc_rate}

\subsubsection{Luminosity and Warm Region in Hot Corino}
\label{sec:disc_rate_luminosity}

\begin{figure}[htb!]
    \gridline{\fig{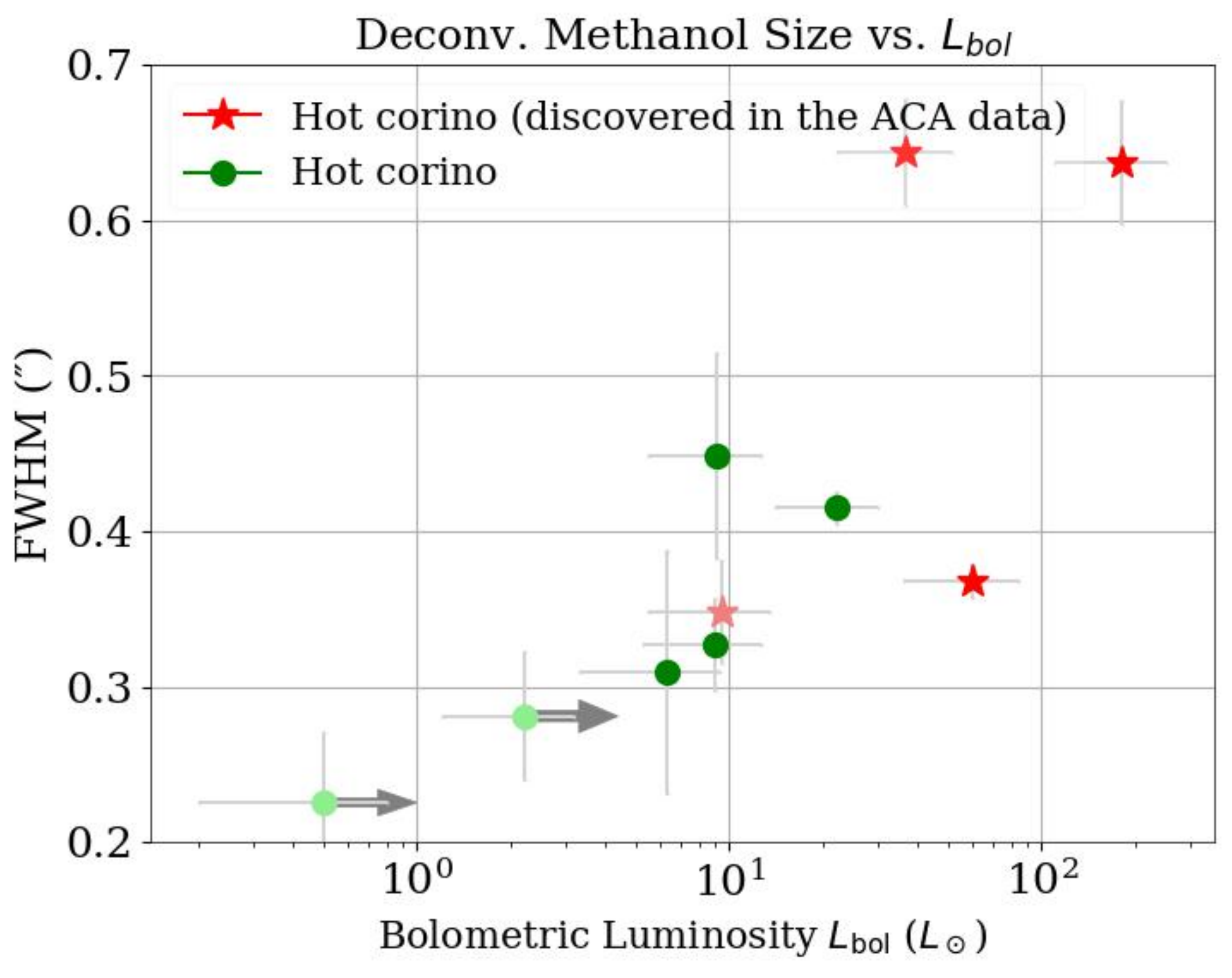}{0.45\textwidth}{}}
\caption{\label{fig:FWHM_Lbol} 
The deconvolved size of the CH$_3$OH--46K transition vs. the bolometric luminosity (\Lbol) adapted from \citet{2020Dutta_ALMASOP} of the hot corino sources discovered in the ACA--only data \citep{Hsu2020_ALMASOP} and the combined data (this study).
Sources without photometric data points at 100 \micron\ are marked in light colors.
The arrows in x-axis mark the sources with \Lbol~lower limits due to their lacking of [70 --- 160] \micron~photometric data points.
G196A is excluded from the plot due to its lack of photometric data in the IR band.
}
\end{figure}

\begin{figure}[htb!]
    \gridline{\fig{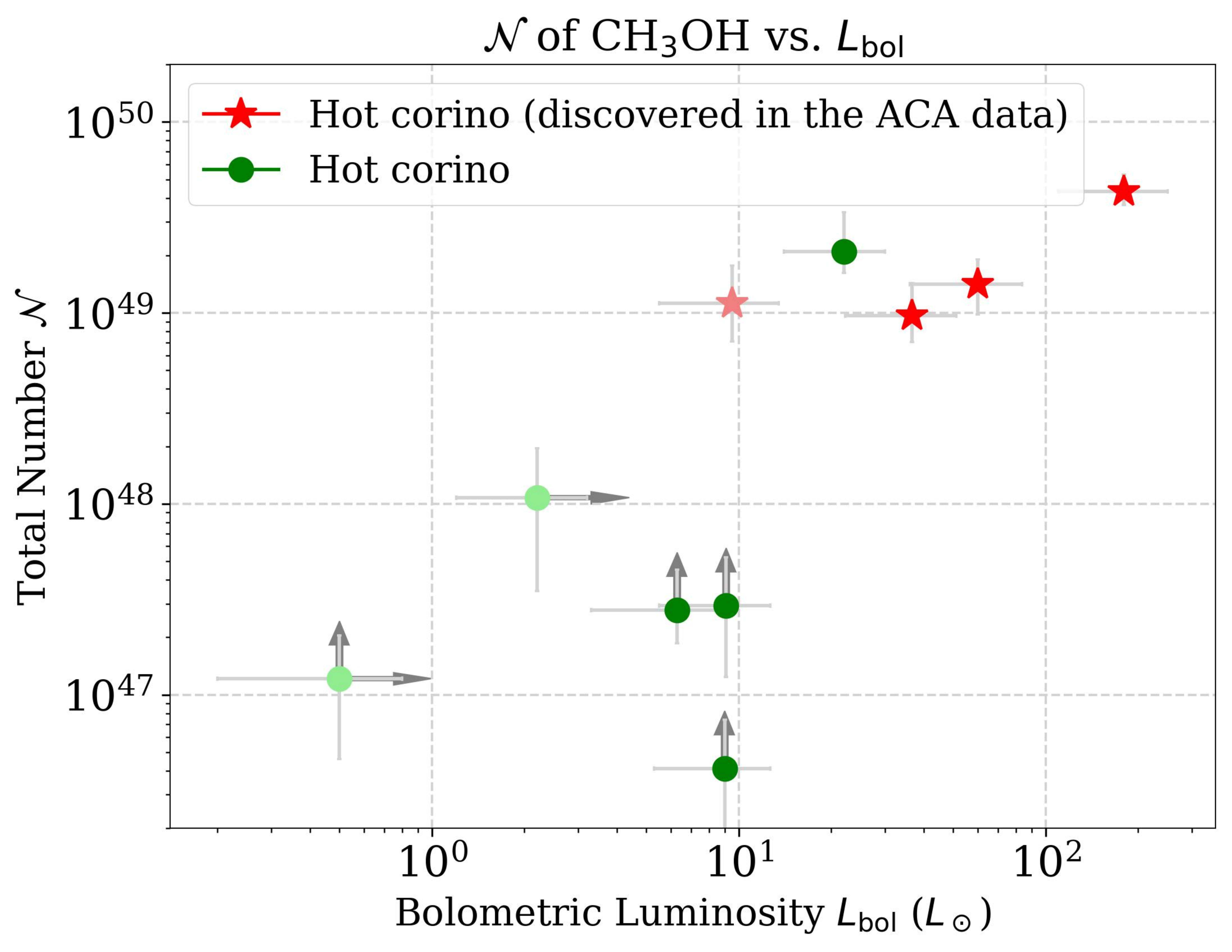}{0.45\textwidth}{}}
\caption{\label{fig:Number_Lbol} 
The total number of methanol within the source size versus bolometric luminosity (\Lbol) adapted from  \citet{2020Dutta_ALMASOP} of the hot corino sources discovered in the ACA--only data \citep{Hsu2020_ALMASOP} and the combined data (this study).
Sources without photometric data points at 100 \micron\ are marked in light colors.
The arrows in x-axis mark the sources with \Lbol~lower limits due to their lacking of [70 --- 160] \micron\ photometric data points.
The methanol column densities are derived by [$^{13}$CH$_3$OH]$\times$\CCIsoOrion.
For the sources without $^{13}$CH$_3$OH, we use [CH$_3$OH] and mark them by the arrows in y-axis.
G196A is excluded from the plot due to its lack of photometric data in the IR band.
}
\end{figure}

In the contemporary paradigm of hot corinos, the COMs frozen in grain mantles get thermally desorbed due to ice sublimation at $\sim$ 100 K \citep[see for example, ][]{2006Garrod_3phase, 2008Garrod_3phase, 2009Herbst_COM_review}.
One may expect that a protostellar core with a higher luminosity may warm up a broader region, which leads to a wider extent of COMs.
Fig. \ref{fig:FWHM_Lbol} displays the bolometric luminosity (\Lbol) versus the deconvolved size of the CH$_3$OH--46K transition.
The extent of methanol, the simplest COM, indeed correlates positively with the bolometric luminosity.

Simulations have suggested that methanol mainly forms during the prestellar phase \citep{2015Drozdovskaya_prestellar_CH3OH, 2020Coutens_CH3CN_CH3OH}.
If the fractional abundance of methanol is comparable among the hot corinos, one would also expect that the total number of gas-phase methanol positively correlates with the bolometric luminosity based on the correlation between the extent of methanol and the bolometric luminosity.
We derive the total number of molecule ($\mathcal{N}$) within the emission extent with: $\mathcal{N} = N \times (\theta_S \times D)^{2}$, where $N$ is the column density, $\theta_S$ is the source size, and $D$ is the distance.
Fig. \ref{fig:Number_Lbol} presents the total number of methanol derived from the (rescaled) column densities and the bolometric luminosities of the hot corino sources.
We use $^{13}$CH$_3$OH with \CCIso\ $=$ \CCIsoOrion\ \citep{2018Kahane_13C12C_Orion} to derive the methanol column density, given the main isotopologue (CH$_3$OH) is likely optically thick
(see Sect. \ref{sec:disc_mol_iso} for more explanations).
The total number of methanol displays a positive correlation with the bolometric luminosity.
The three sources (G203W2, G208E, and G209N1B) with lower limits in \Lbol\ and/or \Ntot\ (illustrated by arrows in Fig. \ref{fig:Number_Lbol}) are consistent with the trend as well.
\subsubsection{Sensitivity and Hot Corino Detection}
\label{sec:disc_rate_sensitivity}

\begin{figure}[htb!]
     \gridline{\fig{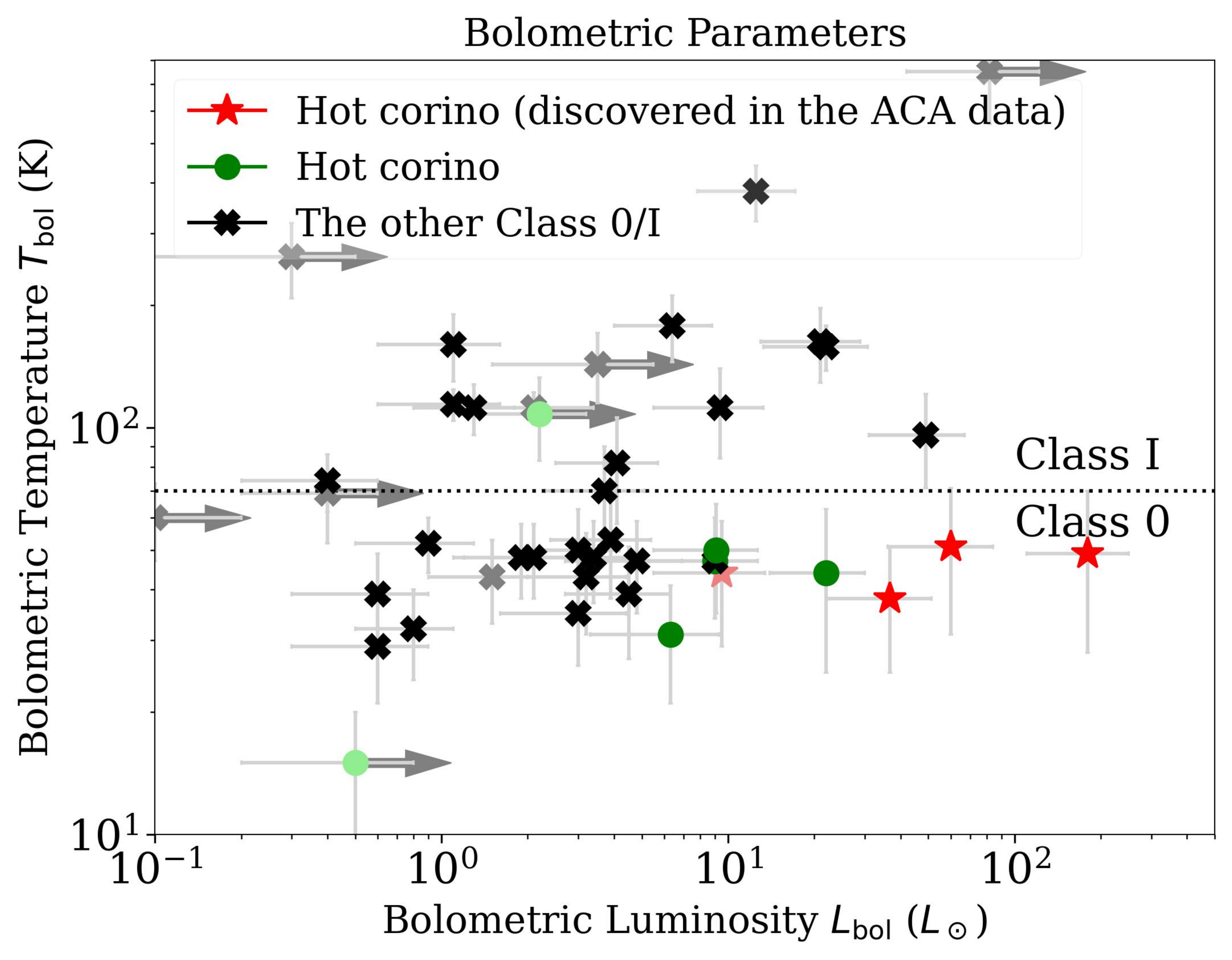}{0.45\textwidth}{}}
\caption{\label{fig:bol_ALMASOP} Bolometric temperature (\Tbol) versus bolometric luminosity (\Lbol) adapted from \citet{2020Dutta_ALMASOP} of protostellar cores.
The sample includes hot corino sources discovered in the ACA--only data \citep{Hsu2020_ALMASOP} and the combined data (this study) and the other Class 0/I protostellar cores imaged by ALMASOP.
Sources without photometric data points at 100 \micron~are illustrated by light colors.
The markers with arrows represent the lower limits of the sources lacking [70 --- 160] \micron~photometric data points.
G196A is excluded due to the lack of photometric data in IR band.
}
\end{figure}

We further plot the bolometric luminosity (\Lbol) and the bolometric temperature (\Tbol) adapted from \citet{2020Dutta_ALMASOP} in Fig.~\ref{fig:bol_ALMASOP}.
The bolometric luminosities of Class 0 hot corino sources are relatively high compared to the other Class 0 protostellar cores at similar bolometric temperature.
The two sources G203W2 and G208E with lower limits in their bolometric luminosity are consistent with this trend as well.

In \citet{Hsu2020_ALMASOP}, we reported the detection of four hot corino sources based on the presence of warm methanol with the ACA data of ALMASOP.
These hot corino sources in general have high bolometric luminosities ($\gtrsim$ 30 L$_\odot$; red asterisks in Fig.~\ref{fig:bol_ALMASOP} ).
In this study with the combined data, we detect warm methanol toward seven additional protostellar cores, which have intermediate luminosities ($\sim 10 - 30$ L$_\odot$; green dots in Fig. \ref{fig:bol_ALMASOP}).
The increase of detected hot corino sources may be mainly attributed to the sensitivity of the observations.
Since \citet{Hsu2020_ALMASOP} and this study use the same set of methanol transitions for detection, we can simply compare their sensitivities by the threshold in brightness temperature and the beam filling factor ($\eta$) defined by the FWHM of the source size ($\theta_\mathrm{S}$) and the HPBW of the synthesized beam ($\theta_\mathrm{A}$) via:

\begin{equation}
    \eta = \frac{\theta_\mathrm{S}^2}{\theta_\mathrm{A}^2+\theta_\mathrm{S}^2}
\end{equation}

The extent of the methanol component in ACA data is unresolved so the source size in ACA data is assumed to be as compact as that in the combined data.
The current threshold in brightness temperature is about 40 times higher than that of \citet{Hsu2020_ALMASOP}, which requires a higher molecular abundance for the detection.
In contrast, the beam filling factor of this study is about 300 times higher than that of \citet{Hsu2020_ALMASOP}, leading to much less beam dilution and much ease for the detection.
The combined effect results in an increased detection rate of warm methanol from 8\% in \citet{Hsu2020_ALMASOP} to 20\% in this work.

Based on the above premise, the number of detected hot corino sources and the detection rates may simply reflect the line sensitivity (for detecting COM emission) and its level as compared with the continuum sensitivity (for detecting protostellar cores) of the observations.
For the other Class 0 protostellar cores which have relatively low luminosities ($\sim 1 - 10$ L$_\odot$; blue crosses), their warm regions may be relatively small and hence their COM emission resulting from thermal desorption would be just too weak to be detected even with our combined data.
It is possible that we may find low--luminosity sources harboring COMs when observed with higher sensitivity observations.
An example is B335 \citep{2016Imai_B335}, a recently reported hot corino source with a relatively low luminosity \citep[$\sim$ 0.7 \Lsun\ at 100 pc, ][]{2015Evans_B335_luminosity} as compared to our hot corino sample.
Note that its bolometric luminosity will be rescaled to 1.9 \Lsun\ if the distance reported by \citet{2020Watson_B335_distance}, 164.5 pc, is adopted.
With deeper observations, there are indeed COM emissions detected in B335 \citep{2016Imai_B335}.

\input{tab_rate}

\subsubsection{Comparisons between ALMASOP and PEACHES}
\label{sec:disc_rate_PEACEHS}

We further compare the detection rate of warm methanol in ALMASOP with that found by PEACHES, which is a census of COMs conducted with ALMA toward embedded protostars in the Perseus cloud by \citet{2021Yang_PEACHES}.
As indicated in Table \ref{tab:COMs_rate}, the apparent detection rate of warm methanol in PEACHES is roughly 2.5 times higher than that in ALMASOP.
Here we discuss two factors affecting the detection rate of warm methanol: 

\begin{itemize}
    \item \textit{Methanol line detection criteria}:
    Since ALMASOP has a similar sensitivity to that of PEACHES, which is $\sim$ 0.5 K, the signal--to--noise ratio (SNR) threshold for a line detection in ALMASOP (5~$\sigma$) is higher than that (3~$\sigma$) in PEACHES.
    A weak methanol detection in PEACHES may be considered as a non-detection in this study.

    \item \textit{Sample selection}:
    The sample selection of PEACHES follows \cite{2018Higuchi_survey_Perseus}, which excludes YSOs with $L_\mathrm{bol}<1~L_\odot$.
    In ALMASOP, the hot corinos are the Class 0 protostellar cores with high luminosity. Applying the same sample selection method, the total number of protostellar cores in ALMASOP will decrease and the detection rate of hot corinos will increase correspondingly.

\end{itemize}

To clarify the influence of these two factors, we also append the hot corino candidates under the same line detection criteria and apply the same sample selection method as PEACHES. 
We find nine additional Class 0/I protostellar cores where at least one methanol transition is detected at $3\sigma$ (see Table \ref{tab:candCoord} for the list).
Including those ``potential hot corinos'' and leaving out all sources fainter than 1 \Lsun\ in the parent sample for calculating the statistics, the detection rate of hot corinos in ALMASOP becomes 15 out of 37, which remains less than the value (28 out of 50) in PEACHES.
The distance difference between the two clouds may contribute to the small difference in the detection rate.
Overall speaking, the two rates are statistically similar due to the limited sample size.
Observing the same set of methanol (or other COM) transitions will mitigate the problem and enable a fair comparison of the detection rate, which may be indicative of the cloud properties and/or evolutionary stage.


Finally, we note that the dust continuum opacity may prevent the detection of molecular line emissions \citep[e.g., ][]{2019Sahu_IRAS4A1, 2020DeSimone_dust_opacity}.
The detection rates of hot corinos in both ALMASOP and PEACHES are therefore possibly underestimated. 

%% file: tab_rate.tex
\begin{deluxetable}{lccc}
\tablecaption{\label{tab:COMs_rate} {Observational parameters and detection rates}}
\tablehead{
\colhead{} & \multicolumn{2}{c}{ALMASOP} & \colhead{PEACHES} }
\startdata
Cloud & \multicolumn{2}{c}{Orion} & Perseus \\
Distance & \multicolumn{2}{c}{$\sim$400 pc} & $\sim$300 pc \\
Beam size & $\sim$6\farcs{} & $\sim$0\farcs{35} & $\sim$0\farcs{45} \\
Beam size (au) & $\sim$2500 & $\sim$145 & $\sim$135\\
Sensitivity (mJy beam$^{-1}$) & 28 & 2.5 & 6 \\
Sensitivity (K) & 0.02 & 0.5 & 0.5 \\
{Detection SNR} & 3 & 5 & 3 \\
{Threshold (K)} & 0.06 & 2.5 & 1.5 \\
Beam filling factor & 0.0046 & 0.58 & 0.55 \\
\#(Class 0/I) & 48 & 56 (37) & 50\\
\#(CH$_3$OH) & 4 & 11 (15) & 28\\
$P$(CH$_3$OH) & $8\%$ & $20\%$ ($41\%$) & $56\%$\\
\enddata
\tablecomments{
Number of sources under PEACHES sample selection method (\Lbol~$>1~L_\odot$) and line detection criteria (3 $\sigma$) is given in parentheses.
}
\tablerefs{
PEACHES: \citet{2021Yang_PEACHES}; Distance: \citet{2018Kounkel_Orion_distance, 2018Ortiz_Perseus_distance, 2020Zucker_Perseus_distance}
}
\end{deluxetable}


%% file: disc_COMs.tex
\subsection{Complex Organic Molecules in Protostellar Cores}
\label{sec:disc_mol}


\begin{figure*}
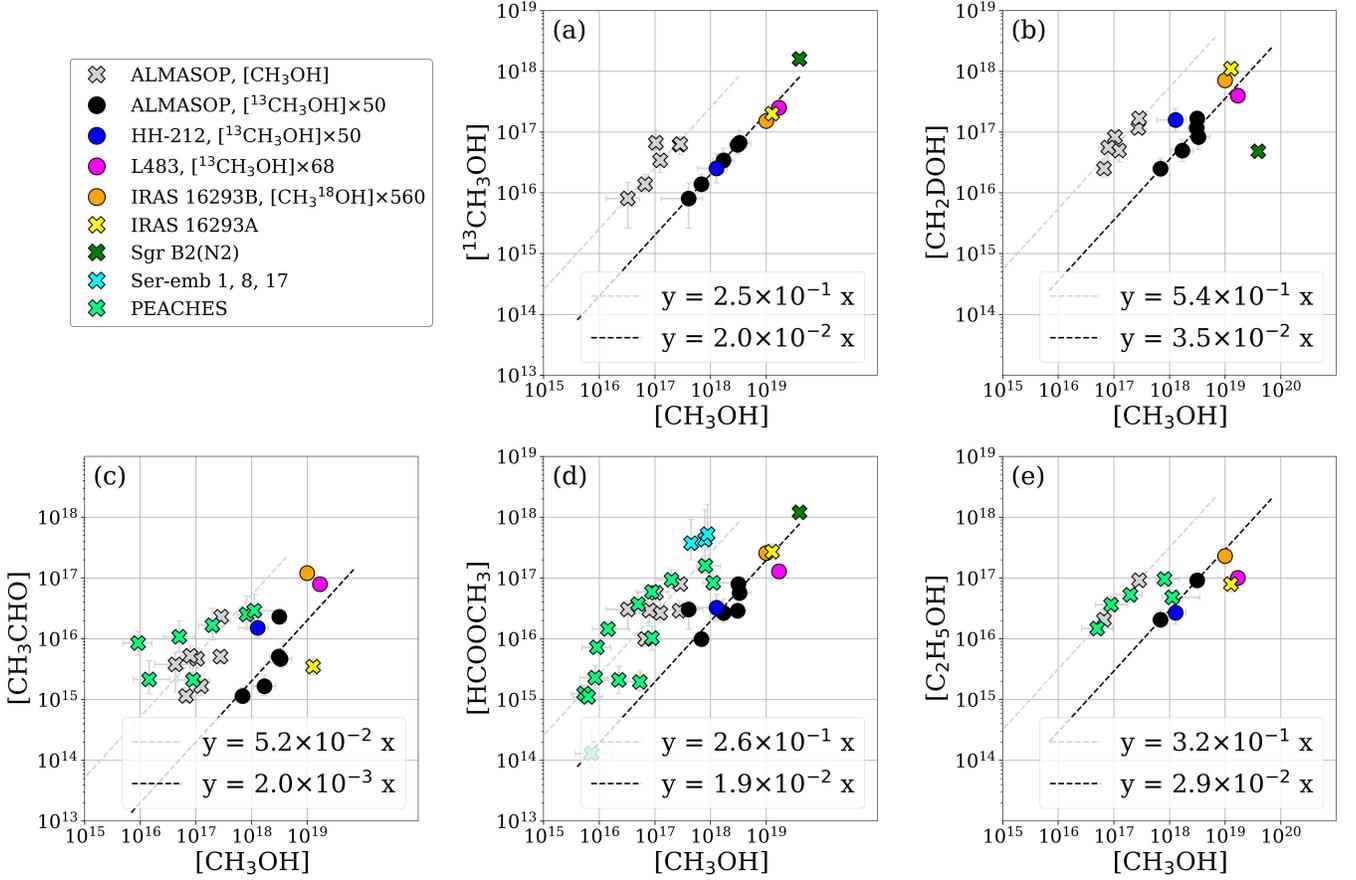

\gridline{\fig{scatter_Ntot_COMs_13.pdf}{0.99\textwidth}{}}
\caption{\label{fig:scatter_Ntot_COM_CH3OH} Selected molecular column densities with respect to methanol column densities.
Note that the circles, namely for HH-212, L483, and IRAS 16293B, denote that the [CH$_3$OH] are derived by the corresponding authors from the rare isotopologues.
Also note that Sgr B2 (N2) is a hot core, which is a massive protostellar core, and its chemistry mechanism may be different from hot corino chemistry.
The black and grey dashed lines are the fit of ALMASOP data (black and grey) by linear proportional function. The units of both x-axis and y-axis are cm$^{-2}$. \\\\
\textbf{References}: HH-212 \citep{2019Lee_HH212}, IRAS 16293A \citep{2020Manigand_IRAS16293-2422-A_COMs}, IRAS 16293 B \citep{2018Jorgensen_IRAS16293B_COM, 2018Persson_IRAS16293B_H2CO}, L483 \citep{2019Jacobsen_L483_COM}, Ser-emb 1, 8 and 17 \citep{2019Bergner_Ser-emb_COM}, Sgr B2(N2) \citep{2016Muller_SgrB2N2_13C12C} and 27 sources in Perseus cloud from PEACHES \citep{2021Yang_PEACHES}}
\end{figure*}



\subsubsection{Isotope Ratios of Methanol: \texorpdfstring{\CCIso}{Lg}, \texorpdfstring{\OOIso}{Lg}, and D/H, }
\label{sec:disc_mol_iso}

In this section we explore the possible correlations between the column densities of COMs in the hot corinos. 
A good estimation of the methanol column density is important for performing the comparison with other COMs.
Fig. \ref{fig:scatter_Ntot_COM_CH3OH}-(a) shows the column densities of CH$_3$OH and $^{13}$CH$_3$OH. The isotope \CCIso\ ratio derived from fitting the observed positive correlation with a linear proportional function is 4, an order of magnitude lower than the nominal local ISM value of $\sim$ 70 \citep{2011Wirstrom_12C13C_70} and $\sim$ 50 in the Orion cloud \citep{2018Kahane_13C12C_Orion}.
The low \CCIso~ratio for our hot corino sources most likely suggests high optical depths of the main CH$_3$OH isotopologue, similar to those reported in the literature (\cite{2013Zapata_IRAS16293B_depth} for IRAS~16293B, \cite{2019Lee_HH212} for HH-212, and \cite{Hsu2020_ALMASOP} for G211S).

A high optical depth of CH$_3$OH may have also led to the low [CH$_3$OH]/[CH$_3^{18}$OH] value of 11.9$^{+3.6}_{-3.5}$ seen in G208N1, which is the only source with CH$_3^{18}$OH detected among the 11 hot corinos.
This [CH$_3$OH]/[CH$_3^{18}$OH] value in G208N1 is significantly lower than $562 \pm 221$ in the hot corino source IRAS 16293A \citep{2020Manigand_IRAS16293-2422-A_COMs}, 181 in the hot core sources Sgr B2(N2) \citep{2016Muller_SgrB2N2_13C12C}, and $560\pm25$ in the local interstellar medium \citep[ISM, ][]{1994Wilson_isotope}.
Considering CH$_3$OH being indeed optically thick and factoring the \CCIso\ ratio ranging between 50 and 77 \citet{1994Wilson_isotope, 2011Wirstrom_12C13C_70, 2018Kahane_13C12C_Orion}, we scale the column density of $^{13}$CH$_3$OH accordingly and find that the \OOIso\ ratio of methanol in G208N1 becomes 127 -- 190.
It appears much closer, though still less than the local ISM value, implying possibly an underestimate of the $^{13}$CH$_3$OH column density.

\citet{2020Bianchi_L1551-IRS5} has reported high opacity in a $^{13}$C substituted methanol transition ($^{13}$CH$_3$OH) at \Eu\ $=$ 48 K toward the Class I hot corino L1551 IRS5.
We consider that the column densities of $^{13}$CH$_3$OH derived in this study are not severely affected by the opacity because, 
first, the [$^{13}$CH$_3$OH] in this study are derived from transitions including those at high upper energies (i.e., \Eu\ $=$ 162 and 254 K) which are less affected by the opacity, and second, the ratio between the column densities of $^{13}$CH$_3$OH and CH$_2$DOH appears to be similar among the hot corino sources in this study and in the literature (Fig. \ref{fig:scatter_Ntot_COM_CH3OH}).
Furthermore, the hot corinos in this study are Class 0, which are different from the case in \citet{2020Bianchi_L1551-IRS5} (see Sect. \ref{sec:disc_class}).

We also detect deuterated methanol (CH$_2$DOH), another isotopologue of methanol, in multiple hot corino source and Fig. \ref{fig:scatter_Ntot_COM_CH3OH}-(b) shows their column densities. 
Fig. \ref{fig:scatter_Ntot_COM_CH3OH} also shows the data points toward other sources, including hot corino sources HH-212 \citep{2019Lee_HH212}, L483 \citep{2019Jacobsen_L483_COM}, IRAS 16293-2422 A \citep[hereafter IRAS 16293A ][]{2020Manigand_IRAS16293-2422-A_COMs}, IRAS 16293-2422 B \citep[hereafter IRAS 16293B ][]{2018Jorgensen_IRAS16293B_COM}, and Ser-emb 1, 8, and 17 \citep{2019Bergner_Ser-emb_COM}, and a high-mass hot core source Sgr B2(N2) \citep{2016Muller_SgrB2N2_13C12C, 2016Belloche_SgrB2N2_DH}, all observed with ALMA.
Note that the [CH$_3$OH] are in the case of L483 inferred from [$^{13}$CH$_3$OH] assuming \CCIso $=68$ \citep{2019Jacobsen_L483_COM}, and in the case of IRAS 16293B calculated based on [CH$_3\,^{18}$OH] assuming \OOIso $=560$ \citep{2018Jorgensen_IRAS16293B_COM}
Applying \CCIso~$ = 50$ \citep{2018Kahane_13C12C_Orion}, the D/H ratio of methanol in the ALMASOP hot corinos are overall consistent with the literature.
The D/H ratios in the hot corino sources are an order of magnitude higher than that in Sgr B2(N2), as is found by \citet{2019Taquet_DHratio}.



\subsubsection{Complex Organic Molecules (COMs)}
\label{sec:disc_mol_COMs}

We explore possible correlations between the column densities of CH$_3$OH, CH$_3$CHO, HCOOCH$_3$, and C$_2$H$_5$OH in Fig. \ref{fig:scatter_Ntot_COM_CH3OH} and fit the observed positive correlation trend by a linear proportional function.
Possibly resulting from the opacity of CH$_3$OH, the ratios of [CH$_3$CHO]/[CH$_3$OH], [HCOOCH$_3$]/[CH$_3$OH], and [C$_2$H$_5$OH]/[CH$_3$OH] in our sources (grey squares in Fig. \ref{fig:scatter_Ntot_COM_CH3OH}) are higher than those of the individual sources from the literature derived from the rare isotopologues of methanol rather than the main one.
Assuming \CCIso $=50$ \citep{2018Kahane_13C12C_Orion}, the column density ratios of CH$_3$CHO, HCOOCH$_3$, and C$_2$H$_5$OH with respect to that of (rescaled) CH$_3$OH in our study, which are respectively 0.20\%, 1.9\%, and 2.9\% (black markers in Fig. \ref{fig:scatter_Ntot_COM_CH3OH}), become comparable to the value of the above individual sources.

\citet{2020vanGelder_COMs} compared the relative abundance of O-bearing COMs with respect to CH$_3$OH for star-forming regions in four different clouds (Perseus, Serpens, Ophiuchus, Orion) and concluded that the abundance ratios of most O-bearing COMs (e.g., HCOOCH$_3$) are similar among different clouds while some COMs (e.g., CH$_3$CHO and C$_2$H$_5$OH) are not.
The former may form at an earlier evolutionary stage (i.e., cold prestellar phase) and the latter are more affected by the individual local properties.
We compare our results with the literature by using the other isotopologues of methanol rather than the main one.
The ratio of [HCOOCH$_3$]/[CH$_3$OH] appears to be similar while that of [CH$_3$CHO]/[CH$_3$OH] is different.
This disparity agrees with what \citet{2020vanGelder_COMs} suggested.
The small number (two) of C$_2$H$_5$OH sources in our sample prevents us from examining the diversity in [C$_2$H$_5$OH]/[CH$_3$OH] ratios indicated by \citet{2020vanGelder_COMs}.

Using the main isotopologue of methanol (CH$_3$OH), we also compare our results with the column densities of the corresponding species in the Perseus protostars from PEACHES \citep{2021Yang_PEACHES} and the three Serpens hot corinos from  \citet{2019Bergner_Ser-emb_COM}.
As shown by the grey dashed lines in Fig. \ref{fig:scatter_Ntot_COM_CH3OH}, the column density ratios of CH$_3$CHO, HCOOCH$_3$, and C$_2$H$_5$OH with respect to CH$_3$OH are in general consistent.
This suggests a comparable and non-negligible CH$_3$OH opacity among these sources.


\subsubsection{Pre-biotic Molecule}
\label{sec:disc_mol_NH2COH}

\begin{figure}
\gridline{\fig{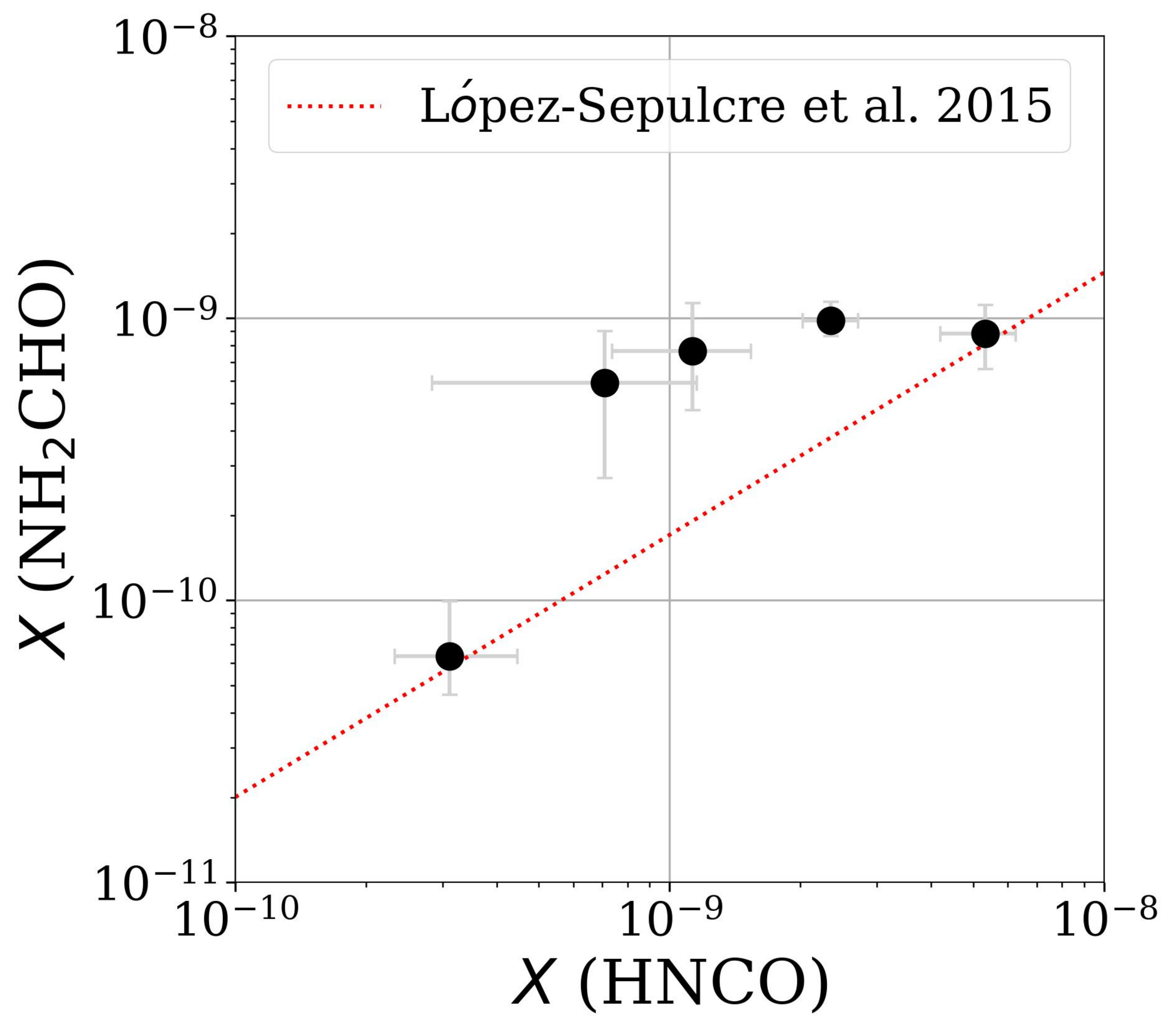}{0.45\textwidth}{}}
\caption{\label{fig:scatter_X_NH2CHO_HNCO} Fractional abundance of NH$_2$CHO vs. that of HNCO. 
The red dashed line represents the empirical correlation inferred by \citet{2015Lopez-Sepulcre_NH2CHO}.
}
\end{figure}

Formamide (NH$_2$CHO) is one of the main components in both (pre)genetic and (pre)metabolic processes, which makes it a potential key species in pre-biotic evolution \citep{2012Saladino_NH2CHO, 2019Lopez-Sepulcre_NH2CHO}. 
There are indications that NH$_2$CHO is chemically related to HNCO.
For example, \citet{2015Lopez-Sepulcre_NH2CHO} found a tight empirical correlation between their fractional abundances, $X($NH$_2$CHO$)= 0.04 \times X(\mathrm{HNCO})^{0.93}$, in star-forming regions with H$_2$CO detection.
On the one hand, \cite{2016Coutens_IRAS19293B_NHDCHO} came across a comparable D/H ratio of NH$_2$CHO and HNCO toward IRAS 16293B, suggesting that NH$_2$CHO and HNCO are chemically related through grain-surface formation.
On the other hand, \citet{2018Quenard_NH2CHO_HNCO} demonstrated that the correlation between HNCO and NH$_2$CHO may originate from their similar response to temperature instead of a chemical link.

Our five sources in which NH$_2$CHO is detected all bear HNCO, too.
Their fractional abundances appear positively correlated, although the NH$_2$CHO fractional abundances appear often lying above the empirical correlation (See Fig. \ref{fig:scatter_X_NH2CHO_HNCO}).
For G211S, the only object where NH$_2$CHO and HNCO were detected both in the ACA data and the combined data, $X$(NH$_2$CHO) is respectively 1.9 and 2.5 times higher than the value predicted by the empirical formula, which was based on observations conducted with the IRAM 30-m single dish telescope at a lower angular resolution. 
The disparity we find in higher angular resolution interferometric observations may hint that the spatial distribution of NH$_2$CHO may be slightly more concentrated at smaller scales.

\subsubsection{CH$_3$CN vs CH$_3$OH}
\label{sec:disc_mol_CH3CN}

\begin{figure}
\plotone{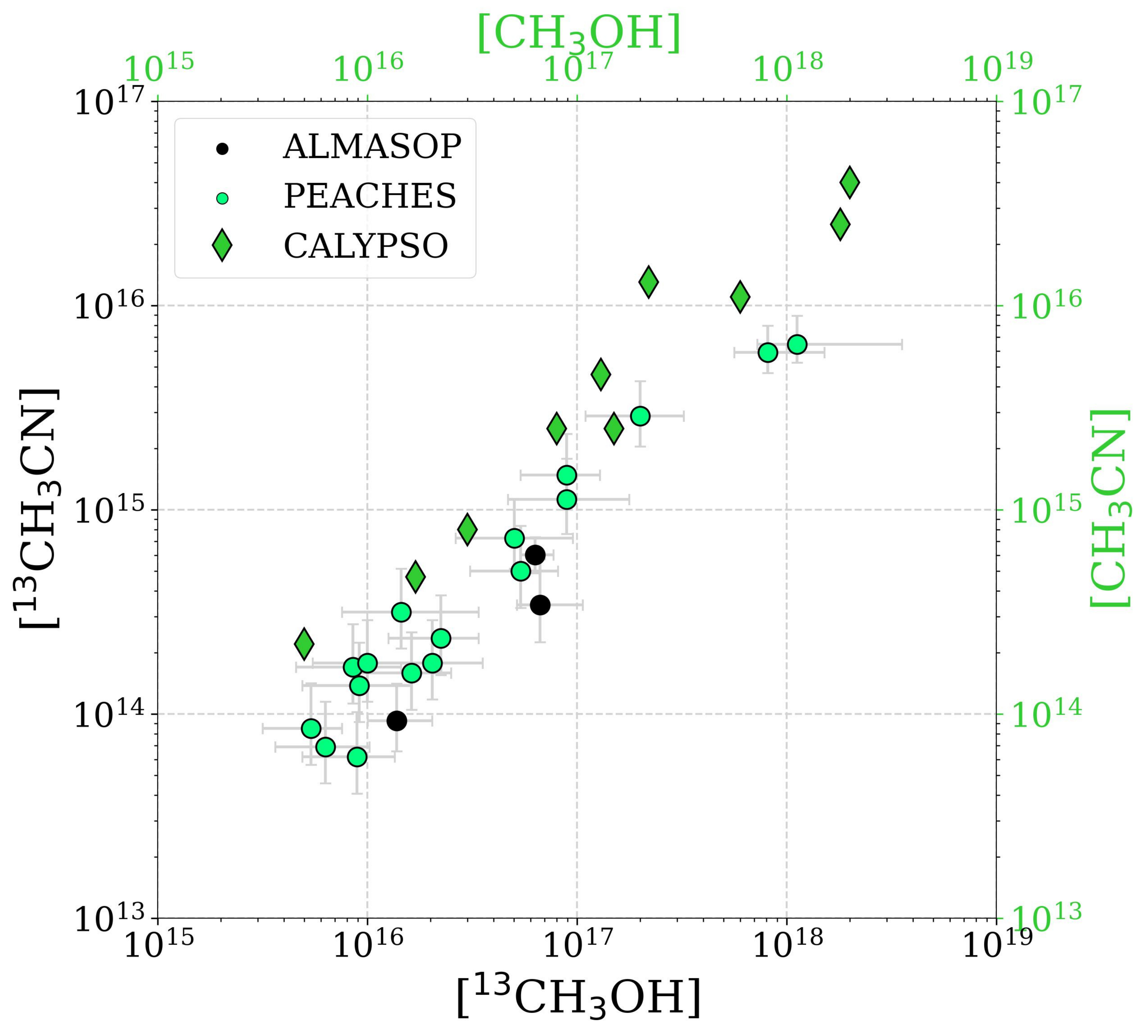}
\caption{\label{fig:scatter_Ntot_CH3CN_CH3OH} Column densities of methyl cyanide versus that of methanol in ALMASOP (black circles), PEACHES \citep[green circles, ][]{2021Yang_PEACHES}, and CALYPSO \citep[green diamonds, ][]{2020Belloche_COM_CALYPSO}.
Note that this study uses $^{13}$CH$_3$CN and $^{13}$CH$_3$OH for ALMASOP and \citet{2021Yang_PEACHES} and \citet{2020Belloche_COM_CALYPSO} use CH$_3$CN and CH$_3$OH for PEACHES and CALYPSO, respectively.
}
\end{figure}

\citet{2021Yang_PEACHES} found a tight abundance correlation between CH$_3$CN and CH$_3$OH in PEACHES.
Both \citet{2017Bergner_COMs} and \citet{2020Belloche_COM_CALYPSO} have shown a similar relation.
Here we examine this abundance correlation among the hot corino sources in ALMASOP.
Since there is no CH$_3$CN transition within our spectral coverage, and to circumvent the potential optical depth issue of CH$_3$OH, we consider the column densities of their $^{13}$C isotopologues, namely $^{13}$CH$_3$CN and $^{13}$CH$_3$OH.
Despite the small number of data points, as shown in Fig. \ref{fig:scatter_Ntot_CH3CN_CH3OH}, the trend of our observational result agrees with \citet{2021Yang_PEACHES}.
Since the CH$_3$OH in PEACHES is likely optically thick (see Sect. \ref{sec:disc_mol_COMs}), this suggests a high opacity of CH$_3$CN as well \citep[IRAS 16293 A and B, ][]{2018Calcutt_CH3CN_opacity} assuming a consistent trend between ALMASOP and PEACHES.

Fig. \ref{fig:scatter_Ntot_CH3CN_CH3OH} also shows the data points in the CALYPSO survey \citep{2020Belloche_COM_CALYPSO}. 
The trends of the three data sets (i.e., this study, PEACHES, and CALYPSO) are similar but with small offsets.
If the correlation between methanol and methyl cyanide is tight and consistent among different molecular clouds, the offsets may hint at an underlying influence of the opacity to the corresponding molecules.

Based on the tight correlation, \citet{2021Yang_PEACHES} speculated that there is either a common abundance ratio between these two molecules on the icy grains or a common elemental abundance ratio between O and N in the Perseus cloud.
As pointed out by \citet{2020Belloche_COM_CALYPSO}, however, CH$_3$CN is unlikely chemically-linked to CH$_3$OH since there is no convincing chemical relation between their formation in star-forming regions.
In addition, \citet{2020Coutens_CH3CN_CH3OH} model the chemical evolution from the prestellar phase to the formation of the disk and show that a significant number of species, including CH$_3$OH and CH$_3$CN, have similar initial and final abundances.
The similarity suggests that these molecules are less sensitive to the history of the physical conditions from the cold core to the disk formation.

%% file: sec_Conclusion.tex
\section{Conclusion}
\label{sec:Conclusions}

\begin{enumerate}
  \item 
  We report the detection of 11 Class 0/I hot corino sources, which harbor compact, warm, and abundant CH$_3$OH as well as other COMs based on the combined data (TM1+TM2+ACA) of ALMASOP. 
  Four of them were previously identified as hot corino sources with the ALMASOP ACA data.

  \item 
  The classification of the only Class I hot corino object in our sample, G208E, may have been based by its high but uncertain bolometric temperature. 
  The reasons are: (1) its lack of photometric data at [70 --- 160] \micron\ bands; (2) its photometric fluxes at other bands are well within the range spanned by Class 0 hot corinos; and (3) the tentative slope of its SED at near-IR band is similar to Class 0 hot corinos.
  All the hot corinos we detect are therefore likely Class 0 protostellar cores.
  
  \item
  We present the YSO models of the hot corino sources inferred from SED fitting.
  The SED data points include the photometric data points at 1.3 mm observed by ALMASOP combined data and ACA data in addition to the archival photometric measurements. 
  The sizes of the warm regions (i.e., $T>100$ K), where the thermal desorption of icy mantles may occur, positively correlate with the observed extent of COMs with a ratio on the order of unity.
  
  \item
  The bolometric luminosity positively correlates with the extent and the total number of gas-phase methanol.
  Both of these correlations are consistent with the thermal desorption paradigm for hot corinos.

  \item 
  The detection rate of hot corinos may reflect the sensitivities of the underlying observations. 
  The combined data of ALMASOP results in an increased detection rate of hot corinos from 8\% in the ACA-only data \citet{Hsu2020_ALMASOP} to 20\% in this work.  
  We may find lower luminosity sources harboring COMs with higher sensitivity observations.
  
  \item 
  The detection rate of warm methanol toward the Orion cloud obtained from this study (ALMASOP) is smaller than that toward the Perseus cloud studied by PEACHES.
  The distance difference between the two clouds may contribute to the difference while the two rates are statistically similar due to the limited sample size.
  Observing the same set of COM transitions would enable a fair comparison between cloud properties.

  \item 
  As is found in much of the literature on hot corinos, the \CCIso\ ratio of methanol indicates the high opacity of the main isotopologue (CH$_3$OH).
  Assuming \CCIso$=50$, the fractional column densities of CH$_3$CHO, HCOOCH$_3$, C$_2$H$_5$OH, and CH$_2$DOH with respect to (rescaled) CH$_3$OH are 0.019, 0.02, 0.029, and 0.035, respectively.
  Together with the data toward the hot corinos, the ratio of  [CH$_3$CHO]/[CH$_3$OH] seems to be more diverse than that of [HCOOCH$_3$]/[CH$_3$OH].
  
  \item 
  The ratio of [$^{13}$CH$_3$CN]/[$^{13}$CH$_3$OH] agrees with the empirical correlation for [CH$_3$CN]/[CH$_3$OH] based on the literature, while there is still no clear chemical connection between these two molecules.
  The three data sets (i.e., ALMASOP, PEACHES and CALYPSO) of methanol and methyl cyanide exhibit similar trends with small offsets.
  Assuming the existence of a tight and consistent correlation between methanol and methyl cyanide among different clouds, the offsets may hint at the influence of opacity to the corresponding molecules.

\end{enumerate}

%% file: appx_source.tex
\subsection{Literature Review}

There are 7 HOPS objects \citep{2016Furlan_HOPS_SED} in the 11 hot corinos.
Six of them are included in the VLA/ALMA Nascent Disk and Multiplicity (VANDAM) Survey of Orion Protostars \citep{2020Tobin_VANDAM-II}.
Four of them are included in the CARMA–NRO Orion Survey \citep{2018Kong_CARMA-NRO_intro,2020Feddersen_CARMA-NRO_orion}. 
Table \ref{tab:survey} lists the surveys and the related sources.

\input{tab_survey}

Not all of the 11 hot corino sources are known or bright YSOs. 
Here we summarize the literature related to 11 sources individually:
\begin{itemize}
    \item \textit{G192.12--11.10 (G192)}: 
    First detected as a hot corino in the ACA data of ALMASOP \citep{Hsu2020_ALMASOP}.
    \citet{2018Yi_PGCC_Orion} has reported it as a starless core.
    Individual studies for this source are rare.
    \item \textit{G196.92--10.37--A (G196A)}: 
    G196A was identified as a YSO candidate in \citet{2008Dunham_G196a_SSTc2d} based on the \textit{Spitzer Space Telescope} Legacy Project, ``From Molecular Cores to Planet Forming Disks'' \citep{2003Ii_SSTc2d}.
    \item \textit{G203.21--11.20W2 (G203W2)}: 
    There are not many studies toward this source, even though it is in the Orion B cloud.
    It is not within the spatial coverage of the HOPS catalog.    
    \item \textit{G205.46--14.56S1--A (G205S1A)}: 
    This source shows apparent CO $J=1-0$, $J=4-3$, and $J=3-2$ outflow which are suspected to contaminate the field of HOPS--401 \citep{2020Nagy_APEXSurvey_outflow}. 
    In addition, G205S1A showed a declining light curve in 850 \micron~for 16 months in the JCMT Transient Survey \citep{2018Mairs_HOPS358_850um}.
    \citet{2021Lee_JCMTTransient}, the four-year summary of the survey, further reported a linearly declining light curve for this source.
    \item \textit{G206.93--16.61W2 (G206W2)}: 
    Similar to G205S1A, the apparent CO $J=1-0$, $J=4-3$, and $J=3-2$ outflow is suspected to contaminate the field of HOPS--272 \citep{2020Nagy_APEXSurvey_outflow}. 
    \item \textit{G208.68--19.20N1 (G208N1)}: 
    First detected as a hot corino in the ACA data of ALMASOP \citep{Hsu2020_ALMASOP}.
    G208N1, coincided with OMC3/MMS6, has a very young outflow indicating that this source is at its very earliest stage of evolution \citep{1997Chini_HOPS87_OMC3-MMS6, 2012Takahashi_HOPS87_outflow, 2019Takahashi_HOPS87_young}.
    \item \textit{G208.89--20.04E (G208E)}: 
    This source is one of the targets in \citet{2012Megeath_MGM2012} but not included in the HOPS catalog due to the lack of 24 \micron~detection.
    \item \textit{G209.55--19.68N1-B (G209N1B)}: 
    CARMA–NRO Orion Survey \citep{2018Kong_CARMA-NRO_intro} reported a clear blue lobe of the CO outflow in G209N1B.
    Although the outflow is not obvious in its CO $J=2-1$ moment 0 image (Fig. \ref{fig:mom0_srcs_cavity}), the large linewidth of the CO $J=2-1$ transition implies the existence of the outflow. 
    \item \textit{G209.55--19.68S1 (G209S1)}: 
    G209S1 has been involved in a survey of companions around YSO conducted with \textit{HST} by  \citet{2016Kounkel_G209S1}, which presented that the companion fraction in the high stellar density region is higher than that in the low stellar density region.
    \item \textit{G210.49--19.79W--A (G210WA)}: 
    First detected as a hot corino in the ACA data of ALMASOP \citep{Hsu2020_ALMASOP}.
    This source is associated to HH 1-–2 VLA 3, which is reported to be in an active state of mass infall and accretion \citep{2010Fischer_HOPS168_Infall}.
    It also coincides to the CH$_3$OH maser KLC--2 \citep[e.g., ][]{1975Lo_HOPS168_maser,2013Kang_HOPS168_maser}.
    \item \textit{G211.47--19.27S (G211S)}: 
    First detected as a hot corino in the ACA data of ALMASOP \citep{Hsu2020_ALMASOP}.
    This is the most line rich source, which meanwhile has the highest luminosity, among the 11 hot corinos reported in this study.
\end{itemize}

\subsection{2D Gaussian Fitting}

\input{tab_CH3OH}

Based on the 2D Gaussian fitting in CASA to the CH$_3$OH--46K transition moment--0 images, we infer the extent of the methanol emission and calculate the molecular hydrogen column density $N(\mathrm{H_2})$ using the 1.3 continuum flux within the extent.
We use the formulae in \citet{Hsu2020_ALMASOP} and \citet{2008Kauffmann_CONTFormula} for optically thin continuum emission. 
The dust opacity is in the form of $\kappa_\nu = 0.1(\nu / 1\,\mathrm{THz})^\beta \mathrm{cm}^2 \mathrm{g}^{-1}$.
The dust opacity index $\beta$ was assigned to be 1.70, which is a typical value for cold clumps in the submillimeter band \citep{1990Beckwith_DustOpacity, 2018Juvela_dustIndex}.
We assumed a (dust) temperature of 100 K, which is the typical value for hot corinos.
Table \ref{tab:continuum} briefly summarizes the 2D Gaussian fitting result of the 1.3 mm continuum images, including the half-power-beam-width (HPBW) of the observation, the full-width-half-maximum (FWHM) of the source angular size, and the integrated flux density. 
The HPBW is around 0\farcs{45}. 
The deconvolved FWHM is ranging from 0\farcs{23} to 0\farcs{64}.
The column density of molecular hydrogen is ranging from $1.5 \times 10^{23}$ cm$^{-2}$ to $2.2 \times 10^{25}$ cm$^{-2}$. 

\input{tab_srcCoord_potential}  

\subsection{Potential Hot Corinos}
Table \ref{tab:candCoord} shows the list of the potential hot corinos. 
The potential hot corinos are, following the criteria of PEACHES, the protostellar cores where at least one methanol transition was detected at 3$\sigma$.
See Sect. \ref{sec:disc_rate_PEACEHS} for more discussions.

%% file: tab_survey.tex
\begin{deluxetable}{l|rccc}
\tablecaption{\label{tab:survey} Surveys related to the sources. }
\tablehead{
\colhead{} & \colhead{HOPS} & \colhead{VANDAM} & \colhead{CARMA-NRO} 
}
\startdata
G192    &     &   &   &  \\
G196A   &     &   &   &  \\
G203W2  &     &   &   &  \\
G205S1A & 358 &   &   &  \\
G206W2  & 399 & V &   &  \\
G208N1  &  87 & V & V &  \\
G208E   &     &   &   &  \\
G209N1B &  12 & V & V &  \\
G209S1  &  11 & V & V &  \\
G210WA  & 168 & V & V &  \\
G211S   & 288 & V &   &  \\
\enddata
\tablerefs{
HOPS: \citet{2016Furlan_HOPS_SED}; VANDAM Survey of Orion Protostars: \citet{2020Tobin_VANDAM-II};  CARMA–NRO Orion Survey : \citet{2018Kong_CARMA-NRO_intro, 2020Feddersen_CARMA-NRO_orion}
}
\end{deluxetable}

%% file: tab_CH3OH.tex
\begin{deluxetable}{lccc}
\tablecaption{\label{tab:continuum} Results of 2D gaussian fitting.}
\tablewidth{0pt}
\tablehead{
\colhead{Name} & \colhead{HPBW} & \colhead{Source Size (CH$_3$OH)}  & \colhead{$N(\mathrm{H_2})$} \\ \colhead{} & \colhead{($''$)} & \colhead{($''$)} & \colhead{(cm$^{-2}$)}  
}
\startdata
G192    & 0.41 & 0.35$\pm$0.03 & 2.1 $\times$ 10$^{24}$ \\
G196A   & 0.44 & 0.36$\pm$0.07 & 4.3 $\times$ 10$^{23}$ \\
G203W2  & 0.43 & 0.23$\pm$0.05 & 1.9 $\times$ 10$^{23}$ \\
G205S1A & 0.43 & 0.41$\pm$0.01 & 1.2 $\times$ 10$^{24}$ \\
G206W2  & 0.41 & 0.31$\pm$0.08 & 4.7 $\times$ 10$^{24}$ \\
G208N1  & 0.41 & 0.64$\pm$0.03 & 8.2 $\times$ 10$^{24}$ \\
G208E   & 0.42 & 0.28$\pm$0.04 & 5.8 $\times$ 10$^{23}$ \\
G209N1B & 0.42 & 0.33$\pm$0.03 & 5.0 $\times$ 10$^{23}$ \\
G209S1  & 0.41 & 0.45$\pm$0.07 & 1.7 $\times$ 10$^{24}$ \\
G210WA  & 0.42 & 0.37$\pm$0.01 & 1.5 $\times$ 10$^{24}$ \\
G211S   & 0.46 & 0.64$\pm$0.04 & 3.5 $\times$ 10$^{24}$ \\
\enddata
\tablecomments{
HPBW is the half-power-beam-width of the synthesized beam.
The source size is the geometric mean of the full-width-half-maximum (FWHM) sizes along the major and the minor axes of the deconvolved CH$_3$OH--46 moment--0 image.
$N_\mathrm{H_2}$ is the column density of the molecular hydrogen within the source size over which the dust temperature is assumed to be 100 K.
See \citet{Hsu2020_ALMASOP} for the formula and the parameters.
}
\end{deluxetable}

%% file: tab_srcCoord_potential.tex
\begin{deluxetable}{llrrrrrcl}[htb!]
\tablecaption{\label{tab:candCoord}
Information of potential hot corinos}
\tablewidth{2pt}
\tablehead{\colhead{Name} & \colhead{Cloud} & \colhead{$\mathrm{\alpha_{J2000}}$} & \colhead{$\mathrm{\delta_{J2000}}$} & 
}
\startdata
G205.46-14.56M2-D & Orion B & 05h46m08.43s & -00d10m00.5s \\
G205.46-14.56N2 & Orion B & 05h46m07.72s & -00d12m21.27s \\
G205.46-14.56S1-B & Orion B & 05h46m07.33s & -00d13m43.49s \\
G205.46-14.56S2 & Orion B & 05h46m04.77s & -00d14m16.67s \\
G208.68-19.20N3-A & Orion A & 05h35m18.06s & -05d00m18.19s \\
G208.68-19.20S-A & Orion A & 05h35m26.56s & -05d03m55.11s \\
G208.68-19.20S-B & Orion A & 05h35m26.54s & -05d03m55.71s \\
G209.55-19.68N1-A & Orion A & 05h35m08.95s & -05d55m54.98s \\
G209.55-19.68N1-C & Orion A & 05h35m08.57s & -05d55m54.54s \\
\enddata
\tablecomments{
$\mathrm{\alpha_{J2000}}$ and $\mathrm{\delta_{J2000}}$ are the right ascension and declination, respectively, of the peak position in our combined 1.3~mm continuum observations.
}
\end{deluxetable}

%% file: appx_xclass.tex
\subsection{Molecular parameters of hot corino sources}
Tables \ref{tab:molfit_G192}-\ref{tab:molfit_G211S} show the parameters of the molecular components optimized by MAGIX for each source.
The parameters \Tex, \Ntot, \deltav, and \vLSR~are the excitation temperature, the total column density, the line width, and the local-standard-of-rest velocity of the component, respectively.
To derive the fractional abundance $X$ for each molecule component, please refer the column density of the molecular hydrogen in Table \ref{tab:continuum}.
We fix the emission size of all species to be the (2D Gaussian) extent of the CH$_3$OH–46K transition deconvolved by the beam size (Table \ref{tab:continuum}).
The errors estimated by MAGIX are also presented.
The excitation temperature \Tex\ of the species detected with only single transition are fixed to be 100 K.
The CO $J=2-1$ transition is detected in all of the sources but not fitted by XCLASS due to the non-Gaussian line profile.

\subsection{Spectra of hot corino sources}

\begin{figure*}
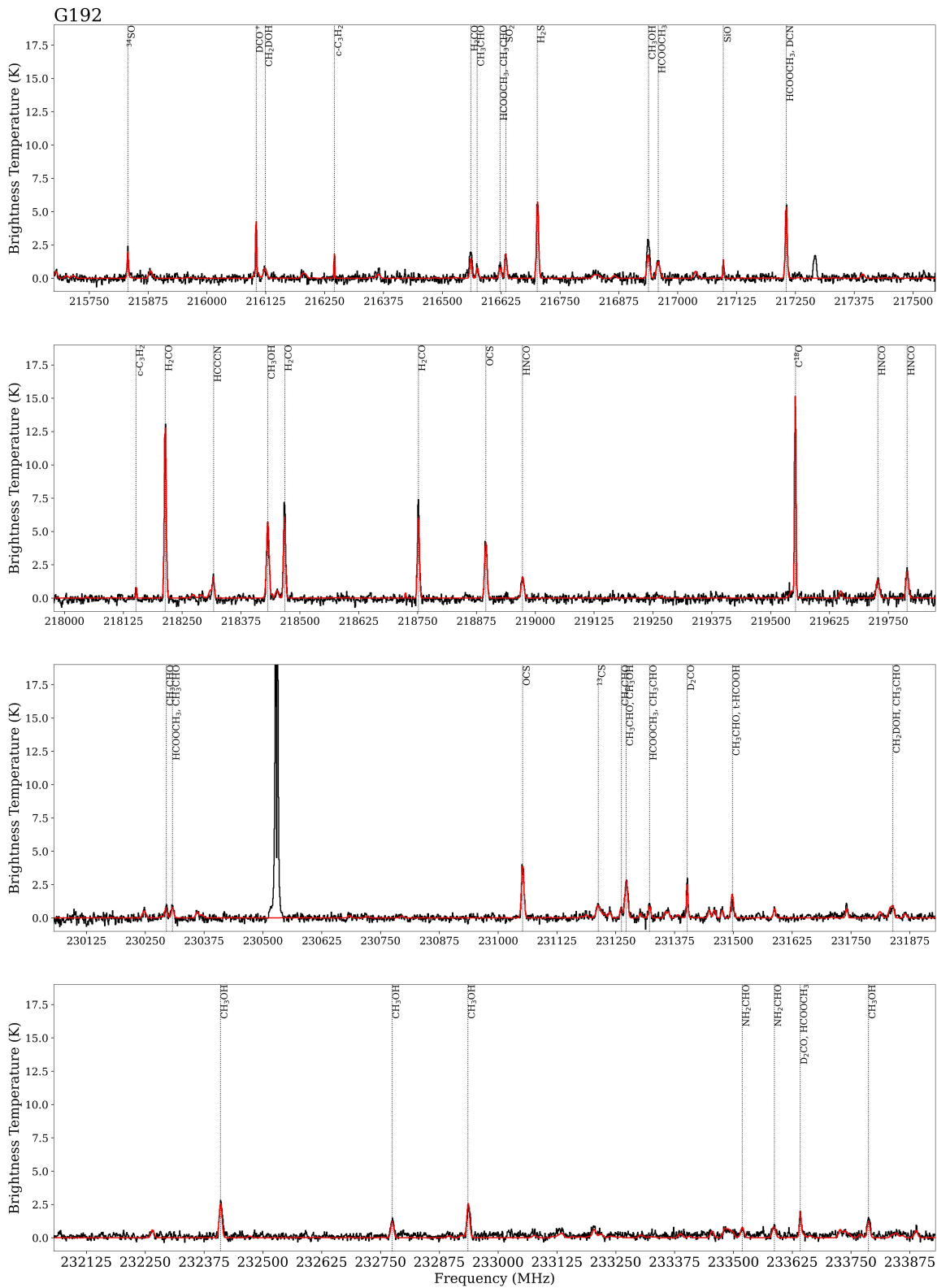

\gridline{\fig{spec_G192_0.pdf}{0.85\textwidth}{}}
\caption{\label{fig:spec_xclass} Observed spectrum and the XCLASS simulation result. The complete figure set (15 images) is available online.}
\end{figure*}

Fig. set \ref{fig:spec_xclass} shows all the spectra of the hot corino sources (see Fig. \ref{fig:spec_xclass} for an example).
The red curve is the spectrum exported by XCLASS.
The molecules of transitions with their peak brightness temperature higher than 5$\sigma$ are labeled.
Note that the ranges of both x- and y-axis of the spectra for each source are different for better illustrations.

\subsection{Moment-0 Images in the Hot Corinos}
\begin{figure*}
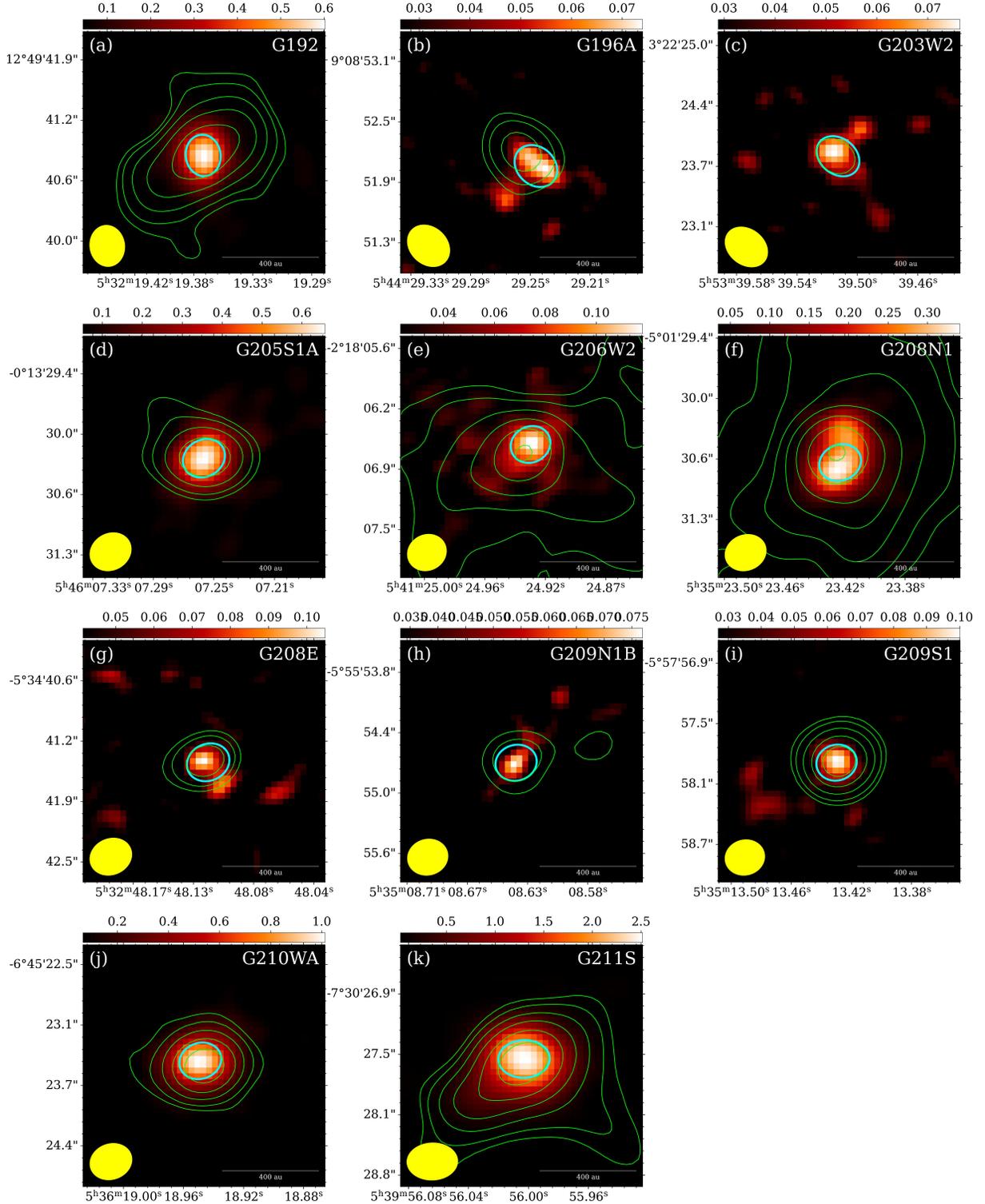

\gridline{\fig{mom0_CH3OH.pdf}{0.9\textwidth}{}}
\caption{\label{fig:mom0_CH3OH} 
Moment-0 images of the CH$_3$OH $10_{2,9}-9_{3,6}$ transition (\Eu=165 K and \frest=231281 MHz).
The green contours show the continuum in steps of [3, 5, 10, 20, 40, 80, 160]$\sigma$. 
The aqua ellipses show the FWHM of the 2D Gaussian fits of the deconvolved methanol emission.
The yellow ellipses show the FWHM of the synthesized beam.
}\end{figure*}

For the 11 hot corinos, Fig. \ref{fig:mom0_CH3OH} shows the moment-0 images of methanol transition: 
    \begin{itemize}
    \item CH$_3$OH: 231281 MHz; 165 K; 
    \begin{sloppypar} rovib $=$ A2; $J=10-9$; $K_a=2-3$; $K_c=9-6$; \end{sloppypar}
    \end{itemize}
The small FWHM, illustrated by the aqua eclipses, implies that the deconvolved source sizes are all compact.

\begin{figure*}
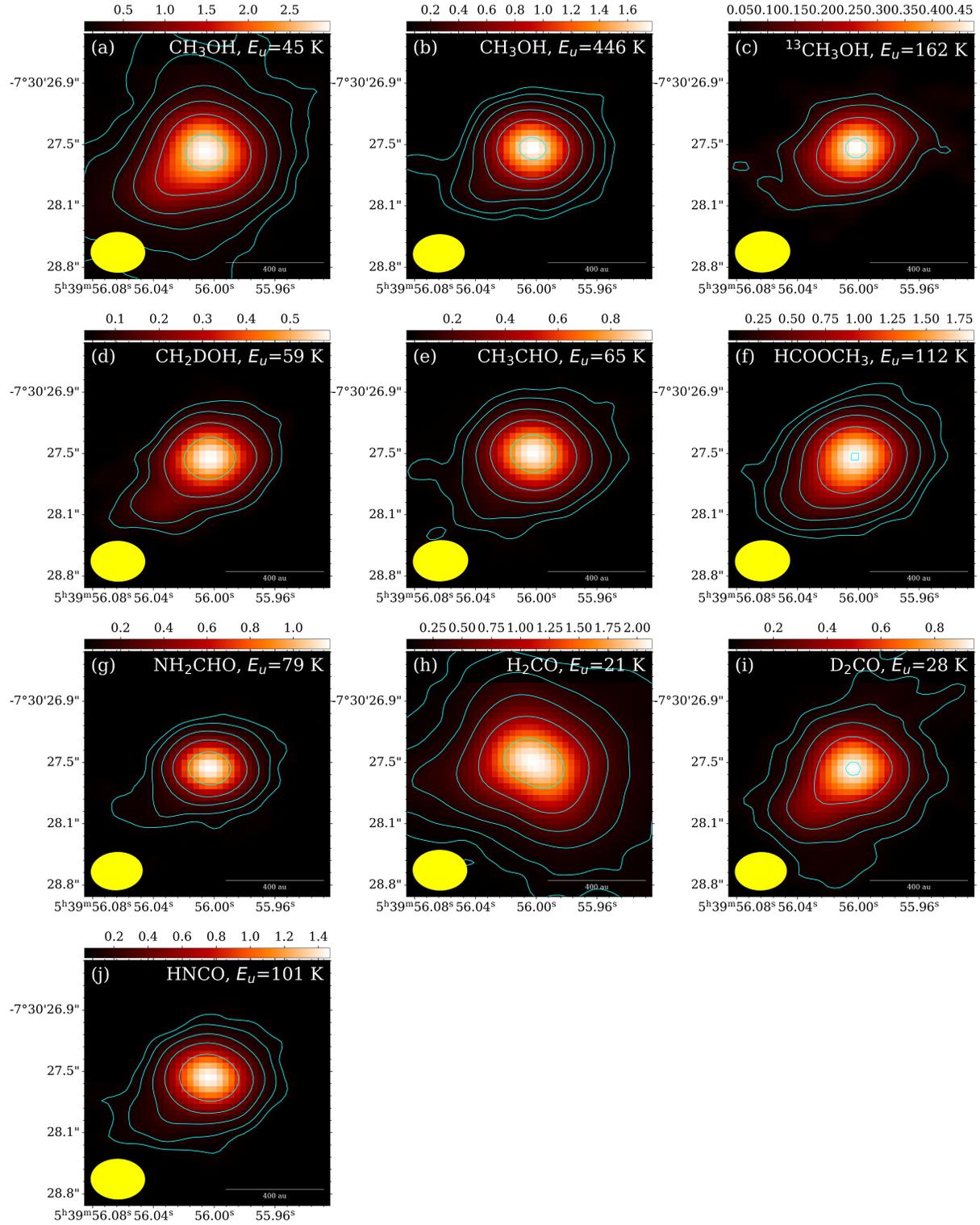

\gridline{\fig{mom0_G211S.pdf}{0.9\textwidth}{}}
\caption{\label{fig:mom0_G211S} 
Moment-0 images of G211S.
The aqua contours are in steps of [3, 5, 10, 20, 40, 80, 160] $\sigma$ for each transition.
See Appendix \ref{appx:xclass} for more information of the transitions.
The yellow ellipses show the FWHM of the synthesized beam.}
\end{figure*}

In addition, Fig. \ref{fig:mom0_G211S} displays the moment-0 images of selected molecules/transitions in G211S, which is the most line-rich hot corino in our sample. 
The transitions from panel (a) to (j) are:
    \begin{enumerate}[label=(\alph*)]
    \item CH$_3$OH: 218440 MHz; 45 K; 
    \begin{sloppypar} rovib=E; $J=5-4$; $K_a=1-2$; $K_c=4-3$;\end{sloppypar}
    \item CH$_3$OH: 232784 MHz; 447 K; 
    \begin{sloppypar} rovib=A2; $J=18-17$; $K_a=3-4$; $K_c=15-14$;\end{sloppypar}
    \item $^{13}$CH$_3$OH: 216370 MHz; 162 K; 
    \begin{sloppypar} $J=10-9$; $K_a=2-3$; $K_c=9-6$; rot=A2;\end{sloppypar}
    \item CH$_2$DOH: 218316 MHz; 59 K; 
    \begin{sloppypar} $J=5$; $K_a=2-1$; $K_c=4-5$; rot=e1;\end{sloppypar}
    \item CH$_3$CHO: 216582 MHz; 65 K; 
    \begin{sloppypar} $J=11-10$; $K_a=1$; $K_c=10-9$; rovib=E;\end{sloppypar}
    \item HCOOCH$_3$: 216966 MHz; 112 K; 
    \begin{sloppypar} $J=20-19$; $K_a=0/1/0-1$; $K_c=20-19$; rovib=E/A;\end{sloppypar}
    \item NH$_2$CHO: 232274 MHz; 79 K; 
    \begin{sloppypar} $J=11-10$; $K_a=2$; $K_c=10-9$;\end{sloppypar}
    \item H$_2$CO: 218222 MHz; 21 K; 
    \begin{sloppypar} $J=3-2$; $K_a=0$; $K_c=3-2$;\end{sloppypar}
    \item D$_2$CO: 231410 MHz; 28 K; 
    \begin{sloppypar} $J=4-3$; $K_a=0$; $K_c=4-3$;\end{sloppypar}
    \item HNCO: 218981 MHz; 101 K; 
    \begin{sloppypar} $J=10-9$; $K_a=1$; $K_c=10-9$;\end{sloppypar} 
    \end{enumerate}
    
\subsection{List of Detected Molecules/Transitions}
Table \ref{tab:trans_all} shows all the detected molecular transitions.
The data are exported from XCLASS and adapted from CDMS \citep{2005CDMS} and JPL \citep{1988JPL}.

\onecolumngrid 

\twocolumngrid 

\newpage

\clearpage
\input{tab_srcs_molfit}

\clearpage
\input{tab_trans_all_short}

%% file: tab_srcs_molfit.tex
\begin{deluxetable}{lcrrr}
\tablecaption{\label{tab:molfit_G192} The molecular parameters of G192.}
\tabletypesize{\scriptsize}
\tablehead{
\colhead{Species} & \colhead{\Tex} & \colhead{\Ntot} & \colhead{\deltav} & \colhead{\vLSR} \\
\colhead{} & \colhead{K} & \colhead{$\log_{10}$(cm$^{-2}$)} & \colhead{km s$^{-1}$} & \colhead{km s$^{-1}$}
}
\startdata
CH$_3$OH;v=0; & 194$^{+9}_{-8}$ & 17.10$^{+0.09}_{-0.11}$ & 8.9$^{+0.7}_{-0.6}$ & 10.2$^{+0.3}_{-0.4}$ \\
$^{13}$CH$_3$OH;v=0; & 374$^{+11}_{-7}$ & 16.53$^{+0.20}_{-0.20}$ & 11.4$^{+0.7}_{-0.5}$ & 8.0$^{+0.3}_{-0.3}$ \\
CH$_2$DOH;v=0; & 74$^{+9}_{-7}$ & 16.70$^{+0.19}_{-0.19}$ & 11.0$^{+0.7}_{-0.5}$ & 7.6$^{+0.3}_{-0.3}$ \\
CH$_3\,^{18}$OH;v=0; & \nodata & \nodata & \nodata & \nodata \\
C$_2$H$_5$OH;v=0; & \nodata & \nodata & \nodata & \nodata \\
CH$_3$CHO;v=0; & 110$^{+9}_{-9}$ & 15.22$^{+0.19}_{-0.15}$ & 7.2$^{+0.8}_{-0.6}$ & 9.4$^{+0.3}_{-0.3}$ \\
CH$_3$CHO;v15=1; & \nodata & \nodata & \nodata & \nodata \\
HCOOCH$_3$;v=0; & 480$^{+9}_{-8}$ & 16.42$^{+0.23}_{-0.16}$ & 11.6$^{+0.9}_{-0.5}$ & 9.6$^{+0.3}_{-0.4}$ \\
HCOOCH$_3$;v18=1; & \nodata & \nodata & \nodata & \nodata \\
CH$_3$OCH$_3$;v=0; & \nodata & \nodata & \nodata & \nodata \\
CH$_2$(OH)CHO;v=0; & \nodata & \nodata & \nodata & \nodata \\
aGg'-(CH$_2$OH)$_2$;v=0; & \nodata & \nodata & \nodata & \nodata \\
H$_2$CO;v=0; & 239$^{+11}_{-6}$ & 16.29$^{+0.06}_{-0.07}$ & 6.3$^{+0.7}_{-0.5}$ & 10.4$^{+0.3}_{-0.2}$ \\
D$_2$CO;v=0; & 254$^{+7}_{-11}$ & 15.19$^{+0.23}_{-0.15}$ & 5.0$^{+0.8}_{-0.5}$ & 9.9$^{+0.4}_{-0.2}$ \\
t-HCOOH;v=0; & (100) & 15.43$^{+0.16}_{-0.24}$ & 8.1$^{+0.6}_{-0.7}$ & 10.2$^{+0.2}_{-0.4}$ \\
NH$_2$CHO;v=0; & 395$^{+10}_{-7}$ & 15.22$^{+0.17}_{-0.21}$ & 10.3$^{+0.7}_{-0.7}$ & 10.1$^{+0.3}_{-0.3}$ \\
HNCO;v=0; & 205$^{+7}_{-8}$ & 15.38$^{+0.13}_{-0.19}$ & 9.0$^{+0.7}_{-0.7}$ & 10.1$^{+0.4}_{-0.2}$ \\
C$_2$H$_5$CN;v=0; & \nodata & \nodata & \nodata & \nodata \\
$^{13}$CH$_3$CN;v=0; & \nodata & \nodata & \nodata & \nodata \\
DCN;v=0; & (100) & 14.21$^{+0.14}_{-0.14}$ & 6.5$^{+0.7}_{-0.6}$ & 10.1$^{+0.3}_{-0.3}$ \\
HCCCN;v=0; & (100) & 13.88$^{+0.22}_{-0.19}$ & 5.5$^{+0.7}_{-0.7}$ & 10.6$^{+0.4}_{-0.2}$ \\
$^{13}$CS;v=0; & (100) & 13.98$^{+0.25}_{-0.19}$ & 10.0$^{+0.8}_{-0.4}$ & 10.2$^{+0.4}_{-0.3}$ \\
H$_2$S;v=0; & (100) & 15.89$^{+0.10}_{-0.17}$ & 6.8$^{+0.6}_{-0.6}$ & 10.3$^{+0.3}_{-0.3}$ \\
OCS;v=0; & 63$^{+10}_{-8}$ & 15.83$^{+0.13}_{-0.13}$ & 8.0$^{+0.7}_{-0.6}$ & 10.6$^{+0.4}_{-0.3}$ \\
O$^{13}$CS;v=0; & \nodata & \nodata & \nodata & \nodata \\
$^{34}$SO;v=0; & (100) & 14.65$^{+0.24}_{-0.16}$ & 4.7$^{+0.7}_{-0.6}$ & 10.1$^{+0.4}_{-0.2}$ \\
SO$_2$;v=0; & (100) & 15.90$^{+0.15}_{-0.23}$ & 6.7$^{+0.7}_{-0.6}$ & 10.8$^{+0.3}_{-0.3}$ \\
CCD;v=0; & \nodata & \nodata & \nodata & \nodata \\
c-C$_3$H$_2$;v=0; & 27$^{+7}_{-8}$ & 14.18$^{+0.29}_{-0.14}$ & 2.7$^{+0.6}_{-0.7}$ & 10.3$^{+0.3}_{-0.4}$ \\
DCO$^+$;v=0; & (100) & 13.60$^{+0.16}_{-0.20}$ & 3.5$^{+0.7}_{-0.7}$ & 10.4$^{+0.4}_{-0.3}$ \\
N2D$^+$;v=0; & \nodata & \nodata & \nodata & \nodata \\
C$^{18}$O;v=0; & (100) & 17.40$^{+0.09}_{-0.07}$ & 4.6$^{+0.6}_{-0.5}$ & 10.1$^{+0.3}_{-0.2}$ \\
SiO;v=0; & (100) & 13.32$^{+0.23}_{-0.19}$ & 3.2$^{+0.7}_{-0.7}$ & 10.0$^{+0.3}_{-0.3}$ \\
\enddata
\end{deluxetable}

\begin{deluxetable}{lcrrr}
\tablecaption{\label{tab:molfit_G196A} The molecular parameters of G196A.}
\tabletypesize{\scriptsize}
\tablehead{
\colhead{Species} & \colhead{\Tex} & \colhead{\Ntot} & \colhead{\deltav} & \colhead{\vLSR} \\
\colhead{} & \colhead{K} & \colhead{$\log_{10}$(cm$^{-2}$)} & \colhead{km s$^{-1}$} & \colhead{km s$^{-1}$}
}
\startdata
CH$_3$OH;v=0; & 190$^{+13}_{-24}$ & 16.57$^{+0.33}_{-0.22}$ & 9.2$^{+1.4}_{-1.2}$ & 9.4$^{+0.6}_{-0.8}$ \\
$^{13}$CH$_3$OH;v=0; & \nodata & \nodata & \nodata & \nodata \\
CH$_2$DOH;v=0; & \nodata & \nodata & \nodata & \nodata \\
CH$_3\,^{18}$OH;v=0; & \nodata & \nodata & \nodata & \nodata \\
C$_2$H$_5$OH;v=0; & \nodata & \nodata & \nodata & \nodata \\
CH$_3$CHO;v=0; & \nodata & \nodata & \nodata & \nodata \\
CH$_3$CHO;v15=1; & \nodata & \nodata & \nodata & \nodata \\
HCOOCH$_3$;v=0; & \nodata & \nodata & \nodata & \nodata \\
HCOOCH$_3$;v18=1; & \nodata & \nodata & \nodata & \nodata \\
CH$_3$OCH$_3$;v=0; & \nodata & \nodata & \nodata & \nodata \\
CH$_2$(OH)CHO;v=0; & \nodata & \nodata & \nodata & \nodata \\
aGg'-(CH$_2$OH)$_2$;v=0; & \nodata & \nodata & \nodata & \nodata \\
H$_2$CO;v=0; & 175$^{+19}_{-13}$ & 15.74$^{+0.11}_{-0.21}$ & 4.6$^{+1.3}_{-1.1}$ & 10.3$^{+0.5}_{-0.5}$ \\
D$_2$CO;v=0; & \nodata & \nodata & \nodata & \nodata \\
t-HCOOH;v=0; & \nodata & \nodata & \nodata & \nodata \\
NH$_2$CHO;v=0; & \nodata & \nodata & \nodata & \nodata \\
HNCO;v=0; & \nodata & \nodata & \nodata & \nodata \\
C$_2$H$_5$CN;v=0; & \nodata & \nodata & \nodata & \nodata \\
$^{13}$CH$_3$CN;v=0; & \nodata & \nodata & \nodata & \nodata \\
DCN;v=0; & (100) & 13.73$^{+0.36}_{-0.31}$ & 5.1$^{+1.3}_{-1.1}$ & 10.9$^{+0.8}_{-0.5}$ \\
HCCCN;v=0; & (100) & 13.78$^{+0.28}_{-0.52}$ & 4.7$^{+1.5}_{-1.2}$ & 11.3$^{+0.9}_{-0.6}$ \\
$^{13}$CS;v=0; & (100) & 13.60$^{+0.40}_{-0.40}$ & 5.9$^{+1.5}_{-1.5}$ & 10.3$^{+0.6}_{-0.6}$ \\
H$_2$S;v=0; & (100) & 15.32$^{+0.28}_{-0.46}$ & 7.6$^{+1.2}_{-1.6}$ & 10.9$^{+0.6}_{-0.8}$ \\
OCS;v=0; & 303$^{+16}_{-18}$ & 15.42$^{+0.33}_{-0.38}$ & 10.5$^{+1.4}_{-1.4}$ & 12.1$^{+0.8}_{-0.6}$ \\
O$^{13}$CS;v=0; & \nodata & \nodata & \nodata & \nodata \\
$^{34}$SO;v=0; & (100) & 14.29$^{+0.39}_{-0.38}$ & 4.5$^{+1.6}_{-1.2}$ & 10.6$^{+0.6}_{-0.7}$ \\
SO$_2$;v=0; & (100) & 15.82$^{+0.41}_{-0.43}$ & 14.4$^{+1.5}_{-1.5}$ & 12.4$^{+0.6}_{-0.7}$ \\
CCD;v=0; & 104$^{+21}_{-14}$ & 14.68$^{+0.27}_{-0.51}$ & 3.1$^{+1.2}_{-1.5}$ & 10.8$^{+0.8}_{-0.5}$ \\
c-C$_3$H$_2$;v=0; & 24$^{+18}_{-12}$ & 14.41$^{+0.31}_{-0.46}$ & 2.4$^{+1.3}_{-1.3}$ & 11.1$^{+0.5}_{-0.7}$ \\
DCO$^+$;v=0; & (100) & 13.29$^{+0.34}_{-0.42}$ & 2.9$^{+1.1}_{-1.7}$ & 10.8$^{+0.6}_{-0.6}$ \\
N2D$^+$;v=0; & \nodata & \nodata & \nodata & \nodata \\
C$^{18}$O;v=0; & (100) & 17.07$^{+0.10}_{-0.19}$ & 3.0$^{+1.0}_{-1.0}$ & 10.8$^{+0.5}_{-0.3}$ \\
SiO;v=0; & (100) & 13.35$^{+0.29}_{-0.55}$ & 4.2$^{+1.2}_{-1.5}$ & 10.5$^{+0.6}_{-0.8}$ \\
\enddata
\end{deluxetable}

\begin{deluxetable}{lcrrr}
\tablecaption{\label{tab:molfit_G203W2} The molecular parameters of G203W2.}
\tabletypesize{\scriptsize}
\tablehead{
\colhead{Species} & \colhead{\Tex} & \colhead{\Ntot} & \colhead{\deltav} & \colhead{\vLSR} \\
\colhead{} & \colhead{K} & \colhead{$\log_{10}$(cm$^{-2}$)} & \colhead{km s$^{-1}$} & \colhead{km s$^{-1}$}
}
\startdata
CH$_3$OH;v=0; & 229$^{+12}_{-18}$ & 16.82$^{+0.23}_{-0.42}$ & 6.7$^{+1.6}_{-1.1}$ & 9.3$^{+0.7}_{-0.6}$ \\
$^{13}$CH$_3$OH;v=0; & \nodata & \nodata & \nodata & \nodata \\
CH$_2$DOH;v=0; & \nodata & \nodata & \nodata & \nodata \\
CH$_3\,^{18}$OH;v=0; & \nodata & \nodata & \nodata & \nodata \\
C$_2$H$_5$OH;v=0; & \nodata & \nodata & \nodata & \nodata \\
CH$_3$CHO;v=0; & \nodata & \nodata & \nodata & \nodata \\
CH$_3$CHO;v15=1; & \nodata & \nodata & \nodata & \nodata \\
HCOOCH$_3$;v=0; & \nodata & \nodata & \nodata & \nodata \\
HCOOCH$_3$;v18=1; & \nodata & \nodata & \nodata & \nodata \\
CH$_3$OCH$_3$;v=0; & \nodata & \nodata & \nodata & \nodata \\
CH$_2$(OH)CHO;v=0; & \nodata & \nodata & \nodata & \nodata \\
aGg'-(CH$_2$OH)$_2$;v=0; & \nodata & \nodata & \nodata & \nodata \\
H$_2$CO;v=0; & 186$^{+20}_{-13}$ & 15.85$^{+0.19}_{-0.19}$ & 3.0$^{+1.7}_{-0.9}$ & 9.6$^{+0.5}_{-0.6}$ \\
D$_2$CO;v=0; & 29$^{+13}_{-20}$ & 14.24$^{+0.44}_{-0.29}$ & 3.0$^{+1.8}_{-1.2}$ & 9.7$^{+0.7}_{-0.7}$ \\
t-HCOOH;v=0; & \nodata & \nodata & \nodata & \nodata \\
NH$_2$CHO;v=0; & \nodata & \nodata & \nodata & \nodata \\
HNCO;v=0; & 323$^{+14}_{-17}$ & 15.30$^{+0.32}_{-0.39}$ & 5.9$^{+1.3}_{-1.3}$ & 9.3$^{+0.5}_{-0.8}$ \\
C$_2$H$_5$CN;v=0; & \nodata & \nodata & \nodata & \nodata \\
$^{13}$CH$_3$CN;v=0; & \nodata & \nodata & \nodata & \nodata \\
DCN;v=0; & (100) & 13.92$^{+0.26}_{-0.49}$ & 2.7$^{+1.3}_{-1.3}$ & 9.9$^{+0.7}_{-0.7}$ \\
HCCCN;v=0; & (100) & 13.88$^{+0.49}_{-0.38}$ & 3.8$^{+1.0}_{-1.6}$ & 9.1$^{+0.9}_{-0.7}$ \\
$^{13}$CS;v=0; & \nodata & \nodata & \nodata & \nodata \\
H$_2$S;v=0; & (100) & 15.39$^{+0.35}_{-0.43}$ & 3.4$^{+1.2}_{-1.3}$ & 9.0$^{+0.7}_{-0.7}$ \\
OCS;v=0; & 273$^{+25}_{-11}$ & 15.65$^{+0.35}_{-0.35}$ & 5.6$^{+1.7}_{-1.2}$ & 9.9$^{+0.6}_{-0.8}$ \\
O$^{13}$CS;v=0; & \nodata & \nodata & \nodata & \nodata \\
$^{34}$SO;v=0; & \nodata & \nodata & \nodata & \nodata \\
SO$_2$;v=0; & \nodata & \nodata & \nodata & \nodata \\
CCD;v=0; & \nodata & \nodata & \nodata & \nodata \\
c-C$_3$H$_2$;v=0; & \nodata & \nodata & \nodata & \nodata \\
DCO$^+$;v=0; & (100) & 13.76$^{+0.38}_{-0.33}$ & 1.2$^{+0.9}_{-0.7}$ & 9.8$^{+0.7}_{-0.5}$ \\
N2D$^+$;v=0; & (100) & 13.22$^{+0.46}_{-0.38}$ & 0.5$^{+0.4}_{-0.3}$ & 10.0$^{+0.7}_{-0.5}$ \\
C$^{18}$O;v=0; & (100) & 17.14$^{+0.23}_{-0.25}$ & 2.0$^{+1.4}_{-0.9}$ & 9.6$^{+0.6}_{-0.5}$ \\
SiO;v=0;\_1 & (100) & 14.13$^{+0.31}_{-0.38}$ & 14.5$^{+1.7}_{-1.1}$ & 6.8$^{+0.8}_{-0.6}$ \\
SiO;v=0;\_2 & (100) & 14.47$^{+0.20}_{-0.34}$ & 18.2$^{+1.5}_{-1.3}$ & -11.9$^{+23.3}_{-24.7}$ \\
\enddata
\end{deluxetable}

\begin{deluxetable}{lcrrr}
\tablecaption{\label{tab:molfit_G205S1A} The molecular parameters of G205S1A.}
\tabletypesize{\scriptsize}
\tablehead{
\colhead{Species} & \colhead{\Tex} & \colhead{\Ntot} & \colhead{\deltav} & \colhead{\vLSR} \\
\colhead{} & \colhead{K} & \colhead{$\log_{10}$(cm$^{-2}$)} & \colhead{km s$^{-1}$} & \colhead{km s$^{-1}$}
}
\startdata
CH$_3$OH;v=0; & 142$^{+9}_{-7}$ & 17.02$^{+0.06}_{-0.11}$ & 8.3$^{+0.7}_{-0.6}$ & 10.4$^{+0.3}_{-0.3}$ \\
$^{13}$CH$_3$OH;v=0; & 207$^{+8}_{-8}$ & 16.82$^{+0.21}_{-0.11}$ & 15.8$^{+0.8}_{-0.5}$ & 8.5$^{+0.3}_{-0.3}$ \\
CH$_2$DOH;v=0; & 87$^{+8}_{-8}$ & 16.92$^{+0.11}_{-0.20}$ & 14.8$^{+0.6}_{-0.6}$ & 9.8$^{+0.4}_{-0.3}$ \\
CH$_3\,^{18}$OH;v=0; & \nodata & \nodata & \nodata & \nodata \\
C$_2$H$_5$OH;v=0; & \nodata & \nodata & \nodata & \nodata \\
CH$_3$CHO;v=0; & 163$^{+8}_{-8}$ & 15.67$^{+0.17}_{-0.14}$ & 15.1$^{+0.7}_{-0.7}$ & 11.0$^{+0.4}_{-0.3}$ \\
CH$_3$CHO;v15=1; & \nodata & \nodata & \nodata & \nodata \\
HCOOCH$_3$;v=0; & 399$^{+8}_{-8}$ & 16.75$^{+0.12}_{-0.14}$ & 13.2$^{+0.6}_{-0.8}$ & 10.4$^{+0.4}_{-0.2}$ \\
HCOOCH$_3$;v18=1; & \nodata & \nodata & \nodata & \nodata \\
CH$_3$OCH$_3$;v=0; & 295$^{+8}_{-8}$ & 16.51$^{+0.17}_{-0.17}$ & 12.2$^{+0.7}_{-0.5}$ & 9.9$^{+0.4}_{-0.3}$ \\
CH$_2$(OH)CHO;v=0; & \nodata & \nodata & \nodata & \nodata \\
aGg'-(CH$_2$OH)$_2$;v=0; & \nodata & \nodata & \nodata & \nodata \\
H$_2$CO;v=0; & 276$^{+10}_{-6}$ & 16.54$^{+0.04}_{-0.07}$ & 7.8$^{+0.8}_{-0.3}$ & 10.5$^{+0.3}_{-0.3}$ \\
D$_2$CO;v=0; & 344$^{+9}_{-8}$ & 15.39$^{+0.19}_{-0.19}$ & 8.3$^{+0.6}_{-0.6}$ & 10.6$^{+0.3}_{-0.4}$ \\
t-HCOOH;v=0; & \nodata & \nodata & \nodata & \nodata \\
NH$_2$CHO;v=0; & \nodata & \nodata & \nodata & \nodata \\
HNCO;v=0; & 119$^{+9}_{-7}$ & 15.30$^{+0.17}_{-0.14}$ & 13.0$^{+0.9}_{-0.6}$ & 9.9$^{+0.4}_{-0.4}$ \\
C$_2$H$_5$CN;v=0; & \nodata & \nodata & \nodata & \nodata \\
$^{13}$CH$_3$CN;v=0; & 427$^{+7}_{-9}$ & 14.53$^{+0.22}_{-0.18}$ & 10.4$^{+0.6}_{-0.6}$ & 11.4$^{+0.4}_{-0.3}$ \\
DCN;v=0; & (100) & 14.32$^{+0.12}_{-0.14}$ & 8.4$^{+0.7}_{-0.7}$ & 10.2$^{+0.3}_{-0.3}$ \\
HCCCN;v=0; & (100) & 14.54$^{+0.14}_{-0.17}$ & 9.5$^{+0.5}_{-0.8}$ & 10.8$^{+0.3}_{-0.3}$ \\
$^{13}$CS;v=0; & (100) & 14.28$^{+0.19}_{-0.19}$ & 12.3$^{+0.6}_{-0.8}$ & 9.5$^{+0.3}_{-0.4}$ \\
H$_2$S;v=0; & (100) & 15.90$^{+0.15}_{-0.12}$ & 7.9$^{+0.6}_{-0.8}$ & 10.2$^{+0.4}_{-0.3}$ \\
OCS;v=0; & 65$^{+9}_{-6}$ & 15.95$^{+0.09}_{-0.14}$ & 9.6$^{+0.7}_{-0.7}$ & 10.4$^{+0.4}_{-0.3}$ \\
O$^{13}$CS;v=0; & 407$^{+8}_{-9}$ & 15.23$^{+0.24}_{-0.22}$ & 9.4$^{+0.7}_{-0.8}$ & 12.5$^{+0.4}_{-0.3}$ \\
$^{34}$SO;v=0; & (100) & 14.90$^{+0.20}_{-0.20}$ & 16.9$^{+0.8}_{-0.4}$ & 9.7$^{+0.3}_{-0.3}$ \\
SO$_2$;v=0; & (100) & 16.07$^{+0.22}_{-0.18}$ & 15.5$^{+0.6}_{-0.7}$ & 12.4$^{+0.4}_{-0.3}$ \\
CCD;v=0; & \nodata & \nodata & \nodata & \nodata \\
c-C$_3$H$_2$;v=0; & \nodata & \nodata & \nodata & \nodata \\
DCO$^+$;v=0; & (100) & 13.63$^{+0.16}_{-0.20}$ & 4.9$^{+0.6}_{-0.6}$ & 10.7$^{+0.4}_{-0.3}$ \\
N2D$^+$;v=0; & \nodata & \nodata & \nodata & \nodata \\
C$^{18}$O;v=0; & (100) & 17.05$^{+0.14}_{-0.13}$ & 4.4$^{+0.8}_{-0.5}$ & 10.3$^{+0.4}_{-0.3}$ \\
SiO;v=0;\_1 & (100) & 13.36$^{+0.17}_{-0.26}$ & 8.6$^{+0.6}_{-0.6}$ & 24.3$^{+0.3}_{-0.4}$ \\
SiO;v=0;\_2 & (100) & 13.62$^{+0.27}_{-0.18}$ & 12.5$^{+0.6}_{-0.7}$ & 1.9$^{+0.3}_{-0.3}$ \\
\enddata
\end{deluxetable}

\begin{deluxetable}{lcrrr}
\tablecaption{\label{tab:molfit_G206W2} The molecular parameters of G206W2.}
\tabletypesize{\scriptsize}
\tablehead{
\colhead{Species} & \colhead{\Tex} & \colhead{\Ntot} & \colhead{\deltav} & \colhead{\vLSR} \\
\colhead{} & \colhead{K} & \colhead{$\log_{10}$(cm$^{-2}$)} & \colhead{km s$^{-1}$} & \colhead{km s$^{-1}$}
}
\startdata
CH$_3$OH;v=0; & 349$^{+10}_{-8}$ & 16.90$^{+0.21}_{-0.17}$ & 3.1$^{+0.8}_{-0.7}$ & 8.9$^{+0.5}_{-0.3}$ \\
$^{13}$CH$_3$OH;v=0; & \nodata & \nodata & \nodata & \nodata \\
CH$_2$DOH;v=0; & 78$^{+8}_{-10}$ & 16.75$^{+0.21}_{-0.21}$ & 7.8$^{+0.9}_{-0.5}$ & 8.7$^{+0.3}_{-0.4}$ \\
CH$_3\,^{18}$OH;v=0; & \nodata & \nodata & \nodata & \nodata \\
C$_2$H$_5$OH;v=0; & \nodata & \nodata & \nodata & \nodata \\
CH$_3$CHO;v=0; & 328$^{+8}_{-8}$ & 15.72$^{+0.16}_{-0.28}$ & 5.6$^{+0.8}_{-0.7}$ & 9.0$^{+0.4}_{-0.3}$ \\
CH$_3$CHO;v15=1; & 435$^{+9}_{-8}$ & 16.26$^{+0.15}_{-0.22}$ & 11.2$^{+0.8}_{-0.8}$ & 8.5$^{+0.3}_{-0.4}$ \\
HCOOCH$_3$;v=0; & 396$^{+10}_{-8}$ & 16.47$^{+0.14}_{-0.26}$ & 9.3$^{+0.8}_{-0.6}$ & 9.2$^{+0.4}_{-0.3}$ \\
HCOOCH$_3$;v18=1; & 344$^{+7}_{-11}$ & 16.69$^{+0.19}_{-0.19}$ & 11.5$^{+0.8}_{-0.6}$ & 8.9$^{+0.4}_{-0.3}$ \\
CH$_3$OCH$_3$;v=0; & \nodata & \nodata & \nodata & \nodata \\
CH$_2$(OH)CHO;v=0; & \nodata & \nodata & \nodata & \nodata \\
aGg'-(CH$_2$OH)$_2$;v=0; & \nodata & \nodata & \nodata & \nodata \\
H$_2$CO;v=0; & 281$^{+9}_{-9}$ & 15.98$^{+0.11}_{-0.16}$ & 3.6$^{+0.7}_{-0.7}$ & 8.5$^{+0.3}_{-0.3}$ \\
D$_2$CO;v=0; & 430$^{+10}_{-8}$ & 15.46$^{+0.19}_{-0.23}$ & 3.1$^{+0.8}_{-0.7}$ & 9.1$^{+0.3}_{-0.3}$ \\
t-HCOOH;v=0; & \nodata & \nodata & \nodata & \nodata \\
NH$_2$CHO;v=0; & \nodata & \nodata & \nodata & \nodata \\
HNCO;v=0; & 321$^{+11}_{-8}$ & 15.41$^{+0.19}_{-0.23}$ & 10.9$^{+0.7}_{-0.7}$ & 9.5$^{+0.4}_{-0.3}$ \\
C$_2$H$_5$CN;v=0; & \nodata & \nodata & \nodata & \nodata \\
$^{13}$CH$_3$CN;v=0; & \nodata & \nodata & \nodata & \nodata \\
DCN;v=0; & (100) & 14.10$^{+0.18}_{-0.18}$ & 6.8$^{+0.8}_{-0.8}$ & 9.6$^{+0.5}_{-0.3}$ \\
HCCCN;v=0; & (100) & 13.88$^{+0.25}_{-0.21}$ & 3.7$^{+0.7}_{-0.7}$ & 9.1$^{+0.4}_{-0.3}$ \\
$^{13}$CS;v=0; & (100) & 14.09$^{+0.22}_{-0.22}$ & 6.1$^{+0.8}_{-0.7}$ & 10.4$^{+0.4}_{-0.4}$ \\
H$_2$S;v=0; & (100) & 15.78$^{+0.15}_{-0.15}$ & 2.1$^{+0.7}_{-0.6}$ & 9.0$^{+0.3}_{-0.3}$ \\
OCS;v=0; & 180$^{+9}_{-9}$ & 15.52$^{+0.15}_{-0.18}$ & 2.4$^{+0.7}_{-0.7}$ & 8.9$^{+0.3}_{-0.3}$ \\
O$^{13}$CS;v=0; & \nodata & \nodata & \nodata & \nodata \\
$^{34}$SO;v=0; & \nodata & \nodata & \nodata & \nodata \\
SO$_2$;v=0; & (100) & 15.83$^{+0.29}_{-0.19}$ & 10.3$^{+0.7}_{-0.7}$ & 10.4$^{+0.4}_{-0.4}$ \\
CCD;v=0; & \nodata & \nodata & \nodata & \nodata \\
c-C$_3$H$_2$;v=0; & \nodata & \nodata & \nodata & \nodata \\
DCO$^+$;v=0; & (100) & 13.37$^{+0.23}_{-0.21}$ & 1.1$^{+0.4}_{-0.3}$ & 9.1$^{+0.3}_{-0.3}$ \\
N2D$^+$;v=0; & \nodata & \nodata & \nodata & \nodata \\
C$^{18}$O;v=0; & (100) & 17.11$^{+0.11}_{-0.16}$ & 1.5$^{+0.6}_{-0.3}$ & 8.8$^{+0.2}_{-0.3}$ \\
SiO;v=0;\_1 & (100) & 15.34$^{+0.05}_{-0.05}$ & 50.9$^{+0.7}_{-0.9}$ & 56.4$^{+0.3}_{-0.4}$ \\
SiO;v=0;\_2 & (100) & 15.38$^{+0.05}_{-0.05}$ & 47.1$^{+0.7}_{-0.6}$ & -36.4$^{+72.5}_{-73.2}$ \\
\enddata
\end{deluxetable}

\begin{deluxetable}{lcrrr}
\tablecaption{\label{tab:molfit_G208N1} The molecular parameters of G208N1.}
\tabletypesize{\scriptsize}
\tablehead{
\colhead{Species} & \colhead{\Tex} & \colhead{\Ntot} & \colhead{\deltav} & \colhead{\vLSR} \\
\colhead{} & \colhead{K} & \colhead{$\log_{10}$(cm$^{-2}$)} & \colhead{km s$^{-1}$} & \colhead{km s$^{-1}$}
}
\startdata
CH$_3$OH;v=0; & 277$^{+7}_{-5}$ & 16.83$^{+0.11}_{-0.11}$ & 3.7$^{+0.5}_{-0.4}$ & 11.3$^{+0.2}_{-0.2}$ \\
$^{13}$CH$_3$OH;v=0; & 223$^{+6}_{-6}$ & 16.14$^{+0.17}_{-0.14}$ & 2.3$^{+0.5}_{-0.5}$ & 11.3$^{+0.3}_{-0.3}$ \\
CH$_2$DOH;v=0; & 44$^{+8}_{-5}$ & 16.40$^{+0.17}_{-0.14}$ & 1.8$^{+0.5}_{-0.4}$ & 11.4$^{+0.3}_{-0.2}$ \\
CH$_3\,^{18}$OH;v=0; & 116$^{+7}_{-5}$ & 15.75$^{+0.19}_{-0.16}$ & 1.6$^{+0.4}_{-0.4}$ & 11.3$^{+0.2}_{-0.3}$ \\
C$_2$H$_5$OH;v=0; & 342$^{+6}_{-6}$ & 16.31$^{+0.16}_{-0.11}$ & 3.6$^{+0.5}_{-0.5}$ & 11.3$^{+0.3}_{-0.2}$ \\
CH$_3$CHO;v=0; & 107$^{+7}_{-6}$ & 15.06$^{+0.14}_{-0.11}$ & 2.0$^{+0.5}_{-0.4}$ & 11.2$^{+0.3}_{-0.2}$ \\
CH$_3$CHO;v15=1; & 199$^{+6}_{-6}$ & 15.27$^{+0.17}_{-0.17}$ & 3.3$^{+0.5}_{-0.6}$ & 11.1$^{+0.3}_{-0.2}$ \\
HCOOCH$_3$;v=0; & 187$^{+8}_{-6}$ & 16.00$^{+0.10}_{-0.12}$ & 1.9$^{+0.4}_{-0.5}$ & 11.4$^{+0.2}_{-0.2}$ \\
HCOOCH$_3$;v18=1; & 262$^{+7}_{-6}$ & 16.46$^{+0.12}_{-0.14}$ & 6.4$^{+0.5}_{-0.4}$ & 11.2$^{+0.3}_{-0.2}$ \\
CH$_3$OCH$_3$;v=0; & 142$^{+6}_{-7}$ & 16.26$^{+0.16}_{-0.13}$ & 5.6$^{+0.5}_{-0.4}$ & 10.8$^{+0.3}_{-0.2}$ \\
CH$_2$(OH)CHO;v=0; & 313$^{+6}_{-6}$ & 15.47$^{+0.15}_{-0.18}$ & 7.6$^{+0.5}_{-0.5}$ & 11.5$^{+0.3}_{-0.3}$ \\
aGg'-(CH$_2$OH)$_2$;v=0; & \nodata & \nodata & \nodata & \nodata \\
H$_2$CO;v=0; & 373$^{+40}_{-49}$ & 16.20$^{+0.10}_{-0.10}$ & 4.5$^{+0.7}_{-0.8}$ & 11.4$^{+0.3}_{-0.2}$ \\
D$_2$CO;v=0; & 179$^{+8}_{-5}$ & 14.63$^{+0.15}_{-0.15}$ & 2.1$^{+0.6}_{-0.5}$ & 11.6$^{+0.3}_{-0.2}$ \\
t-HCOOH;v=0; & \nodata & \nodata & \nodata & \nodata \\
NH$_2$CHO;v=0; & 367$^{+8}_{-6}$ & 14.72$^{+0.19}_{-0.14}$ & 11.3$^{+0.5}_{-0.4}$ & 10.2$^{+0.3}_{-0.2}$ \\
HNCO;v=0; & 317$^{+8}_{-5}$ & 15.41$^{+0.16}_{-0.13}$ & 8.7$^{+0.6}_{-0.5}$ & 11.2$^{+0.3}_{-0.2}$ \\
C$_2$H$_5$CN;v=0; & 98$^{+7}_{-6}$ & 14.71$^{+0.18}_{-0.14}$ & 8.9$^{+0.6}_{-0.5}$ & 11.2$^{+0.3}_{-0.2}$ \\
$^{13}$CH$_3$CN;v=0; & 395$^{+8}_{-5}$ & 13.97$^{+0.18}_{-0.15}$ & 1.6$^{+0.4}_{-0.4}$ & 11.2$^{+0.2}_{-0.3}$ \\
DCN;v=0; & (100) & 13.77$^{+0.16}_{-0.16}$ & 6.2$^{+0.5}_{-0.5}$ & 11.1$^{+0.3}_{-0.2}$ \\
HCCCN;v=0; & (100) & 13.82$^{+0.19}_{-0.16}$ & 6.5$^{+0.5}_{-0.5}$ & 11.1$^{+0.3}_{-0.2}$ \\
$^{13}$CS;v=0; & (100) & 13.85$^{+0.13}_{-0.16}$ & 1.3$^{+0.4}_{-0.3}$ & 11.3$^{+0.3}_{-0.2}$ \\
H$_2$S;v=0; & (100) & 15.84$^{+0.10}_{-0.08}$ & 2.0$^{+0.5}_{-0.3}$ & 11.4$^{+0.2}_{-0.2}$ \\
OCS;v=0; & 98$^{+8}_{-7}$ & 15.37$^{+0.14}_{-0.09}$ & 1.9$^{+0.5}_{-0.3}$ & 11.4$^{+0.2}_{-0.2}$ \\
O$^{13}$CS;v=0; & 20$^{+5}_{-5}$ & 14.52$^{+0.17}_{-0.17}$ & 2.1$^{+0.7}_{-0.4}$ & 12.0$^{+0.2}_{-0.3}$ \\
$^{34}$SO;v=0; & (100) & 14.58$^{+0.17}_{-0.17}$ & 10.9$^{+0.6}_{-0.3}$ & 11.4$^{+0.2}_{-0.3}$ \\
SO$_2$;v=0; & \nodata & \nodata & \nodata & \nodata \\
CCD;v=0; & \nodata & \nodata & \nodata & \nodata \\
c-C$_3$H$_2$;v=0; & \nodata & \nodata & \nodata & \nodata \\
DCO$^+$;v=0; & \nodata & \nodata & \nodata & \nodata \\
N2D$^+$;v=0; & \nodata & \nodata & \nodata & \nodata \\
C$^{18}$O;v=0; & (100) & 16.78$^{+0.12}_{-0.10}$ & 1.8$^{+0.5}_{-0.4}$ & 11.3$^{+0.2}_{-0.2}$ \\
SiO;v=0; & \nodata & \nodata & \nodata & \nodata \\
\enddata
\end{deluxetable}

\begin{deluxetable}{lcrrr}
\tablecaption{\label{tab:molfit_G208E} The molecular parameters of G208E.}
\tabletypesize{\scriptsize}
\tablehead{
\colhead{Species} & \colhead{\Tex} & \colhead{\Ntot} & \colhead{\deltav} & \colhead{\vLSR} \\
\colhead{} & \colhead{K} & \colhead{$\log_{10}$(cm$^{-2}$)} & \colhead{km s$^{-1}$} & \colhead{km s$^{-1}$}
}
\startdata
CH$_3$OH;v=0; & 154$^{+20}_{-16}$ & 16.51$^{+0.21}_{-0.40}$ & 7.5$^{+1.3}_{-1.6}$ & 8.5$^{+0.8}_{-0.6}$ \\
$^{13}$CH$_3$OH;v=0; & 83$^{+16}_{-20}$ & 15.91$^{+0.26}_{-0.49}$ & 11.6$^{+1.5}_{-1.5}$ & 6.7$^{+0.9}_{-0.6}$ \\
CH$_2$DOH;v=0; & \nodata & \nodata & \nodata & \nodata \\
CH$_3\,^{18}$OH;v=0; & \nodata & \nodata & \nodata & \nodata \\
C$_2$H$_5$OH;v=0; & \nodata & \nodata & \nodata & \nodata \\
CH$_3$CHO;v=0; & \nodata & \nodata & \nodata & \nodata \\
CH$_3$CHO;v15=1; & \nodata & \nodata & \nodata & \nodata \\
HCOOCH$_3$;v=0; & 328$^{+10}_{-24}$ & 16.48$^{+0.14}_{-0.33}$ & 10.3$^{+1.9}_{-1.2}$ & 9.8$^{+0.8}_{-0.5}$ \\
HCOOCH$_3$;v18=1; & \nodata & \nodata & \nodata & \nodata \\
CH$_3$OCH$_3$;v=0; & \nodata & \nodata & \nodata & \nodata \\
CH$_2$(OH)CHO;v=0; & \nodata & \nodata & \nodata & \nodata \\
aGg'-(CH$_2$OH)$_2$;v=0; & \nodata & \nodata & \nodata & \nodata \\
H$_2$CO;v=0; & 323$^{+18}_{-22}$ & 16.05$^{+0.10}_{-0.23}$ & 3.1$^{+1.4}_{-1.0}$ & 9.0$^{+0.6}_{-0.5}$ \\
D$_2$CO;v=0; & \nodata & \nodata & \nodata & \nodata \\
t-HCOOH;v=0; & \nodata & \nodata & \nodata & \nodata \\
NH$_2$CHO;v=0; & \nodata & \nodata & \nodata & \nodata \\
HNCO;v=0; & \nodata & \nodata & \nodata & \nodata \\
C$_2$H$_5$CN;v=0; & \nodata & \nodata & \nodata & \nodata \\
$^{13}$CH$_3$CN;v=0; & \nodata & \nodata & \nodata & \nodata \\
DCN;v=0; & (100) & 13.84$^{+0.24}_{-0.44}$ & 3.2$^{+1.7}_{-1.2}$ & 9.1$^{+0.6}_{-0.9}$ \\
HCCCN;v=0; & \nodata & \nodata & \nodata & \nodata \\
$^{13}$CS;v=0; & \nodata & \nodata & \nodata & \nodata \\
H$_2$S;v=0; & \nodata & \nodata & \nodata & \nodata \\
OCS;v=0; & 20$^{+8}_{-14}$ & 16.36$^{+0.42}_{-0.64}$ & 9.3$^{+1.0}_{-1.8}$ & 6.6$^{+0.6}_{-0.9}$ \\
O$^{13}$CS;v=0; & \nodata & \nodata & \nodata & \nodata \\
$^{34}$SO;v=0; & (100) & 14.87$^{+0.40}_{-0.33}$ & 11.4$^{+1.5}_{-1.5}$ & 9.6$^{+0.6}_{-0.8}$ \\
SO$_2$;v=0; & \nodata & \nodata & \nodata & \nodata \\
CCD;v=0; & 238$^{+21}_{-14}$ & 15.18$^{+0.36}_{-0.40}$ & 1.9$^{+1.4}_{-1.1}$ & 8.8$^{+0.7}_{-0.8}$ \\
c-C$_3$H$_2$;v=0; & \nodata & \nodata & \nodata & \nodata \\
DCO$^+$;v=0; & \nodata & \nodata & \nodata & \nodata \\
N2D$^+$;v=0; & \nodata & \nodata & \nodata & \nodata \\
C$^{18}$O;v=0; & (100) & 16.73$^{+0.29}_{-0.29}$ & 3.0$^{+1.4}_{-1.2}$ & 8.8$^{+0.6}_{-0.7}$ \\
SiO;v=0; & (100) & 13.89$^{+0.19}_{-0.45}$ & 5.0$^{+1.3}_{-1.4}$ & 8.6$^{+0.9}_{-0.6}$ \\
\enddata
\end{deluxetable}

\begin{deluxetable}{lcrrr}
\tablecaption{\label{tab:molfit_G209N1B} The molecular parameters of G209N1B.}
\tabletypesize{\scriptsize}
\tablehead{
\colhead{Species} & \colhead{\Tex} & \colhead{\Ntot} & \colhead{\deltav} & \colhead{\vLSR} \\
\colhead{} & \colhead{K} & \colhead{$\log_{10}$(cm$^{-2}$)} & \colhead{km s$^{-1}$} & \colhead{km s$^{-1}$}
}
\startdata
CH$_3$OH;v=0; & 119$^{+19}_{-16}$ & 16.05$^{+0.26}_{-0.38}$ & 3.2$^{+1.5}_{-1.3}$ & 6.9$^{+0.8}_{-0.6}$ \\
$^{13}$CH$_3$OH;v=0; & \nodata & \nodata & \nodata & \nodata \\
CH$_2$DOH;v=0; & \nodata & \nodata & \nodata & \nodata \\
CH$_3\,^{18}$OH;v=0; & \nodata & \nodata & \nodata & \nodata \\
C$_2$H$_5$OH;v=0; & \nodata & \nodata & \nodata & \nodata \\
CH$_3$CHO;v=0; & \nodata & \nodata & \nodata & \nodata \\
CH$_3$CHO;v15=1; & \nodata & \nodata & \nodata & \nodata \\
HCOOCH$_3$;v=0; & \nodata & \nodata & \nodata & \nodata \\
HCOOCH$_3$;v18=1; & \nodata & \nodata & \nodata & \nodata \\
CH$_3$OCH$_3$;v=0; & \nodata & \nodata & \nodata & \nodata \\
CH$_2$(OH)CHO;v=0; & \nodata & \nodata & \nodata & \nodata \\
aGg'-(CH$_2$OH)$_2$;v=0; & \nodata & \nodata & \nodata & \nodata \\
H$_2$CO;v=0; & 133$^{+14}_{-22}$ & 15.48$^{+0.15}_{-0.12}$ & 2.7$^{+1.1}_{-0.6}$ & 7.1$^{+0.4}_{-0.3}$ \\
D$_2$CO;v=0; & \nodata & \nodata & \nodata & \nodata \\
t-HCOOH;v=0; & \nodata & \nodata & \nodata & \nodata \\
NH$_2$CHO;v=0; & \nodata & \nodata & \nodata & \nodata \\
HNCO;v=0; & \nodata & \nodata & \nodata & \nodata \\
C$_2$H$_5$CN;v=0; & \nodata & \nodata & \nodata & \nodata \\
$^{13}$CH$_3$CN;v=0; & \nodata & \nodata & \nodata & \nodata \\
DCN;v=0; & (100) & 13.72$^{+0.18}_{-0.35}$ & 2.7$^{+1.7}_{-0.9}$ & 7.1$^{+0.7}_{-0.6}$ \\
HCCCN;v=0; & (100) & 13.42$^{+0.49}_{-0.42}$ & 2.4$^{+1.7}_{-0.9}$ & 7.6$^{+0.8}_{-0.7}$ \\
$^{13}$CS;v=0; & \nodata & \nodata & \nodata & \nodata \\
H$_2$S;v=0; & (100) & 14.63$^{+0.50}_{-0.41}$ & 1.9$^{+1.2}_{-1.6}$ & 6.6$^{+0.7}_{-0.6}$ \\
OCS;v=0; & 160$^{+26}_{-11}$ & 14.84$^{+0.47}_{-0.38}$ & 3.7$^{+1.2}_{-1.8}$ & 7.2$^{+0.7}_{-0.8}$ \\
O$^{13}$CS;v=0; & \nodata & \nodata & \nodata & \nodata \\
$^{34}$SO;v=0; & \nodata & \nodata & \nodata & \nodata \\
SO$_2$;v=0; & \nodata & \nodata & \nodata & \nodata \\
CCD;v=0; & \nodata & \nodata & \nodata & \nodata \\
c-C$_3$H$_2$;v=0; & \nodata & \nodata & \nodata & \nodata \\
DCO$^+$;v=0; & (100) & 13.39$^{+0.23}_{-0.35}$ & 2.5$^{+1.2}_{-1.5}$ & 6.8$^{+0.8}_{-0.5}$ \\
N2D$^+$;v=0; & \nodata & \nodata & \nodata & \nodata \\
C$^{18}$O;v=0; & (100) & 16.78$^{+0.18}_{-0.18}$ & 1.4$^{+0.9}_{-0.9}$ & 6.9$^{+0.3}_{-0.3}$ \\
SiO;v=0; & \nodata & \nodata & \nodata & \nodata \\
\enddata
\end{deluxetable}

\begin{deluxetable}{lcrrr}
\tablecaption{\label{tab:molfit_G209S1} The molecular parameters of G209S1.}
\tabletypesize{\scriptsize}
\tablehead{
\colhead{Species} & \colhead{\Tex} & \colhead{\Ntot} & \colhead{\deltav} & \colhead{\vLSR} \\
\colhead{} & \colhead{K} & \colhead{$\log_{10}$(cm$^{-2}$)} & \colhead{km s$^{-1}$} & \colhead{km s$^{-1}$}
}
\startdata
CH$_3$OH;v=0; & 384$^{+18}_{-18}$ & 16.63$^{+0.25}_{-0.37}$ & 5.2$^{+1.1}_{-1.4}$ & 6.3$^{+0.4}_{-0.8}$ \\
$^{13}$CH$_3$OH;v=0; & \nodata & \nodata & \nodata & \nodata \\
CH$_2$DOH;v=0; & \nodata & \nodata & \nodata & \nodata \\
CH$_3\,^{18}$OH;v=0; & \nodata & \nodata & \nodata & \nodata \\
C$_2$H$_5$OH;v=0; & \nodata & \nodata & \nodata & \nodata \\
CH$_3$CHO;v=0; & 376$^{+21}_{-14}$ & 15.58$^{+0.20}_{-0.46}$ & 5.7$^{+1.6}_{-1.0}$ & 7.4$^{+0.9}_{-0.5}$ \\
CH$_3$CHO;v15=1; & \nodata & \nodata & \nodata & \nodata \\
HCOOCH$_3$;v=0; & \nodata & \nodata & \nodata & \nodata \\
HCOOCH$_3$;v18=1; & \nodata & \nodata & \nodata & \nodata \\
CH$_3$OCH$_3$;v=0; & \nodata & \nodata & \nodata & \nodata \\
CH$_2$(OH)CHO;v=0; & \nodata & \nodata & \nodata & \nodata \\
aGg'-(CH$_2$OH)$_2$;v=0; & \nodata & \nodata & \nodata & \nodata \\
H$_2$CO;v=0; & 98$^{+15}_{-19}$ & 15.08$^{+0.19}_{-0.19}$ & 3.7$^{+1.8}_{-0.8}$ & 6.8$^{+0.7}_{-0.4}$ \\
D$_2$CO;v=0; & 101$^{+14}_{-21}$ & 14.29$^{+0.25}_{-0.45}$ & 4.3$^{+2.4}_{-1.0}$ & 7.3$^{+0.7}_{-0.7}$ \\
t-HCOOH;v=0; & \nodata & \nodata & \nodata & \nodata \\
NH$_2$CHO;v=0; & 387$^{+20}_{-13}$ & 15.01$^{+0.18}_{-0.34}$ & 7.3$^{+1.5}_{-1.2}$ & 7.3$^{+0.7}_{-0.6}$ \\
HNCO;v=0; & 392$^{+19}_{-16}$ & 15.09$^{+0.21}_{-0.40}$ & 5.8$^{+1.4}_{-1.3}$ & 6.3$^{+0.7}_{-0.6}$ \\
C$_2$H$_5$CN;v=0; & \nodata & \nodata & \nodata & \nodata \\
$^{13}$CH$_3$CN;v=0; & \nodata & \nodata & \nodata & \nodata \\
DCN;v=0; & (100) & 13.45$^{+0.32}_{-0.32}$ & 2.9$^{+1.5}_{-1.2}$ & 6.8$^{+0.6}_{-0.7}$ \\
HCCCN;v=0; & \nodata & \nodata & \nodata & \nodata \\
$^{13}$CS;v=0; & \nodata & \nodata & \nodata & \nodata \\
H$_2$S;v=0; & (100) & 14.99$^{+0.32}_{-0.39}$ & 2.9$^{+1.0}_{-1.5}$ & 6.8$^{+0.9}_{-0.5}$ \\
OCS;v=0; & 47$^{+16}_{-16}$ & 15.03$^{+0.33}_{-0.46}$ & 3.6$^{+1.5}_{-1.2}$ & 7.1$^{+0.7}_{-0.6}$ \\
O$^{13}$CS;v=0; & \nodata & \nodata & \nodata & \nodata \\
$^{34}$SO;v=0; & \nodata & \nodata & \nodata & \nodata \\
SO$_2$;v=0; & \nodata & \nodata & \nodata & \nodata \\
CCD;v=0; & \nodata & \nodata & \nodata & \nodata \\
c-C$_3$H$_2$;v=0; & \nodata & \nodata & \nodata & \nodata \\
DCO$^+$;v=0; & (100) & 13.18$^{+0.22}_{-0.41}$ & 1.5$^{+1.0}_{-1.0}$ & 7.3$^{+0.4}_{-0.7}$ \\
N2D$^+$;v=0; & \nodata & \nodata & \nodata & \nodata \\
C$^{18}$O;v=0; & (100) & 16.64$^{+0.11}_{-0.26}$ & 1.7$^{+0.7}_{-1.1}$ & 7.1$^{+0.3}_{-0.4}$ \\
SiO;v=0; & \nodata & \nodata & \nodata & \nodata \\
\enddata
\end{deluxetable}

\begin{deluxetable}{lcrrr}
\tablecaption{\label{tab:molfit_G210WA} The molecular parameters of G210WA.}
\tabletypesize{\scriptsize}
\tablehead{
\colhead{Species} & \colhead{\Tex} & \colhead{\Ntot} & \colhead{\deltav} & \colhead{\vLSR} \\
\colhead{} & \colhead{K} & \colhead{$\log_{10}$(cm$^{-2}$)} & \colhead{km s$^{-1}$} & \colhead{km s$^{-1}$}
}
\startdata
CH$_3$OH;v=0; & 206$^{+7}_{-5}$ & 17.44$^{+0.06}_{-0.05}$ & 8.8$^{+0.6}_{-0.4}$ & 9.0$^{+0.3}_{-0.2}$ \\
$^{13}$CH$_3$OH;v=0; & 305$^{+7}_{-4}$ & 16.79$^{+0.13}_{-0.16}$ & 12.9$^{+0.6}_{-0.4}$ & 6.5$^{+0.3}_{-0.2}$ \\
CH$_2$DOH;v=0; & 77$^{+7}_{-6}$ & 17.07$^{+0.15}_{-0.10}$ & 13.1$^{+0.5}_{-0.5}$ & 7.2$^{+0.3}_{-0.2}$ \\
CH$_3\,^{18}$OH;v=0; & \nodata & \nodata & \nodata & \nodata \\
C$_2$H$_5$OH;v=0; & \nodata & \nodata & \nodata & \nodata \\
CH$_3$CHO;v=0; & 150$^{+7}_{-5}$ & 15.71$^{+0.10}_{-0.12}$ & 9.8$^{+0.6}_{-0.5}$ & 9.5$^{+0.3}_{-0.2}$ \\
CH$_3$CHO;v15=1; & \nodata & \nodata & \nodata & \nodata \\
HCOOCH$_3$;v=0; & 400$^{+7}_{-6}$ & 16.46$^{+0.16}_{-0.13}$ & 11.6$^{+0.4}_{-0.5}$ & 9.1$^{+0.3}_{-0.2}$ \\
HCOOCH$_3$;v18=1; & \nodata & \nodata & \nodata & \nodata \\
CH$_3$OCH$_3$;v=0; & \nodata & \nodata & \nodata & \nodata \\
CH$_2$(OH)CHO;v=0; & \nodata & \nodata & \nodata & \nodata \\
aGg'-(CH$_2$OH)$_2$;v=0; & \nodata & \nodata & \nodata & \nodata \\
H$_2$CO;v=0; & 339$^{+5}_{-7}$ & 16.74$^{+0.05}_{-0.03}$ & 6.6$^{+0.4}_{-0.4}$ & 9.2$^{+0.2}_{-0.2}$ \\
D$_2$CO;v=0; & 43$^{+7}_{-5}$ & 14.50$^{+0.15}_{-0.12}$ & 5.0$^{+0.5}_{-0.5}$ & 8.9$^{+0.2}_{-0.2}$ \\
t-HCOOH;v=0; & \nodata & \nodata & \nodata & \nodata \\
NH$_2$CHO;v=0; & 131$^{+6}_{-8}$ & 15.11$^{+0.10}_{-0.12}$ & 10.9$^{+0.5}_{-0.5}$ & 7.3$^{+0.3}_{-0.2}$ \\
HNCO;v=0; & 204$^{+6}_{-6}$ & 15.90$^{+0.07}_{-0.10}$ & 11.3$^{+0.5}_{-0.5}$ & 8.2$^{+0.2}_{-0.3}$ \\
C$_2$H$_5$CN;v=0; & \nodata & \nodata & \nodata & \nodata \\
$^{13}$CH$_3$CN;v=0; & \nodata & \nodata & \nodata & \nodata \\
DCN;v=0; & (100) & 14.57$^{+0.07}_{-0.10}$ & 6.9$^{+0.6}_{-0.5}$ & 9.0$^{+0.3}_{-0.2}$ \\
HCCCN;v=0; & (100) & 14.43$^{+0.17}_{-0.09}$ & 7.1$^{+0.6}_{-0.4}$ & 8.1$^{+0.2}_{-0.2}$ \\
$^{13}$CS;v=0; & (100) & 14.35$^{+0.14}_{-0.14}$ & 10.2$^{+0.5}_{-0.4}$ & 8.3$^{+0.3}_{-0.3}$ \\
H$_2$S;v=0; & (100) & 16.00$^{+0.11}_{-0.11}$ & 7.1$^{+0.5}_{-0.4}$ & 9.3$^{+0.2}_{-0.2}$ \\
OCS;v=0; & 96$^{+5}_{-7}$ & 15.99$^{+0.09}_{-0.06}$ & 8.2$^{+0.5}_{-0.4}$ & 9.0$^{+0.3}_{-0.2}$ \\
O$^{13}$CS;v=0; & \nodata & \nodata & \nodata & \nodata \\
$^{34}$SO;v=0; & (100) & 15.10$^{+0.15}_{-0.12}$ & 10.7$^{+0.4}_{-0.5}$ & 7.9$^{+0.3}_{-0.2}$ \\
SO$_2$;v=0; & (100) & 16.46$^{+0.11}_{-0.14}$ & 9.5$^{+0.5}_{-0.5}$ & 8.0$^{+0.2}_{-0.2}$ \\
CCD;v=0; & \nodata & \nodata & \nodata & \nodata \\
c-C$_3$H$_2$;v=0; & \nodata & \nodata & \nodata & \nodata \\
DCO$^+$;v=0; & (100) & 13.64$^{+0.17}_{-0.14}$ & 4.0$^{+0.5}_{-0.4}$ & 8.5$^{+0.3}_{-0.2}$ \\
N2D$^+$;v=0; & \nodata & \nodata & \nodata & \nodata \\
C$^{18}$O;v=0; & (100) & 16.98$^{+0.11}_{-0.13}$ & 4.0$^{+0.5}_{-0.4}$ & 8.5$^{+0.2}_{-0.2}$ \\
SiO;v=0; & (100) & 13.99$^{+0.15}_{-0.12}$ & 5.9$^{+0.5}_{-0.5}$ & 8.6$^{+0.2}_{-0.2}$ \\
\enddata
\end{deluxetable}

\begin{deluxetable}{lcrrr}
\tablecaption{\label{tab:molfit_G211S} The molecular parameters of G211S.}
\tabletypesize{\scriptsize}
\tablehead{
\colhead{Species} & \colhead{\Tex} & \colhead{\Ntot} & \colhead{\deltav} & \colhead{\vLSR} \\
\colhead{} & \colhead{K} & \colhead{$\log_{10}$(cm$^{-2}$)} & \colhead{km s$^{-1}$} & \colhead{km s$^{-1}$}
}
\startdata
CH$_3$OH;v=0; & 204$^{+4}_{-3}$ & 17.46$^{+0.03}_{-0.04}$ & 5.7$^{+0.3}_{-0.2}$ & 2.8$^{+0.1}_{-0.1}$ \\
$^{13}$CH$_3$OH;v=0; & 223$^{+3}_{-3}$ & 16.80$^{+0.09}_{-0.07}$ & 7.3$^{+0.3}_{-0.2}$ & 3.2$^{+0.1}_{-0.1}$ \\
CH$_2$DOH;v=0; & 70$^{+4}_{-3}$ & 17.22$^{+0.06}_{-0.06}$ & 7.7$^{+0.3}_{-0.2}$ & 2.3$^{+0.1}_{-0.1}$ \\
CH$_3\,^{18}$OH;v=0; & \nodata & \nodata & \nodata & \nodata \\
C$_2$H$_5$OH;v=0; & 352$^{+3}_{-4}$ & 16.96$^{+0.07}_{-0.05}$ & 10.5$^{+0.2}_{-0.3}$ & 3.6$^{+0.1}_{-0.1}$ \\
CH$_3$CHO;v=0; & 306$^{+3}_{-3}$ & 16.36$^{+0.04}_{-0.05}$ & 5.4$^{+0.2}_{-0.2}$ & 2.2$^{+0.1}_{-0.1}$ \\
CH$_3$CHO;v15=1; & \nodata & \nodata & \nodata & \nodata \\
HCOOCH$_3$;v=0; & 347$^{+3}_{-3}$ & 16.90$^{+0.05}_{-0.05}$ & 7.6$^{+0.3}_{-0.3}$ & 2.7$^{+0.1}_{-0.1}$ \\
HCOOCH$_3$;v18=1; & 431$^{+3}_{-3}$ & 16.97$^{+0.07}_{-0.07}$ & 10.4$^{+0.3}_{-0.3}$ & 2.7$^{+0.2}_{-0.1}$ \\
CH$_3$OCH$_3$;v=0; & 313$^{+3}_{-3}$ & 16.63$^{+0.10}_{-0.07}$ & 10.5$^{+0.2}_{-0.2}$ & 2.6$^{+0.1}_{-0.1}$ \\
CH$_2$(OH)CHO;v=0; & \nodata & \nodata & \nodata & \nodata \\
aGg'-(CH$_2$OH)$_2$;v=0; & 426$^{+4}_{-3}$ & 16.59$^{+0.06}_{-0.05}$ & 10.3$^{+0.3}_{-0.3}$ & 0.6$^{+0.2}_{-0.1}$ \\
H$_2$CO;v=0; & 415$^{+3}_{-3}$ & 16.81$^{+0.03}_{-0.03}$ & 5.3$^{+0.3}_{-0.2}$ & 2.9$^{+0.2}_{-0.1}$ \\
D$_2$CO;v=0; & 312$^{+4}_{-3}$ & 15.59$^{+0.07}_{-0.08}$ & 6.5$^{+0.3}_{-0.3}$ & 3.5$^{+0.1}_{-0.1}$ \\
t-HCOOH;v=0; & (100) & 15.55$^{+0.08}_{-0.08}$ & 8.3$^{+0.2}_{-0.3}$ & 2.5$^{+0.1}_{-0.1}$ \\
NH$_2$CHO;v=0; & 291$^{+3}_{-3}$ & 15.53$^{+0.07}_{-0.05}$ & 8.1$^{+0.2}_{-0.2}$ & 2.1$^{+0.2}_{-0.1}$ \\
HNCO;v=0; & 307$^{+4}_{-3}$ & 15.91$^{+0.06}_{-0.06}$ & 7.2$^{+0.3}_{-0.3}$ & 2.7$^{+0.1}_{-0.2}$ \\
C$_2$H$_5$CN;v=0; & \nodata & \nodata & \nodata & \nodata \\
$^{13}$CH$_3$CN;v=0; & 429$^{+4}_{-3}$ & 14.78$^{+0.09}_{-0.09}$ & 10.4$^{+0.3}_{-0.2}$ & 2.4$^{+0.1}_{-0.1}$ \\
DCN;v=0; & (100) & 14.43$^{+0.06}_{-0.07}$ & 6.3$^{+0.2}_{-0.3}$ & 3.1$^{+0.1}_{-0.1}$ \\
HCCCN;v=0; & (100) & 14.37$^{+0.08}_{-0.08}$ & 7.9$^{+0.3}_{-0.2}$ & 3.1$^{+0.1}_{-0.1}$ \\
$^{13}$CS;v=0; & (100) & 14.22$^{+0.10}_{-0.07}$ & 6.5$^{+0.3}_{-0.2}$ & 2.0$^{+0.1}_{-0.1}$ \\
H$_2$S;v=0; & (100) & 16.00$^{+0.06}_{-0.07}$ & 5.5$^{+0.3}_{-0.2}$ & 2.7$^{+0.1}_{-0.1}$ \\
OCS;v=0; & 254$^{+4}_{-3}$ & 16.16$^{+0.05}_{-0.05}$ & 6.0$^{+0.3}_{-0.2}$ & 2.7$^{+0.1}_{-0.1}$ \\
O$^{13}$CS;v=0; & 310$^{+3}_{-3}$ & 15.28$^{+0.08}_{-0.08}$ & 7.4$^{+0.3}_{-0.2}$ & 4.3$^{+0.1}_{-0.1}$ \\
$^{34}$SO;v=0; & (100) & 14.94$^{+0.08}_{-0.08}$ & 8.0$^{+0.3}_{-0.2}$ & 2.3$^{+0.2}_{-0.1}$ \\
SO$_2$;v=0; & (100) & 16.03$^{+0.07}_{-0.09}$ & 7.3$^{+0.2}_{-0.3}$ & 3.4$^{+0.1}_{-0.1}$ \\
CCD;v=0; & \nodata & \nodata & \nodata & \nodata \\
c-C$_3$H$_2$;v=0; & \nodata & \nodata & \nodata & \nodata \\
DCO$^+$;v=0; & (100) & 13.48$^{+0.09}_{-0.09}$ & 8.1$^{+0.2}_{-0.2}$ & 2.5$^{+0.2}_{-0.1}$ \\
N2D$^+$;v=0; & \nodata & \nodata & \nodata & \nodata \\
C$^{18}$O;v=0; & (100) & 16.99$^{+0.08}_{-0.07}$ & 4.9$^{+0.3}_{-0.3}$ & 3.3$^{+0.1}_{-0.1}$ \\
SiO;v=0; & (100) & 13.83$^{+0.10}_{-0.06}$ & 14.5$^{+0.2}_{-0.3}$ & 2.0$^{+0.2}_{-0.1}$ \\
\enddata
\end{deluxetable}

%% file: tab_trans_all_short.tex
\startlongtable
\begin{deluxetable*}{llccccll}
\tablecaption{\label{tab:trans_all} The list of all identified transitions.}
\tabletypesize{\small}
\tablehead{
\colhead{Chemical Name} & \colhead{species} & \colhead{\frest} & \colhead{\Aij} & \colhead{\Eu} & \colhead{\gu} & \colhead{Quantum Numbers} & \colhead{Ref.} \\
\colhead{} & \colhead{} & \colhead{MHz} & \colhead{s$^{-1}$} & \colhead{K} & \colhead{} & \colhead{} & \colhead{} 
}
\startdata
Methanol & CH$_3$OH & 216945.6 & 1.2135E-05 & 55.9 & 44 & rovibSym=E; v12=0; J=5-4; Ka=1-2; Kc=4-3 & XCDMS-1370 \\
 & CH$_3$OH & 218440.0 & 4.6863E-05 & 45.5 & 36 & rovibSym=E; v12=0; J=4-3; Ka=2-1; Kc=3-2 & XCDMS-1370 \\
 & CH$_3$OH & 231281.1 & 1.8314E-05 & 165.3 & 84 & rovibSym=A2; v12=0; J=10-9; Ka=2-3; Kc=9-6 & XCDMS-1370 \\
 & CH$_3$OH & 232418.6 & 1.8675E-05 & 165.4 & 84 & rovibSym=A1; v12=0; J=10-9; Ka=2-3; Kc=8-7 & XCDMS-1370 \\
 & CH$_3$OH & 232783.5 & 2.1649E-05 & 446.5 & 148 & rovibSym=A1; v12=0; J=18-17; Ka=3-4; Kc=16-13 & XCDMS-1370 \\
 & CH$_3$OH & 232945.0 & 2.1267E-05 & 190.4 & 84 & rovibSym=E; v12=0; J=10-11; Ka=3-2; Kc=7-9 & XCDMS-1370 \\
 & CH$_3$OH & 233795.8 & 2.1978E-05 & 446.6 & 148 & rovibSym=A2; v12=0; J=18-17; Ka=3-4; Kc=15-14 & XCDMS-1370 \\
\enddata
\tablecomments{
The \Aij~is the Einstein coefficient in s$^{-1}$.
The \Eu~is the upper level of the transition in K.
The \gu~is the degeneracy of the upper state.
The last column is the species ID in the XCLASS.
This table is published in its entirety in machine-readable form.
}
\end{deluxetable*}

%% file: appx_sed.tex
\input{tab_sed_info}

\begin{figure*}
\gridline{\fig{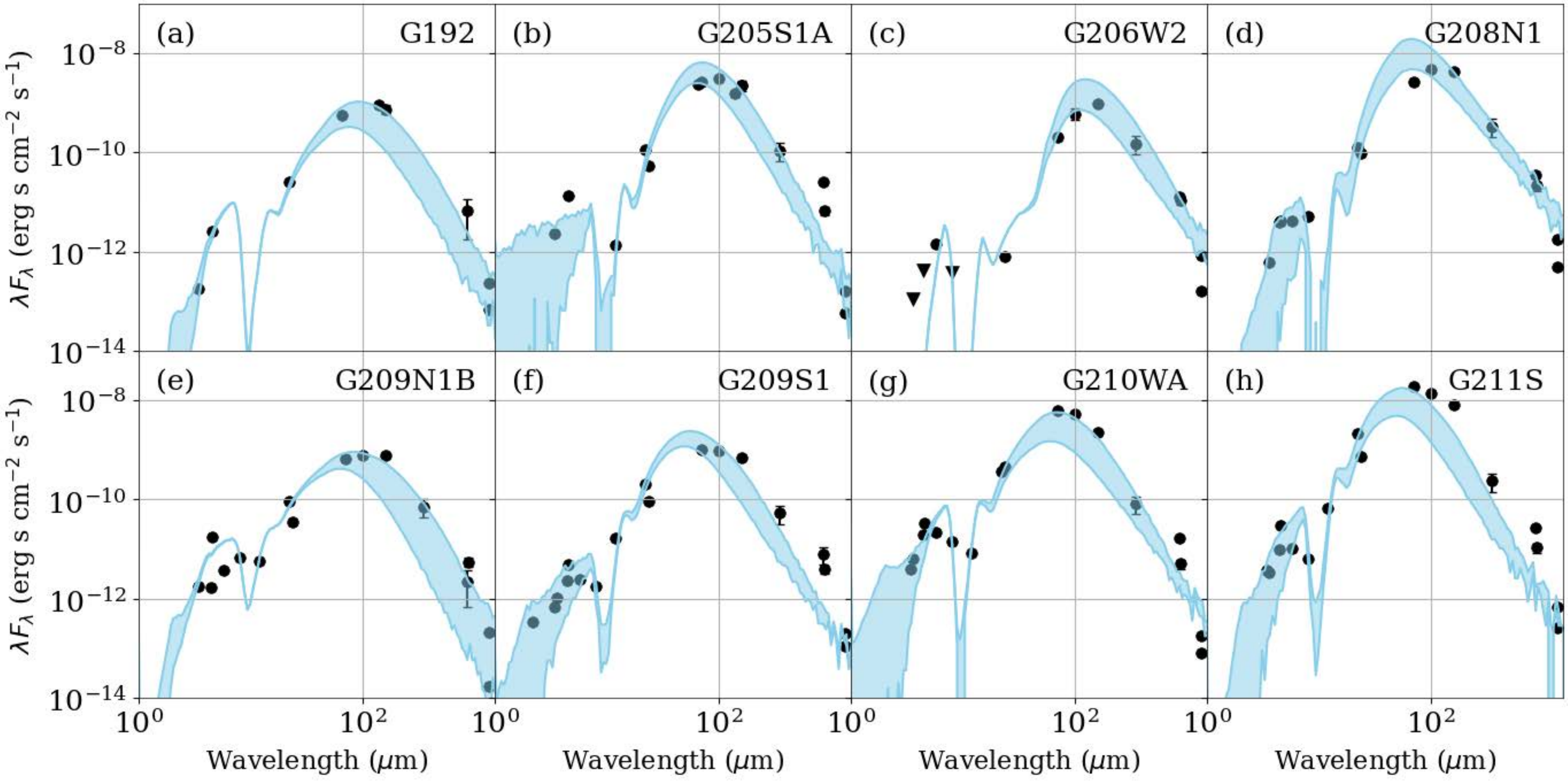}{0.95\textwidth}{}}
\caption{\label{fig:sed_best}
The observed SED data points of the hot corino sources and the best fitted curves obtained from the SED Fitter for the corresponding sources.
The black dots and triangles represent the observed SED data points and the upper limits, respectively, at their corresponding wavelengths.
The blue shades illustrate the SED profile exported by the SED Fitter.
The lower and the upper boundaries of the shades represent the flux observed by the minimum and the maximum apertures, respectively (i.e., 0\arcsec.4 and 19\arcsec, respectively, multiplied by their distance $d\sim$ 400 pc).}
\end{figure*}

All the photometric data points we use for SED fitting can be found in the Table 6 of \citet{2020Dutta_ALMASOP}.
In this section, Table \ref{tab:sed_info} shows the wavelength, the aperture, instrument, and the observatory for each SED photometric data points.
Appendix \ref{appx:sed_cavity} explains how we derive and apply the additional constraint for the cavity of YSO model.
Table \ref{tab:sed_cavity} shows the related cavity parameters (e.g., opening angle and position angle) and Fig. \ref{fig:mom0_srcs_cavity} shows the CO $J=2-1$ outflow as well as the cavity wall exported by the SED Fitter.
Table \ref{tab:sed_fit_top5} displays the parameters of the five best-fit HO--CHUNK YSO models exported by the SED Fitter. 
They can be used as the input parameters for the HO-CHUNK simulation.
There are some limitations in reproducing the original YSO model SEDs, and Appendix \ref{appx:sed_reproduce} explains those limitations.
Fig. \ref{fig:sed_best} shows the SED data points from the observation and the best fitted SED curves obtained by the SED Fitter.
\subsection{Additional Constraint to the Cavity of YSO Model}
\label{appx:sed_cavity}
\begin{figure*}
\plotone{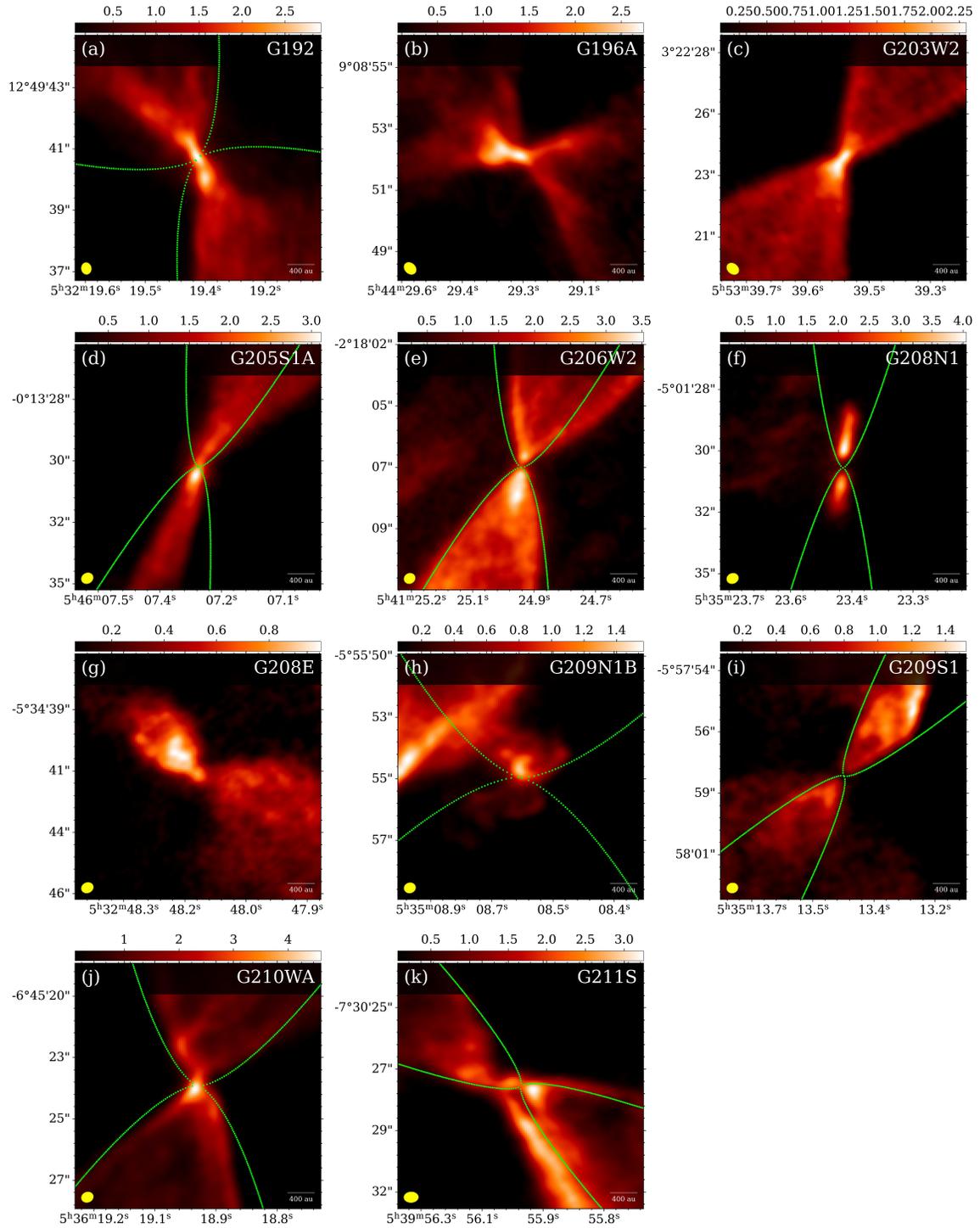}
\caption{\label{fig:mom0_srcs_cavity} 
Moment-0 image of CO $J=2-1$ transition (\Eu=21 K and \frest=231281 MHz).
The green curves represent the cavity walls obtained from the SED Fitter.
}
\end{figure*}

    We aimed to derive the effective opening angle for an additional constraint to the YSO model.
    The shape of the cavity wall is described as:
    \begin{equation}
        \label{eq:cavity_wall}
        z = a\tilde{r}^b + z_0
    \end{equation}
    where $z$ and $\tilde{r}$ are the height and the radial distance in cylindrical coordinates, $z_0$ is the offset of the cavity, $b$ is the polynomial index assumed to be 1.5 and $a$ is a parameter correlated to the opening angle ($\theta$).
    The opening angle $\theta$ of cavity in the HO-CHUNK is defined by:
    \begin{equation}
        \label{eq:cavity_theta}
        \tan\theta \equiv \frac{\tilde{r}_{\max}}{z_{\max}}
    \end{equation}
    where $\tilde{r}_{\max}$ is the outer radius of the envelope and $z_{\max}=z(\tilde{r}_{\max})$.
    Due to the inclination, the effective height $Z$ in observation needs to be corrected by:
    \begin{equation}
    Z = z\cos(90^\circ-\varphi)
    \end{equation}
    where $\varphi$ is the inclination angle, and the effective opening angle ($\Theta$) is therefore:
    \begin{equation}
    \tan\Theta = \frac{\tan\theta}{\cos(90^\circ-\varphi)}
    \end{equation}
    From \citet{2020Dutta_ALMASOP}, we adopt the effective opening angle at 400 au (\ThetaIV) where \ThetaIV~is defined by: 
    \begin{equation}
    \Theta_\mathrm{400} \equiv \tan^{-1}(\frac{\tilde{r}}{Z})\biggr\rvert_{Z=400}
    \end{equation}
    For the two sources whose \ThetaIV were not reported by \citet{2020Dutta_ALMASOP}, namely G208N1 and G211S, we follow the similar procedure of \citet{2020Dutta_ALMASOP} to derive their \ThetaIV.
    We ruled out the YSO models with the effective opening angle 1 degree apart from our observed values (i.e. kept those with $|\Theta^\mathrm{Fitter}_{400}-\Theta^\mathrm{obs}_{400}|<1^\circ$).
    Table \ref{tab:sed_cavity} shows the position angle and the opening angles of the cavities in the hot corino source.
    Fig. \ref{fig:mom0_srcs_cavity} illustrates the bipolar polynomial curves in the form of Eq. \ref{eq:cavity_wall}.

\subsection{Limitation for Reproducing YSO Model SEDs}
\label{appx:sed_reproduce}
The limitations basically come from two reasons.
First, the version of HO--CHUNK code for making the R06 grid is no longer available.
The updates, based on the change log of the version 2008, include updates of: the dust grain model file for correctly extrapolating toward 3.6 mm; and the temperature calculation for the first absorption.
Second, some input parameters for the HO--CHUNK are not directly provided by the R06 grid while some of them are described in \citet{2006Robitaille_sedfitter, 2007Robitaille_sedfitter}.
In the following, we introduce how we derive the input parameters which are not directly provided by the R06 grid.
    \begin{itemize}
      \item Stellar Photosphere Model:
      We interpolate the stellar atmosphere model by the model grid adopted from \citet{2005Brott_PHOENIX_GAIA} to the relevant stellar temperature and surface gravity for each YSO model and the metallicity was assumed to be the solar metallicity.
    The \citet{2005Brott_PHOENIX_GAIA} model grid was created with PHOENIX version AMES-cond-v2.6 \citep{2005Brott_PHOENIX_GAIA, 2013Husser_PHOENIX_HiRes} which could be downloaded via FTP \footnote{ftp.hs.uni-hamburg.de/pub/outgoing/phoenix/GAIA\_I/GAIA\_Grid\_v2.6.1/2A\_Res}.
      \item Dust Grain Models:
      There are four input dust grain models in the HO-CHUNK code, which are that at the inner (denser) region and the outer (less denser) region of the disk, the envelope and the outflow.
      For the inner disk region, the model is the "Model I" in \citet{2002Wood_dust} (or the "Disk midplane" model in the Table 3 of \citet{2003Whitney_SED}) which fits the SED of an edge-on classical T Tauri star, HH 30 IRS.
      The size distribution from 50 \micron~to 1 mm is in the form of:
      \begin{equation}
        \label{eq:dust_model}
        n(a)da = C_i\,a^{-p}\,\exp\bigg[-(\frac{a}{a_{\min}})^q\bigg]da
      \end{equation}
      where $n$ is the distribution function, $a$ is the grain size and $C_i$, $p$ and $q$ are adapted from \citet{2002Wood_dust}.
      For the other three regions, the model is the "Outflow" model in the Table 3 of \citet{2003Whitney_SED}, which has a size distribution fitting a ratio of total-to-selective extinction $R_V$ \citep[i.e. "KMH" model presented by][]{1994Kim_dust_KMH}. 
      The $R_V$ is 3.6 which is slightly larger than that of the canonical sight line ($R_V=3.1$).
      The boundary between the inner and the outer region of the disk is defined with the density of the molecular hydrogen $n(\mathrm{H}_2)>10^{10}\,\mathrm{cm}^{-3}$ (i.e.,  $\rho > 3.35\times 10^{-14}\,\mathrm{g~cm}^{-3}$).
      \item Magnetospheric accretion: 
      The magnetospheric truncation radius $R_\mathrm{trunc}$ and the fractional area of the hot spot ($f_\mathrm{spot}$) are user-defined; however, we do not find the value applied by SED Fitter from \citet{2006Robitaille_sedfitter}.
      We therefore applied the default value of HO-CHUNK 2004 release according to the change log of HO-CHUNK 2008 version: $R_\mathrm{trunc}=5R_{\star}$ and $f_\mathrm{spot}=0.007$.
      Note that the latter is the mean value of \citet{1998Calvet_magnetosphere} which works on the accretion shock models for the infalling accretion flow in magnetosphere surrounding a classical T Tauri stars.
      \item Foreground extinction:
      SED Fitter introduces a foreground extinction $A_V$ and the flux density is affected by:
      \begin{equation}
        \label{eq:extinction}
        \log_{10}F_\lambda^\mathrm{obs} = \log_{10}F_\lambda - 0.4A_V\frac{\kappa_\lambda}{\kappa_V}
      \end{equation}
      where $\kappa_\lambda$ and $\kappa_V$ are the opacity at the examined wavelength and 0.55 $\micron$, respectively. 
      The extinction law model is exported from SED Fitter code and it was derived by fitting a galactic ISM curve modified for the mid-IR band \citep{2005Indebetouw_extinction} with the method in \citet{1994Kim_dust_KMH}.
    \end{itemize}

    

\input{tab_sed_cavity}
\input{tab_sed_fit}

%% file: tab_sed_info.tex
\begin{splitdeluxetable*}{l|c|c|c|c|c|Bl|c|c|c|c|c|c|c}
\tablecaption{The information of the observation for SED. \label{tab:sed_info}}
\tabletypesize{\small}
\tablehead{
\colhead{Observatory} & \colhead{UKIRT} & \colhead{WISE}  & \colhead{\Spitzer} & \colhead{\Spitzer} & \colhead{AKARI PSC} & 
\colhead{Observatory} & \colhead{AKARI BSC} & \colhead{\Herschel} &  \colhead{APEX} & \colhead{JCMT} & \colhead{ALMA}
}
\startdata
Filter or Band & UKIDSS $K$ & W1, W2, W3, W4 & IRAC & MIPS & IRC S9W, L18W & 
Filter or Band & FIS N60, WIDE-L, N160 & PACS & SABOCA, LABOCA & SCUBA2 850 & Band 6 \\
Wavelength (\micron) & 2.2 & 3.4, 4.6, 12, 22 & 3.6, 4.5, 5.8, & 24 & 9, 18 & 
Wavelength (\micron) & 65, 140, 160 & 70, 100, 160 & 350, 870 & 850 & 1300 \\
Aperture & 4\farcs{0} & 6\farcs{0} & 2\farcs{4} & 6\farcs{0} & 5\farcs{5}, 5\farcs{7} & 
Aperture & 7\farcs{5} & 9\farcs{6}, 9\farcs{6}, 12\farcs{8} & 7\farcs{3}, 19\farcs{0} & 14\farcs{0} & $\sim$0\farcs{35}, $\sim$6\farcs{0} \\
\enddata
\tablecomments{
The SED flux data points were collected from \citet{2020Dutta_ALMASOP}.
}
\tablerefs{
UKIRT/UKIDSS: \citet{2007Lawrence_SED_UKIDSS}; 
WISE: \citet{2010Wright_SED_WISE}
\Spitzer:~\citet{2004Werner_Spitzer}; 
IRAC:~\citet{2004Fazio_IRAC}; 
MIPS:~\citet{2004Rieke_MIPS}; 
AKARI PSC: \citet{2010Ishihara_SED_AKARI_IRC};
AKARI BSC: \citet{2010Yamamura_SED_AKARI_FIS};
\Herschel:~\citet{2010Pilbratt_Herschel};
PACS:~\citet{2010Poglitsch_PACS}; 
APEX:~\citet{2006Gusten_APEX}; 
SABOCA:~\citet{2010Siringo_APEX350_SABOCA}; 
LABOCA:~\citet{2009Siringo_APEX870_LABOCA}; 
JCMT/SCUBA2: \citet{2008Francesco_JCMTS_submm}; 
}
\end{splitdeluxetable*}

%% file: tab_sed_cavity.tex
\begin{deluxetable}{lrrrrr}
\tabletypesize{\small}
\tablecaption{\label{tab:sed_cavity} Comparison of the cavity parameters derived from different methods.}
\tablewidth{2pt}
\tablehead{\colhead{Name} & 
\colhead{PA} & 
\colhead{$\theta^\mathrm{Fitter}$} &
\colhead{$\Theta^\mathrm{Fitter}$ } &
\colhead{$\Theta^\mathrm{Fitter}_{400}$ } &
\colhead{$\Theta^\mathrm{Dutta}_{400}$ }
} 
\startdata
G192 & 132\DEG & 23\DEG & 32\DEG & 60\DEG & 61\DEG \\
G196A & \nodata & \nodata & \nodata & \nodata & 54\DEG \\
G203W2 & \nodata & \nodata & \nodata & \nodata & 48\DEG \\
G205S1A & 72\DEG & 7\DEG & 12\DEG & 35\DEG & 34\DEG \\
G206W2 & 76\DEG & 3\DEG & 10\DEG & 39\DEG & 38\DEG \\
G208E & \nodata & \nodata & \nodata & \nodata & 41\DEG \\
G208N1 & 85\DEG & 8\DEG & 11\DEG & 29\DEG & $^{\dagger}$~29\DEG \\
G209N1B & 82\DEG & 18\DEG & 31\DEG & 64\DEG & \nodata \\
G209S1 & 51\DEG & 11\DEG & 11\DEG & 31\DEG & 31\DEG \\
G210WA & 78\DEG & 20\DEG & 26\DEG & 53\DEG & 53\DEG \\
G211S & 146\DEG & 11\DEG & 11\DEG & 35\DEG & $^{\dagger}$~36\DEG \\
\enddata
\tablecomments{
PA is the position angle.
$\theta$ is the opening angle of the cavity.
$\Theta$ and $\Theta_{400}$ are the effective opening angles at $z=z_\mathrm{max}$ and 400 au.
$\Theta^\mathrm{Dutta}$ is the opening angle adapted from \citet{2020Dutta_ALMASOP}.
"Fitter" represents the best-fit model fit by SED Fitter.
$\dagger$: Fitted by this study.
}
\end{deluxetable}

%% file: tab_sed_fit.tex
\clearpage
\begin{deluxetable*}{lrrrrrrrrrrrrrrrrrrrrrrrrrrrrr}
\rotate
\setlength{\tabcolsep}{5pt}
\renewcommand{\arraystretch}{0.95}
\tabletypesize{\scriptsize}
\tablecaption{\label{tab:sed_fit_top5} The parameters of the five best SED Fitter models. }
\tablehead{
\colhead{Source} & \colhead{Model} & \colhead{\#} & \colhead{chi2} & \colhead{Av} & \colhead{$L_\mathrm{tot}$} & \colhead{$t_{\star}$} & \colhead{$M_{\star}$} & \colhead{$R_{\star}$} & \colhead{$T_{\star}$} & \colhead{$M_\mathrm{disk}$} & \colhead{$\dot{M}_\mathrm{disk}$} & \colhead{$R^\mathrm{inner}$} & \colhead{$R^\mathrm{outer}_\mathrm{disk}$} & \colhead{$\dot{M}_\mathrm{env}$} & \colhead{$R^\mathrm{outer}_\mathrm{env}$} & \colhead{$\theta$} & \colhead{$\cos\varphi$} & \colhead{$\rho_\mathrm{cav}$} & \colhead{$\rho_\mathrm{amb}$} \\
\colhead{ } & \colhead{ } & \colhead{ } & \colhead{ } & \colhead{ } & \colhead{ } & \colhead{yr} & \colhead{\Msun} & \colhead{\Rsun} & \colhead{K} & \colhead{\Msun} & \colhead{\Msun yr$^{-1}$} & \colhead{$R_\mathrm{sub}$} & \colhead{au} & \colhead{\Msun yr$^{-1}$} & \colhead{au} & \colhead{$^{\circ}$} & \colhead{ } & \colhead{g cm$^{-3}$} & \colhead{g cm$^{-3}$}
}
\startdata
G192 & 30110098 & 9 & 356.445 & 19.942 & 8.65 & 26800 & 0.736 & 6.324 & 3928 & 1.03E-02 & 3.13E-08 & 1.0 & 73.32 & 1.47E-04 & 5529 & 22.55 & 0.75 & 8.84E-21 & 4.54E-22 \\
 & 30004555 & 9 & 476.747 & 25.581 & 17.46 & 114800 & 1.280 & 7.658 & 4214 & 5.07E-03 & 1.90E-07 & 1.0 & 76.81 & 1.25E-04 & 3399 & 35.42 & 0.45 & 2.98E-21 & 8.21E-22 \\
 & 30111619 & 9 & 504.379 & 30.059 & 6.96 & 23180 & 0.542 & 5.538 & 3776 & 3.79E-03 & 5.07E-07 & 1.0 & 14.12 & 1.43E-04 & 5518 & 19.79 & 0.85 & 3.35E-20 & 1.65E-22 \\
 & 30004556 & 9 & 1017.876 & 38.539 & 17.46 & 114800 & 1.280 & 7.658 & 4214 & 5.07E-03 & 1.90E-07 & 1.0 & 76.81 & 1.25E-04 & 3399 & 35.42 & 0.55 & 2.98E-21 & 8.21E-22 \\
 & 30148869 & 9 & 1314.257 & 16.334 & 5.76 & 9730 & 0.485 & 5.813 & 3689 & 3.04E-02 & 6.92E-08 & 8.99 & 53.09 & 1.29E-04 & 8806 & 15.83 & 0.85 & 1.26E-20 & 1.04E-22 \\
\hline
G205S1A & 30168999 & 15 & 3283.434 & 1.279 & 37.74 & 3505 & 1.351 & 12.340 & 4064 & 9.78E-03 & 1.53E-07 & 1.0 & 18.07 & 1.59E-04 & 2874 & 6.50 & 0.85 & 2.66E-20 & 4.70E-22 \\
 & 300005210 & 15 & 4379.332 & 35.234 & 22.23 & 11260 & 1.017 & 9.814 & 3999 & 6.65E-03 & 4.42E-08 & 1.0 & 18.04 & 1.18E-04 & 3123 & 3.53 & 0.95 & 4.74E-20 & 4.33E-22 \\
 & 301481210 & 15 & 4981.519 & 25.538 & 9.32 & 9795 & 0.629 & 7.018 & 3809 & 3.83E-04 & 6.69E-09 & 1.0 & 21.78 & 5.87E-05 & 2501 & 4.00 & 0.95 & 2.72E-20 & 4.02E-22 \\
 & 30052308 & 15 & 5310.032 & 0.855 & 26.98 & 4341 & 1.038 & 10.880 & 3977 & 3.73E-02 & 1.43E-07 & 1.0 & 18.71 & 1.02E-04 & 1754 & 10.11 & 0.75 & 1.95E-20 & 2.23E-22 \\
 & 30029961 & 15 & 5409.914 & 2.349 & 25.83 & 14180 & 1.414 & 9.678 & 4181 & 2.00E-02 & 3.23E-08 & 1.642 & 52.16 & 6.01E-05 & 2982 & 11.56 & 0.05 & 5.67E-20 & 4.01E-22 \\
\hline
G206W2 & 300261810 & 10 & 3104.624 & 164.322 & 23.03 & 2008 & 0.927 & 10.320 & 3923 & 2.94E-02 & 1.71E-07 & 6.048 & 13.67 & 4.48E-04 & 6461 & 3.17 & 0.95 & 2.27E-20 & 3.54E-22 \\
 & 300775310 & 10 & 4999.664 & 160.318 & 26.76 & 1838 & 1.073 & 10.680 & 4008 & 3.52E-03 & 1.28E-07 & 3.703 & 5.87 & 1.55E-04 & 8718 & 2.51 & 0.95 & 3.03E-20 & 4.57E-22 \\
 & 30068109 & 10 & 5447.386 & 202.478 & 34.30 & 2089 & 0.737 & 8.435 & 3848 & 9.22E-02 & 8.45E-06 & 1.0 & 4.98 & 1.75E-05 & 2031 & 8.77 & 0.85 & 1.97E-20 & 3.70E-22 \\
 & 30091768 & 10 & 5670.928 & 233.681 & 83.12 & 2644 & 1.291 & 14.530 & 3969 & 2.73E-02 & 1.49E-05 & 1.0 & 3.93 & 1.02E-05 & 13520 & 4.37 & 0.75 & 1.75E-20 & 2.77E-22 \\
 & 30075069 & 10 & 5863.251 & 232.446 & 80.85 & 12800 & 2.036 & 14.280 & 4215 & 1.18E-01 & 5.97E-06 & 1.0 & 10.15 & 2.53E-05 & 3522 & 6.69 & 0.85 & 8.94E-21 & 6.09E-22 \\
\hline
G208N1 & 30171007 & 15 & 4551.015 & 22.018 & 130.70 & 1486 & 1.000 & 12.900 & 3884 & 2.68E-02 & 4.55E-05 & 1.0 & 1.59 & 1.59E-04 & 1674 & 8.46 & 0.65 & 7.44E-20 & 6.61E-22 \\
 & 30193948 & 15 & 8440.588 & 40.363 & 112.20 & 3691 & 2.283 & 19.420 & 4137 & 1.56E-01 & 4.07E-06 & 1.0 & 4.29 & 8.80E-05 & 2664 & 6.23 & 0.75 & 1.97E-20 & 1.20E-21 \\
 & 30147519 & 15 & 12921.615 & 13.108 & 53.71 & 2673 & 0.895 & 9.212 & 3937 & 1.04E-01 & 1.33E-05 & 1.0 & 6.57 & 2.02E-04 & 1178 & 6.84 & 0.85 & 7.71E-20 & 4.75E-22 \\
 & 30193247 & 15 & 14046.726 & 34.773 & 76.63 & 1403 & 1.674 & 13.720 & 4134 & 4.87E-02 & 8.44E-06 & 1.389 & 2.42 & 5.48E-05 & 4682 & 5.14 & 0.65 & 4.65E-20 & 4.93E-22 \\
 & 30042819 & 15 & 14241.034 & 16.968 & 38.74 & 2137 & 0.420 & 6.304 & 3573 & 2.33E-02 & 1.77E-05 & 1.0 & 2.49 & 7.44E-05 & 1348 & 7.04 & 0.85 & 1.89E-20 & 2.11E-22 \\
\hline
G209N1B & 30120589 & 16 & 3257.637 & 12.456 & 6.18 & 118300 & 0.748 & 5.151 & 4004 & 6.61E-04 & 1.49E-08 & 1.0 & 42.36 & 1.17E-04 & 8746 & 17.76 & 0.85 & 7.44E-21 & 1.78E-22 \\
 & 301073310 & 16 & 3372.405 & 22.544 & 2.51 & 29560 & 0.348 & 4.359 & 3476 & 3.12E-04 & 9.77E-09 & 1.0 & 29.98 & 1.25E-04 & 3644 & 16.78 & 0.95 & 1.89E-20 & 1.34E-22 \\
 & 30186318 & 16 & 3380.836 & 14.018 & 6.16 & 16960 & 0.542 & 5.842 & 3764 & 9.36E-04 & 2.01E-09 & 1.0 & 86.70 & 6.04E-05 & 4201 & 22.54 & 0.75 & 1.42E-20 & 1.07E-22 \\
 & 301051110 & 16 & 3477.510 & 24.504 & 2.77 & 47640 & 0.417 & 4.222 & 3620 & 6.63E-04 & 6.25E-09 & 1.0 & 78.66 & 1.36E-04 & 3817 & 17.71 & 0.95 & 4.71E-21 & 1.54E-22 \\
 & 30110939 & 16 & 3502.317 & 15.991 & 4.49 & 9906 & 0.393 & 5.540 & 3546 & 5.39E-04 & 6.84E-08 & 1.087 & 10.13 & 6.09E-05 & 5026 & 18.41 & 0.85 & 6.06E-20 & 1.10E-22 \\
\hline
G209S1 & 30119282 & 18 & 3415.220 & 5.028 & 18.98 & 5267 & 0.465 & 6.087 & 3651 & 2.28E-02 & 6.44E-06 & 1.754 & 8.25 & 2.37E-05 & 2419 & 10.67 & 0.15 & 1.48E-20 & 1.84E-22 \\
 & 30083783 & 18 & 3420.874 & 6.633 & 13.86 & 33880 & 0.834 & 6.391 & 4012 & 4.66E-02 & 1.33E-06 & 8.829 & 17.64 & 2.34E-05 & 2411 & 10.17 & 0.25 & 2.49E-20 & 2.09E-22 \\
 & 30059808 & 18 & 3682.985 & 6.121 & 17.96 & 27780 & 1.145 & 8.213 & 4125 & 1.30E-03 & 1.16E-07 & 1.0 & 16.98 & 7.88E-05 & 1413 & 9.39 & 0.75 & 4.23E-20 & 2.52E-22 \\
 & 30083782 & 18 & 3769.373 & 5.959 & 13.86 & 33880 & 0.834 & 6.391 & 4012 & 4.66E-02 & 1.33E-06 & 8.829 & 17.64 & 2.34E-05 & 2411 & 10.17 & 0.15 & 2.49E-20 & 2.09E-22 \\
 & 30119283 & 18 & 3905.069 & 5.476 & 18.98 & 5267 & 0.465 & 6.087 & 3651 & 2.28E-02 & 6.44E-06 & 1.754 & 8.25 & 2.37E-05 & 2419 & 10.67 & 0.25 & 1.48E-20 & 1.84E-22 \\
\hline
G210WA & 30174457 & 17 & 8439.560 & 6.493 & 42.89 & 26680 & 1.975 & 11.860 & 4276 & 8.33E-02 & 1.55E-07 & 4.538 & 104.40 & 1.50E-04 & 4044 & 19.93 & 0.65 & 1.44E-20 & 8.54E-22 \\
 & 30040215 & 17 & 9277.259 & 7.012 & 55.95 & 10530 & 1.315 & 11.610 & 4072 & 4.64E-02 & 7.51E-06 & 1.348 & 10.63 & 6.96E-05 & 8874 & 16.90 & 0.45 & 7.29E-20 & 3.36E-22 \\
 & 30179766 & 17 & 9523.582 & 20.464 & 23.33 & 262500 & 2.471 & 7.392 & 4591 & 1.07E-01 & 1.81E-07 & 2.871 & 127.40 & 5.05E-05 & 2872 & 25.48 & 0.55 & 2.95E-21 & 1.33E-21 \\
 & 30129478 & 17 & 9821.316 & 0.000 & 38.81 & 6727 & 1.181 & 11.170 & 4032 & 2.66E-02 & 3.41E-06 & 4.227 & 6.00 & 1.08E-04 & 11500 & 11.00 & 0.75 & 6.50E-20 & 2.37E-22 \\
 & 30040214 & 17 & 10155.993 & 4.526 & 55.95 & 10530 & 1.315 & 11.610 & 4072 & 4.64E-02 & 7.51E-06 & 1.348 & 10.63 & 6.96E-05 & 8874 & 16.90 & 0.35 & 7.29E-20 & 3.36E-22 \\
\hline
G211S & 30126313 & 17 & 6055.181 & 12.285 & 141.20 & 8642 & 2.829 & 19.120 & 4242 & 6.66E-02 & 9.30E-06 & 5.658 & 42.33 & 9.40E-05 & 3817 & 10.58 & 0.25 & 1.28E-20 & 1.05E-21 \\
 & 30172775 & 17 & 7948.439 & 15.047 & 129.70 & 7524 & 3.368 & 20.640 & 4290 & 5.27E-03 & 4.59E-08 & 5.479 & 9.75 & 9.84E-05 & 3375 & 10.96 & 0.45 & 4.76E-20 & 1.95E-21 \\
 & 30126315 & 17 & 8019.270 & 15.346 & 141.20 & 8642 & 2.829 & 19.120 & 4242 & 6.66E-02 & 9.30E-06 & 5.658 & 42.33 & 9.40E-05 & 3817 & 10.58 & 0.45 & 1.28E-20 & 1.05E-21 \\
 & 30126314 & 17 & 8119.271 & 13.117 & 141.20 & 8642 & 2.829 & 19.120 & 4242 & 6.66E-02 & 9.30E-06 & 5.658 & 42.33 & 9.40E-05 & 3817 & 10.58 & 0.35 & 1.28E-20 & 1.05E-21 \\
 & 30080897 & 17 & 10187.777 & 16.364 & 166.90 & 2401 & 3.483 & 23.660 & 4242 & 1.06E-02 & 1.06E-06 & 1.0 & 3.16 & 1.06E-04 & 11730 & 4.93 & 0.65 & 3.19E-20 & 7.33E-22 \\
\enddata
\tablecomments{
The \# is the number of the SED data points. 
The $A_\mathrm{V}$ is the foreground extinction.
The $L_\mathrm{tot}$ is the total luminosity.
The $t_{\star}$, $M_{\star}$, $R_{\star}$ and $T_{\star}$ are the age, the mass, the radius and the temperature of central protostar, respectively.
The $M_\mathrm{disk}$ is the total mass of disk.
The $\dot{M}_\mathrm{disk}$ and $\dot{M}_\mathrm{env}$ are the mass accretion rate of disk and the in--fall rate of envelope, respectively.
The $R^{\mathrm{inner}}$ is the inner radius of both envelope and disk.
The $R_\mathrm{sub}$ is sublimation radius.
The $R^\mathrm{outer}_\mathrm{disk}$ and $R^\mathrm{outer}_\mathrm{env}$ are respectively the outer radius of disk and envelope.
The $\theta$ and $\varphi$ are the opening angle and the inclination of cavity. 
The $\rho_\mathrm{cav}$ and the $\rho_\mathrm{amb}$ are respectively cavity density and ambient density.
}
\end{deluxetable*}

%% file: ms.bbl
\begin{thebibliography}{}
\expandafter\ifx\csname natexlab\endcsname\relax\def\natexlab#1{#1}\fi
\providecommand{\url}[1]{\href{#1}{#1}}
\providecommand{\dodoi}[1]{doi:~\href{http://doi.org/#1}{\nolinkurl{#1}}}
\providecommand{\doeprint}[1]{\href{http://ascl.net/#1}{\nolinkurl{http://ascl.net/#1}}}
\providecommand{\doarXiv}[1]{\href{https://arxiv.org/abs/#1}{\nolinkurl{https://arxiv.org/abs/#1}}}

\bibitem[{Arce {et~al.}(2008)Arce, Santiago-García, Jørgensen, Tafalla, \&
  Bachiller}]{2008Arce_L1157_COMs_outflow}
Arce, H.~G., Santiago-García, J., Jørgensen, J.~K., Tafalla, M., \&
  Bachiller, R. 2008, The Astrophysical Journal, 681, L21,
  \dodoi{10.1086/590110}

\bibitem[{{Astropy Collaboration} {et~al.}(2013){Astropy Collaboration},
  {Robitaille}, {Tollerud}, {Greenfield}, {Droettboom}, {Bray}, {Aldcroft},
  {Davis}, {Ginsburg}, {Price-Whelan}, {Kerzendorf}, {Conley}, {Crighton},
  {Barbary}, {Muna}, {Ferguson}, {Grollier}, {Parikh}, {Nair}, {Unther},
  {Deil}, {Woillez}, {Conseil}, {Kramer}, {Turner}, {Singer}, {Fox}, {Weaver},
  {Zabalza}, {Edwards}, {Azalee Bostroem}, {Burke}, {Casey}, {Crawford},
  {Dencheva}, {Ely}, {Jenness}, {Labrie}, {Lim}, {Pierfederici}, {Pontzen},
  {Ptak}, {Refsdal}, {Servillat}, \& {Streicher}}]{astropy:2013}
{Astropy Collaboration}, {Robitaille}, T.~P., {Tollerud}, E.~J., {et~al.} 2013,
  \aap, 558, A33, \dodoi{10.1051/0004-6361/201322068}

\bibitem[{Bacmann {et~al.}(2012)Bacmann, Taquet, Faure, Kahane, \&
  Ceccarelli}]{2012Bacmann_COM_prestellar}
Bacmann, A., Taquet, V., Faure, A., Kahane, C., \& Ceccarelli, C. 2012,
  Astronomy and Astrophysics, 541, L12, \dodoi{10.1051/0004-6361/201219207}

\bibitem[{{Beckwith} {et~al.}(1990){Beckwith}, {Sargent}, {Chini}, \&
  {Guesten}}]{1990Beckwith_DustOpacity}
{Beckwith}, S. V.~W., {Sargent}, A.~I., {Chini}, R.~S., \& {Guesten}, R. 1990,
  \aj, 99, 924, \dodoi{10.1086/115385}

\bibitem[{Belloche {et~al.}(2016)Belloche, Müller, Garrod, \&
  Menten}]{2016Belloche_SgrB2N2_DH}
Belloche, A., Müller, H. S.~P., Garrod, R.~T., \& Menten, K.~M. 2016,
  Astronomy and Astrophysics, 587, A91, \dodoi{10.1051/0004-6361/201527268}

\bibitem[{Belloche {et~al.}(2020)Belloche, Maury, Maret, Anderl, Bacmann,
  André, Bontemps, Cabrit, Codella, Gaudel, Gueth, Lefèvre, Lefloch, Podio,
  \& Testi}]{2020Belloche_COM_CALYPSO}
Belloche, A., Maury, A.~J., Maret, S., {et~al.} 2020, Astronomy and
  Astrophysics, 635, A198, \dodoi{10.1051/0004-6361/201937352}

\bibitem[{Bergner {et~al.}(2019)Bergner, Martín-Doménech, Öberg, Jørgensen,
  Artur de~la Villarmois, \& Brinch}]{2019Bergner_Ser-emb_COM}
Bergner, J.~B., Martín-Doménech, R., Öberg, K.~I., {et~al.} 2019, ACS Earth
  and Space Chemistry, 3, 1564, \dodoi{10.1021/acsearthspacechem.9b00059}

\bibitem[{Bergner {et~al.}(2017)Bergner, Öberg, Garrod, \&
  Graninger}]{2017Bergner_COMs}
Bergner, J.~B., Öberg, K.~I., Garrod, R.~T., \& Graninger, D.~M. 2017, The
  Astrophysical Journal, 841, 120, \dodoi{10.3847/1538-4357/aa72f6}

\bibitem[{Bianchi {et~al.}(2019)Bianchi, Codella, Ceccarelli, Vazart,
  Bachiller, Balucani, Bouvier, De~Simone, Enrique-Romero, Kahane, Lefloch,
  López-Sepulcre, Ospina-Zamudio, Podio, \& Taquet}]{2019Bianchi_SVS13A}
Bianchi, E., Codella, C., Ceccarelli, C., {et~al.} 2019, Monthly Notices of the
  Royal Astronomical Society, 483, 1850, \dodoi{10.1093/mnras/sty2915}

\bibitem[{Bianchi {et~al.}(2020)Bianchi, Chandler, Ceccarelli, Codella, Sakai,
  López-Sepulcre, Maud, Moellenbrock, Svoboda, Watanabe, Sakai, Ménard,
  Aikawa, Alves, Balucani, Bouvier, Caselli, Caux, Charnley, Choudhury,
  De~Simone, Dulieu, Durán, Evans, Favre, Fedele, Feng, Fontani, Francis,
  Hama, Hanawa, Herbst, Hirota, Imai, Isella, Jiménez-Serra, Johnstone,
  Kahane, Lefloch, Loinard, Maureira, Mercimek, Miotello, Mori, Nakatani,
  Nomura, Oba, Ohashi, Okoda, Ospina-Zamudio, Oya, Pineda, Podio, Rimola, Cox,
  Shirley, Taquet, Testi, Vastel, Viti, Watanabe, Witzel, Xue, Zhang, Zhao, \&
  Yamamoto}]{2020Bianchi_L1551-IRS5}
Bianchi, E., Chandler, C.~J., Ceccarelli, C., {et~al.} 2020, Monthly Notices of
  the Royal Astronomical Society, 498, L87, \dodoi{10.1093/mnrasl/slaa130}

\bibitem[{Bottinelli {et~al.}(2004)Bottinelli, Ceccarelli, Neri, Williams,
  Caux, Cazaux, Lefloch, Maret, \& Tielens}]{2004Bottinelli_IRAS16293A_COM}
Bottinelli, S., Ceccarelli, C., Neri, R., {et~al.} 2004, The Astrophysical
  Journal, 617, L69, \dodoi{10.1086/426964}

\bibitem[{Brott \& Hauschildt(2005)}]{2005Brott_PHOENIX_GAIA}
Brott, I., \& Hauschildt, P.~H. 2005, in A PHOENIX Model Atmosphere Grid for
  Gaia, Vol. 576, eprint: arXiv:astro-ph/0503395, 565.
\newblock \url{https://ui.adsabs.harvard.edu/abs/2005ESASP.576..565B}

\bibitem[{Calcutt {et~al.}(2018)Calcutt, Jørgensen, Müller, Kristensen,
  Coutens, Bourke, Garrod, Persson, van~der Wiel, van Dishoeck, \&
  Wampfler}]{2018Calcutt_CH3CN_opacity}
Calcutt, H., Jørgensen, J.~K., Müller, H. S.~P., {et~al.} 2018, Astronomy and
  Astrophysics, 616, A90, \dodoi{10.1051/0004-6361/201732289}

\bibitem[{Calvet \& Gullbring(1998)}]{1998Calvet_magnetosphere}
Calvet, N., \& Gullbring, E. 1998, The Astrophysical Journal, 509, 802,
  \dodoi{10.1086/306527}

\bibitem[{Cazaux {et~al.}(2003)Cazaux, Tielens, Ceccarelli, Castets, Wakelam,
  Caux, Parise, \& Teyssier}]{2003Cazaux_IRAS16293-2422}
Cazaux, S., Tielens, A. G. G.~M., Ceccarelli, C., {et~al.} 2003, The
  Astrophysical Journal, 593, L51, \dodoi{10.1086/378038}

\bibitem[{Ceccarelli(2004)}]{2004Ceccarelli_HotCorino}
Ceccarelli, C. 2004, in Star Formation in the Interstellar Medium: In Honor of
  David Hollenbach, Chris McKee and Frank Shu, ASP Conference Proceeding, ed.
  D.~Johnstone, F.~Adams, D.~Lin, D.~Neufeld, \& E.~Ostriker, Vol. 323, San
  Francisco, Astronomical Society of the Pacific (ASP Conference Proceedings)

\bibitem[{Chini {et~al.}(1997)Chini, Reipurth, Ward-Thompson, Bally, Nyman,
  Sievers, \& Billawala}]{1997Chini_HOPS87_OMC3-MMS6}
Chini, R., Reipurth, B., Ward-Thompson, D., {et~al.} 1997, The Astrophysical
  Journal, 474, L135, \dodoi{10.1086/310436}

\bibitem[{Codella {et~al.}(2016)Codella, Ceccarelli, Cabrit, Gueth, Podio,
  Bachiller, Fontani, Gusdorf, Lefloch, Leurini, \&
  Tafalla}]{2016Codella_HH212}
Codella, C., Ceccarelli, C., Cabrit, S., {et~al.} 2016, Astronomy and
  Astrophysics, 586, L3, \dodoi{10.1051/0004-6361/201527424}

\bibitem[{Coutens {et~al.}(2020)Coutens, Commerçon, \&
  Wakelam}]{2020Coutens_CH3CN_CH3OH}
Coutens, A., Commerçon, B., \& Wakelam, V. 2020, Astronomy and Astrophysics,
  643, A108, \dodoi{10.1051/0004-6361/202038437}

\bibitem[{Coutens {et~al.}(2016)Coutens, Jørgensen, van~der Wiel, Müller,
  Lykke, Bjerkeli, Bourke, Calcutt, Drozdovskaya, Favre, Fayolle, Garrod,
  Jacobsen, Ligterink, Öberg, Persson, van Dishoeck, \&
  Wampfler}]{2016Coutens_IRAS19293B_NHDCHO}
Coutens, A., Jørgensen, J.~K., van~der Wiel, M. H.~D., {et~al.} 2016,
  Astronomy and Astrophysics, 590, L6, \dodoi{10.1051/0004-6361/201628612}

\bibitem[{De~Simone {et~al.}(2020)De~Simone, Ceccarelli, Codella, Svoboda,
  Chandler, Bouvier, Yamamoto, Sakai, Caselli, Favre, Loinard, Lefloch, Liu,
  López-Sepulcre, Pineda, Taquet, \& Testi}]{2020DeSimone_dust_opacity}
De~Simone, M., Ceccarelli, C., Codella, C., {et~al.} 2020, The Astrophysical
  Journal, 896, L3, \dodoi{10.3847/2041-8213/ab8d41}

\bibitem[{Di~Francesco {et~al.}(2008)Di~Francesco, Johnstone, Kirk, MacKenzie,
  \& Ledwosinska}]{2008Francesco_JCMTS_submm}
Di~Francesco, J., Johnstone, D., Kirk, H., MacKenzie, T., \& Ledwosinska, E.
  2008, The Astrophysical Journal Supplement Series, 175, 277,
  \dodoi{10.1086/523645}

\bibitem[{Drozdovskaya {et~al.}(2015)Drozdovskaya, Walsh, Visser, Harsono, \&
  van Dishoeck}]{2015Drozdovskaya_prestellar_CH3OH}
Drozdovskaya, M.~N., Walsh, C., Visser, R., Harsono, D., \& van Dishoeck, E.~F.
  2015, Monthly Notices of the Royal Astronomical Society, 451, 3836,
  \dodoi{10.1093/mnras/stv1177}

\bibitem[{Dunham {et~al.}(2008)Dunham, Crapsi, Evans~Ii, Bourke, Huard, Myers,
  \& Kauffmann}]{2008Dunham_G196a_SSTc2d}
Dunham, M.~M., Crapsi, A., Evans~Ii, N.~J., {et~al.} 2008, The Astrophysical
  Journal Supplement Series, 179, 249, \dodoi{10.1086/591085}

\bibitem[{Dutta {et~al.}(2020)Dutta, Lee, Liu, Hirano, Liu, Tatematsu, Kim,
  Shang, Sahu, Kim, Moraghan, Jhan, Hsu, Evans, Johnstone, Ward-Thompson, Kuan,
  Lee, Lee, Traficante, Juvela, Vastel, Zhang, Sanhueza, Soam, Kwon, Bronfman,
  Eden, Goldsmith, He, Wu, Pelkonen, Qin, Li, \& Li}]{2020Dutta_ALMASOP}
Dutta, S., Lee, C.-F., Liu, T., {et~al.} 2020, The Astrophysical Journal
  Supplement Series, 251, 20, \dodoi{10.3847/1538-4365/abba26}

\bibitem[{Eden {et~al.}(2019)Eden, Liu, Kim, Juvela, Liu, Tatematsu, Francesco,
  Wang, Wu, Thompson, Fuller, Li, Ristorcelli, Kang, Hirano, Johnstone, Lin,
  He, Koch, Sanhueza, Qin, Zhang, Goldsmith, Evans, Yuan, Zhang, White, Choi,
  Lee, Toth, Mairs, Yi, Tang, Soam, Peretto, Samal, Fich, Parsons, Malinen,
  Bendo, Rivera-Ingraham, Liu, Wouterloot, Li, Qian, Rawlings, Rawlings, Feng,
  Wang, Li, Liu, Luo, Marston, Pattle, Pelkonen, Rigby, Zahorecz, Zhang,
  Bőgner, Aikawa, Akhter, Alina, Bell, Bernard, Blain, Bronfman, Byun,
  Chapman, Chen, Chen, Chen, Chen, Chen, Chrysostomou, Chu, Chung, Cornu,
  Cosentino, Cunningham, Demyk, Drabek-Maunder, Doi, Eswaraiah, Falgarone,
  Fehér, Fraser, Friberg, Garay, Ge, Gear, Greaves, Guan, Harvey-Smith,
  Hasegawa, He, Henkel, Hirota, Holland, Hughes, Jarken,
  {et~al.}}]{2019Eden_SCOPE}
Eden, D.~J., Liu, T., Kim, K.-T., {et~al.} 2019, Monthly Notices of the Royal
  Astronomical Society, 485, 2895, \dodoi{10.1093/mnras/stz574}

\bibitem[{Evans {et~al.}(2015)Evans, Francesco, Lee, J{\o}rgensen, Choi, Myers,
  \& Mardones}]{2015Evans_B335_luminosity}
Evans, N.~J., Francesco, J.~D., Lee, J.-E., {et~al.} 2015, The Astrophysical
  Journal, 814, 22, \dodoi{10.1088/0004-637x/814/1/22}

\bibitem[{Evans~Ii {et~al.}(2003)Evans~Ii, Allen, Blake, Boogert, Bourke,
  Harvey, Kessler, Koerner, Lee, Mundy, Myers, Padgett, Pontoppidan, Sargent,
  Stapelfeldt, van Dishoeck, Young, \& Young}]{2003Ii_SSTc2d}
Evans~Ii, N., Allen, L., Blake, G., {et~al.} 2003, Publications of the
  Astronomical Society of the Pacific, 115, 965, \dodoi{10.1086/376697}

\bibitem[{Fazio {et~al.}(2004)Fazio, Hora, Allen, Ashby, Barmby, Deutsch,
  Huang, Kleiner, Marengo, Megeath, Melnick, Pahre, Patten, Polizotti, Smith,
  Taylor, Wang, Willner, Hoffmann, Pipher, Forrest, McMurty, McCreight,
  McKelvey, McMurray, Koch, Moseley, Arendt, Mentzell, Marx, Losch, Mayman,
  Eichhorn, Krebs, Jhabvala, Gezari, Fixsen, Flores, Shakoorzadeh, Jungo,
  Hakun, Workman, Karpati, Kichak, Whitley, Mann, Tollestrup, Eisenhardt,
  Stern, Gorjian, Bhattacharya, Carey, Nelson, Glaccum, Lacy, Lowrance, Laine,
  Reach, Stauffer, Surace, Wilson, Wright, Hoffman, Domingo, \&
  Cohen}]{2004Fazio_IRAC}
Fazio, G.~G., Hora, J.~L., Allen, L.~E., {et~al.} 2004, The Astrophysical
  Journal Supplement Series, 154, 10, \dodoi{10.1086/422843}

\bibitem[{Feddersen {et~al.}(2020)Feddersen, Arce, Kong, Suri, Sánchez-Monge,
  Ossenkopf-Okada, Dunham, Nakamura, Shimajiri, \&
  Bally}]{2020Feddersen_CARMA-NRO_orion}
Feddersen, J.~R., Arce, H.~G., Kong, S., {et~al.} 2020, The Astrophysical
  Journal, 896, 11, \dodoi{10.3847/1538-4357/ab86a9}

\bibitem[{Fischer {et~al.}(2010)Fischer, Megeath, Ali, Tobin, Osorio, Allen,
  Kryukova, Stanke, Stutz, Bergin, Calvet, di~Francesco, Furlan, Hartmann,
  Henning, Krause, Manoj, Maret, Muzerolle, Myers, Neufeld, Pontoppidan,
  Poteet, Watson, \& Wilson}]{2010Fischer_HOPS168_Infall}
Fischer, W.~J., Megeath, S.~T., Ali, B., {et~al.} 2010, Astronomy and
  Astrophysics, 518, L122, \dodoi{10.1051/0004-6361/201014636}

\bibitem[{Fischer {et~al.}(2013)Fischer, Megeath, Stutz, Tobin, Ali, Stanke,
  Osorio, Furlan, Team, \& Orion~Protostar}]{2012Fischer_HOPS}
Fischer, W.~J., Megeath, S.~T., Stutz, A.~M., {et~al.} 2013, Astronomische
  Nachrichten, 334, 53, \dodoi{10.1002/asna.201211761}

\bibitem[{Furlan {et~al.}(2016)Furlan, Fischer, Ali, Stutz, Stanke, Tobin,
  Megeath, Osorio, Hartmann, Calvet, Poteet, Booker, Manoj, Watson, \&
  Allen}]{2016Furlan_HOPS_SED}
Furlan, E., Fischer, W.~J., Ali, B., {et~al.} 2016, The Astrophysical Journal
  Supplement Series, 224, 5, \dodoi{10.3847/0067-0049/224/1/5}

\bibitem[{Garrod \& Herbst(2006)}]{2006Garrod_3phase}
Garrod, R.~T., \& Herbst, E. 2006, Astronomy and Astrophysics, 457, 927,
  \dodoi{10.1051/0004-6361:20065560}

\bibitem[{Garrod {et~al.}(2008)Garrod, Widicus~Weaver, \&
  Herbst}]{2008Garrod_3phase}
Garrod, R.~T., Widicus~Weaver, S.~L., \& Herbst, E. 2008, The Astrophysical
  Journal, 682, 283, \dodoi{10.1086/588035}

\bibitem[{{G{\"u}sten} {et~al.}(2006){G{\"u}sten}, {Nyman}, {Schilke},
  {Menten}, {Cesarsky}, \& {Booth}}]{2006Gusten_APEX}
{G{\"u}sten}, R., {Nyman}, L.~{\r{A}}., {Schilke}, P., {et~al.} 2006, Astronomy
  and Astrophysics, 454, L13, \dodoi{10.1051/0004-6361:20065420}

\bibitem[{Herbst \& van Dishoeck(2009)}]{2009Herbst_COM_review}
Herbst, E., \& van Dishoeck, E.~F. 2009, \araa, 47, 427,
  \dodoi{10.1146/annurev-astro-082708-101654}

\bibitem[{Higuchi {et~al.}(2018)Higuchi, Sakai, Watanabe,
  {et~al.}}]{2018Higuchi_survey_Perseus}
Higuchi, A.~E., Sakai, N., Watanabe, Y., {et~al.} 2018, \apjs, 236, 52,
  \dodoi{10.3847/1538-4365/aabfe9}

\bibitem[{Hsu {et~al.}(2020)Hsu, Liu, Liu, Sahu, Hirano, Lee, Tatematsu, Kim,
  Juvela, Sanhueza, He, Johnstone, Qin, Bronfman, Chen, Dutta, Eden, Jhan, Kim,
  Kuan, Kwon, Lee, Lee, Moraghan, Rawlings, Shang, Soam, Thompson, Traficante,
  Wu, Yang, \& Zhang}]{Hsu2020_ALMASOP}
Hsu, S.-Y., Liu, S.-Y., Liu, T., {et~al.} 2020, The Astrophysical Journal, 898,
  107, \dodoi{10.3847/1538-4357/ab9f3a}

\bibitem[{Huang {et~al.}(2005)Huang, Kuan, Charnley, Hirano, Takakuwa, \&
  Bourke}]{2005Huang_IRAS16293B_COM}
Huang, H.-C., Kuan, Y.-J., Charnley, S.~B., {et~al.} 2005, Advances in Space
  Research, 36, 146, \dodoi{10.1016/j.asr.2005.03.115}

\bibitem[{Husser {et~al.}(2013)Husser, Wende-von Berg, Dreizler, Homeier,
  Reiners, Barman, \& Hauschildt}]{2013Husser_PHOENIX_HiRes}
Husser, T.~O., Wende-von Berg, S., Dreizler, S., {et~al.} 2013, Astronomy and
  Astrophysics, 553, A6, \dodoi{10.1051/0004-6361/201219058}

\bibitem[{Imai {et~al.}(2016)}]{2016Imai_B335}
Imai, M., {et~al.} 2016, \apj, 830, L37, \dodoi{10.3847/2041-8205/830/2/l37}

\bibitem[{Indebetouw {et~al.}(2005)Indebetouw, Mathis, Babler, Meade, Watson,
  Whitney, Wolff, Wolfire, Cohen, Bania, Benjamin, Clemens, Dickey, Jackson,
  Kobulnicky, Marston, Mercer, Stauffer, Stolovy, \&
  Churchwell}]{2005Indebetouw_extinction}
Indebetouw, R., Mathis, J.~S., Babler, B.~L., {et~al.} 2005, The Astrophysical
  Journal, 619, 931, \dodoi{10.1086/426679}

\bibitem[{Ishihara {et~al.}(2010)Ishihara, Onaka, Kataza, Salama, Alfageme,
  Cassatella, Cox, García-Lario, Stephenson, Cohen, Fujishiro, Fujiwara,
  Hasegawa, Ita, Kim, Matsuhara, Murakami, Müller, Nakagawa, Ohyama, Oyabu,
  Pyo, Sakon, Shibai, Takita, Tanabé, Uemizu, Ueno, Usui, Wada, Watarai,
  Yamamura, \& Yamauchi}]{2010Ishihara_SED_AKARI_IRC}
Ishihara, D., Onaka, T., Kataza, H., {et~al.} 2010, Astronomy and Astrophysics,
  514, A1, \dodoi{10.1051/0004-6361/200913811}

\bibitem[{Jacobsen {et~al.}(2019)Jacobsen, Jørgensen, Di~Francesco, Evans,
  Choi, \& Lee}]{2019Jacobsen_L483_COM}
Jacobsen, S.~K., Jørgensen, J.~K., Di~Francesco, J., {et~al.} 2019, Astronomy
  and Astrophysics, 629, A29, \dodoi{10.1051/0004-6361/201833214}

\bibitem[{Juvela {et~al.}(2018)Juvela, {He, Jinhua}, {Pattle, Katherine}, {Liu,
  Tie}, {Bendo, George}, {Eden, David J.}, {Feh\'er, Orsolya}, {Michel, Fich},
  {Fuller, Gary}, {Hirano, Naomi}, {Kim, Kee-Tae}, {Li, Di}, {Liu, Sheng-Yuan},
  {Malinen, Johanna}, {Marshall, Douglas J.}, {Paradis, Deborah}, {Parsons,
  Harriet}, {Pelkonen, Veli-Matti}, {Rawlings, Mark G.}, {Ristorcelli,
  Isabelle}, {Samal, Manash R.}, {Tatematsu, Ken\'{}ichi}, {Thompson, Mark},
  {Traficante, Alessio}, {Wang, Ke}, {Ward-Thompson, Derek}, {Wu, Yuefang},
  {Yi, Hee-Weon}, \& {Yoo, Hyunju}}]{2018Juvela_dustIndex}
Juvela, M., {He, Jinhua}, {Pattle, Katherine}, {et~al.} 2018, A\&A, 612, A71,
  \dodoi{10.1051/0004-6361/201731921}

\bibitem[{Jørgensen {et~al.}(2018)Jørgensen, Müller, Calcutt, Coutens,
  Drozdovskaya, Öberg, Persson, Taquet, van Dishoeck, \&
  Wampfler}]{2018Jorgensen_IRAS16293B_COM}
Jørgensen, J.~K., Müller, H. S.~P., Calcutt, H., {et~al.} 2018, Astronomy and
  Astrophysics, 620, A170, \dodoi{10.1051/0004-6361/201731667}

\bibitem[{Kahane {et~al.}(2018)Kahane, Jaber Al-Edhari, Ceccarelli,
  López-Sepulcre, Fontani, \& Kama}]{2018Kahane_13C12C_Orion}
Kahane, C., Jaber Al-Edhari, A., Ceccarelli, C., {et~al.} 2018, The
  Astrophysical Journal, 852, 130, \dodoi{10.3847/1538-4357/aa9e88}

\bibitem[{Kang {et~al.}(2013)Kang, Lee, Choi, Choi, Kim, Di~Francesco, \&
  Park}]{2013Kang_HOPS168_maser}
Kang, M., Lee, J.-E., Choi, M., {et~al.} 2013, The Astrophysical Journal
  Supplement Series, 209, 25, \dodoi{10.1088/0067-0049/209/2/25}

\bibitem[{Kauffmann {et~al.}(2008)Kauffmann, {Bertoldi, F.}, {Bourke, T. L.},
  {Evans, N. J.}, \& {Lee, C. W.}}]{2008Kauffmann_CONTFormula}
Kauffmann, J., {Bertoldi, F.}, {Bourke, T. L.}, {Evans, N. J.}, \& {Lee, C. W.}
  2008, \aap, 487, 993, \dodoi{10.1051/0004-6361:200809481}

\bibitem[{Kim {et~al.}(2020)Kim, Tatematsu, Liu, Yi, He, Hirano, Liu, Choi,
  Sanhueza, Tóth, Evans~Ii, Feng, Juvela, Kim, Vastel, Lee,
  Nguyễn~Lu’o’, Kang, Ristorcelli, Fehér, Wu, Ohashi, Wang, Kandori,
  Hirota, Sakai, Lu, Thompson, Fuller, Li, Shinnaga, \&
  Kim}]{2020Kim_ALMASOP_Nobeyama}
Kim, G., Tatematsu, K., Liu, T., {et~al.} 2020, The Astrophysical Journal
  Supplement Series, 249, 33, \dodoi{10.3847/1538-4365/aba746}

\bibitem[{Kim {et~al.}(1994)Kim, Martin, \& Hendry}]{1994Kim_dust_KMH}
Kim, S.-H., Martin, P.~G., \& Hendry, P.~D. 1994, The Astrophysical Journal,
  422, 164, \dodoi{10.1086/173714}

\bibitem[{Kong {et~al.}(2018)Kong, Arce, Feddersen, Carpenter, Nakamura,
  Shimajiri, Isella, Ossenkopf-Okada, Sargent, Sánchez-Monge, Suri, Kauffmann,
  Pillai, Pineda, Koda, Bally, Lis, Padoan, Klessen, Mairs, Goodman, Goldsmith,
  McGehee, Schilke, Teuben, Maureira, Hara, Ginsburg, Burkhart, Smith,
  Schmiedeke, Pineda, Ishii, Sasaki, Kawabe, Urasawa, Oyamada, \&
  Tanabe}]{2018Kong_CARMA-NRO_intro}
Kong, S., Arce, H.~G., Feddersen, J.~R., {et~al.} 2018, The Astrophysical
  Journal Supplement Series, 236, 25, \dodoi{10.3847/1538-4365/aabafc}

\bibitem[{Kounkel {et~al.}(2016)Kounkel, Megeath, Poteet, Fischer, \&
  Hartmann}]{2016Kounkel_G209S1}
Kounkel, M., Megeath, S.~T., Poteet, C.~A., Fischer, W.~J., \& Hartmann, L.
  2016, The Astrophysical Journal, 821, 52, \dodoi{10.3847/0004-637X/821/1/52}

\bibitem[{Kounkel {et~al.}(2018)Kounkel, Covey, Suárez, Román-Zúñiga,
  Hernandez, Stassun, Jaehnig, Feigelson, Peña~Ramírez, Roman-Lopes, Da~Rio,
  Stringfellow, Kim, Borissova, Fernández-Trincado, Burgasser,
  García-Hernández, Zamora, Pan, \& Nitschelm}]{2018Kounkel_Orion_distance}
Kounkel, M., Covey, K., Suárez, G., {et~al.} 2018, The Astronomical Journal,
  156, 84, \dodoi{10.3847/1538-3881/aad1f1}

\bibitem[{Kuan {et~al.}(2004)Kuan, Huang, Charnley, Hirano, Takakuwa, Wilner,
  Liu, Ohashi, Bourke, Qi, \& Zhang}]{2004Kuan_IRAS16293B_COM}
Kuan, Y.-J., Huang, H.-C., Charnley, S.~B., {et~al.} 2004, The Astrophysical
  Journal, 616, L27, \dodoi{10.1086/426315}

\bibitem[{Lawrence {et~al.}(2007)Lawrence, Warren, Almaini, Edge, Hambly,
  Jameson, Lucas, Casali, Adamson, Dye, Emerson, Foucaud, Hewett, Hirst,
  Hodgkin, Irwin, Lodieu, McMahon, Simpson, Smail, Mortlock, \&
  Folger}]{2007Lawrence_SED_UKIDSS}
Lawrence, A., Warren, S.~J., Almaini, O., {et~al.} 2007, Monthly Notices of the
  Royal Astronomical Society, 379, 1599,
  \dodoi{10.1111/j.1365-2966.2007.12040.x}

\bibitem[{Lee {et~al.}(2019{\natexlab{a}})Lee, Codella, Li, \&
  Liu}]{2019Lee_HH212}
Lee, C.-F., Codella, C., Li, Z.-Y., \& Liu, S.-Y. 2019{\natexlab{a}}, The
  Astrophysical Journal, 876, 63, \dodoi{10.3847/1538-4357/ab15db}

\bibitem[{Lee {et~al.}(2017)Lee, Li, Ho, Hirano, Zhang, \&
  Shang}]{2017Lee_HH212}
Lee, C.-F., Li, Z.-Y., Ho, P. T.~P., {et~al.} 2017, The Astrophysical Journal,
  843, 27, \dodoi{10.3847/1538-4357/aa7757}

\bibitem[{Lee {et~al.}(2019{\natexlab{b}})Lee, Lee, Baek, Aikawa, Cieza, Yoon,
  Herczeg, Johnstone, \& Casassus}]{2019Lee_V883Ori_COMs_disk}
Lee, J.-E., Lee, S., Baek, G., {et~al.} 2019{\natexlab{b}}, Nature Astronomy,
  3, 314, \dodoi{10.1038/s41550-018-0680-0}

\bibitem[{Lee {et~al.}(2021)Lee, Johnstone, Lee, Herczeg, Mairs,
  Contreras-Peña, Hatchell, Naylor, Bell, Bourke, Broughton, Francis, Gupta,
  Harsono, Liu, Park, Plovie, Moriarty-Schieven, Scholz, Sharma, Teixeira,
  Wang, Aikawa, Bower, Chen, Bae, Baek, Chapman, Chen, Du, Dutta, Forbrich,
  Guo, Inutsuka, Kang, Kirk, Kuan, Kwon, Lai, Lalchand, Lane, Lee, Liu, Morata,
  Pearson, Pon, Sahu, Shang, Stamatellos, Tang, Xu, \&
  Yoo}]{2021Lee_JCMTTransient}
Lee, Y.-H., Johnstone, D., Lee, J.-E., {et~al.} 2021, arXiv e-prints,
  arXiv:2107.10750.
\newblock \url{https://ui.adsabs.harvard.edu/abs/2021arXiv210710750L}

\bibitem[{Liu {et~al.}(2018)Liu, Kim, Juvela, Wang, Tatematsu, Di~Francesco,
  Liu, Wu, Thompson, Fuller, Eden, Li, Ristorcelli, Kang, Lin, Johnstone, He,
  Koch, Sanhueza, Qin, Zhang, Hirano, Goldsmith, Evans, White, Choi, Lee, Toth,
  Mairs, Yi, Tang, Soam, Peretto, Samal, Fich, Parsons, Yuan, Zhang, Malinen,
  Bendo, Rivera-Ingraham, Liu, Wouterloot, Li, Qian, Rawlings, Rawlings, Feng,
  Aikawa, Akhter, Alina, Bell, Bernard, Blain, Bőgner, Bronfman, Byun,
  Chapman, Chen, Chen, Chen, Chen, Chen, Chrysostomou, Cosentino, Cunningham,
  Demyk, Drabek-Maunder, Doi, Eswaraiah, Falgarone, Fehér, Fraser, Friberg,
  Garay, Ge, Gear, Greaves, Guan, Harvey-Smith, Hasegawa, Hatchell, He, Henkel,
  Hirota, Holland, Hughes, Jarken, Ji, Jimenez-Serra, Kang, Kawabata, Kim, Kim,
  Kim, Kim, Koo, Kwon, Kuan, Lacaille, {et~al.}}]{2018Liu_TOP-SCOPE}
Liu, T., Kim, K.-T., Juvela, M., {et~al.} 2018, The Astrophysical Journal
  Supplement Series, 234, 28, \dodoi{10.3847/1538-4365/aaa3dd}

\bibitem[{Lo {et~al.}(1975)Lo, Burke, \& Haschick}]{1975Lo_HOPS168_maser}
Lo, K.~Y., Burke, B.~F., \& Haschick, A.~D. 1975, The Astrophysical Journal,
  202, 81, \dodoi{10.1086/153954}

\bibitem[{López-Sepulcre {et~al.}(2019)López-Sepulcre, Balucani, Ceccarelli,
  Codella, Dulieu, \& Theulé}]{2019Lopez-Sepulcre_NH2CHO}
López-Sepulcre, A., Balucani, N., Ceccarelli, C., {et~al.} 2019, ACS Earth and
  Space Chemistry, 3, 2122, \dodoi{10.1021/acsearthspacechem.9b00154}

\bibitem[{López-Sepulcre {et~al.}(2015)López-Sepulcre, Jaber, Mendoza,
  Lefloch, Ceccarelli, Vastel, Bachiller, Cernicharo, Codella, Kahane, Kama, \&
  Tafalla}]{2015Lopez-Sepulcre_NH2CHO}
López-Sepulcre, A., Jaber, A.~A., Mendoza, E., {et~al.} 2015, \mnras, 449,
  2438, \dodoi{10.1093/mnras/stv377}

\bibitem[{Mairs {et~al.}(2018)Mairs, Bell, Johnstone, Herczeg, Bower, Aikawa,
  Lee, Chen, Hatchell, Kang, Contreras~Pena, Scholz, \&
  Naylor}]{2018Mairs_HOPS358_850um}
Mairs, S., Bell, G.~S., Johnstone, D., {et~al.} 2018, The Astronomer's
  Telegram, 11583, 1.
\newblock \url{https://ui.adsabs.harvard.edu/abs/2018ATel11583....1M}

\bibitem[{Manigand {et~al.}(2020)Manigand, Jørgensen, Calcutt, Müller,
  Ligterink, Coutens, Drozdovskaya, van Dishoeck, \&
  Wampfler}]{2020Manigand_IRAS16293-2422-A_COMs}
Manigand, S., Jørgensen, J.~K., Calcutt, H., {et~al.} 2020, Astronomy and
  Astrophysics, 635, A48, \dodoi{10.1051/0004-6361/201936299}

\bibitem[{{McMullin} {et~al.}(2007){McMullin}, Waters, Schiebel, Young, \&
  Golap}]{2007McMullin_CASA}
{McMullin}, J.~P., Waters, B., Schiebel, D., Young, W., \& Golap, K. 2007, in
  Astronomical Data Analysis Software and Systems XVI, ed. R.~A. Shaw, F.~Hill,
  \& D.~J. Bell, Vol. 376, San Francisco, Astronomical Society of the Pacific
  (ASP Conference Series)

\bibitem[{Megeath {et~al.}(2012)Megeath, Gutermuth, Muzerolle, Kryukova,
  Flaherty, Hora, Allen, Hartmann, Myers, Pipher, Stauffer, Young, \&
  Fazio}]{2012Megeath_MGM2012}
Megeath, S.~T., Gutermuth, R., Muzerolle, J., {et~al.} 2012, The Astronomical
  Journal, 144, 192, \dodoi{10.1088/0004-6256/144/6/192}

\bibitem[{M\"oller {et~al.}(2017)M\"oller, {Endres, C.}, \& {Schilke,
  P.}}]{2017Moller_XCLASS}
M\"oller, T., {Endres, C.}, \& {Schilke, P.} 2017, \aap, 598, A7,
  \dodoi{10.1051/0004-6361/201527203}

\bibitem[{Müller {et~al.}(2005)Müller, Schlöder, Stutzki, \&
  Winnewisser}]{2005CDMS}
Müller, H.~S., Schlöder, F., Stutzki, J., \& Winnewisser, G. 2005, Journal of
  Molecular Structure, 742, 215 ,
  \dodoi{https://doi.org/10.1016/j.molstruc.2005.01.027}

\bibitem[{Müller {et~al.}(2016)Müller, Belloche, Xu, Lees, Garrod, Walters,
  van Wijngaarden, Lewen, Schlemmer, \& Menten}]{2016Muller_SgrB2N2_13C12C}
Müller, H. S.~P., Belloche, A., Xu, L.-H., {et~al.} 2016, Astronomy and
  Astrophysics, 587, A92, \dodoi{10.1051/0004-6361/201527470}

\bibitem[{Nagy {et~al.}(2020)Nagy, Menechella, Megeath, Tobin, Booker, Fischer,
  Manoj, Stanke, Stutz, \& Wyrowski}]{2020Nagy_APEXSurvey_outflow}
Nagy, Z., Menechella, A., Megeath, S.~T., {et~al.} 2020, Astronomy and
  Astrophysics, 642, A137, \dodoi{10.1051/0004-6361/201937342}

\bibitem[{Nazari {et~al.}(2021)Nazari, van Gelder, van Dishoeck, Tabone, van't
  Hoff, Ligterink, Beuther, Boogert, Caratti~o Garatti, Klaassen, Linnartz,
  Taquet, \& Tychoniec}]{2021Nazari_COMs}
Nazari, P., van Gelder, M.~L., van Dishoeck, E.~F., {et~al.} 2021, Astronomy
  and Astrophysics, 650, A150, \dodoi{10.1051/0004-6361/202039996}

\bibitem[{Ortiz-León {et~al.}(2018)Ortiz-León, Loinard, Dzib, Galli, Kounkel,
  Mioduszewski, Rodríguez, Torres, Hartmann, Boden, Evans, Briceño, \&
  Tobin}]{2018Ortiz_Perseus_distance}
Ortiz-León, G.~N., Loinard, L., Dzib, S.~A., {et~al.} 2018, The Astrophysical
  Journal, 865, 73, \dodoi{10.3847/1538-4357/aada49}

\bibitem[{Persson {et~al.}(2018)Persson, Jørgensen, Müller, Coutens, van
  Dishoeck, Taquet, Calcutt, van~der Wiel, Bourke, \&
  Wampfler}]{2018Persson_IRAS16293B_H2CO}
Persson, M.~V., Jørgensen, J.~K., Müller, H. S.~P., {et~al.} 2018, Astronomy
  and Astrophysics, 610, A54, \dodoi{10.1051/0004-6361/201731684}

\bibitem[{Pickett {et~al.}(1998)Pickett, Poynter, Cohen, Delitsky, Pearson, \&
  Muller}]{1988JPL}
Pickett, H., Poynter, R., Cohen, E., {et~al.} 1998, Journal of Quantitative
  Spectroscopy and Radiative Transfer, 60, 883 ,
  \dodoi{https://doi.org/10.1016/S0022-4073(98)00091-0}

\bibitem[{Pilbratt {et~al.}(2010)Pilbratt, Riedinger, Passvogel, Crone, Doyle,
  Gageur, Heras, Jewell, Metcalfe, Ott, \& Schmidt}]{2010Pilbratt_Herschel}
Pilbratt, G.~L., Riedinger, J.~R., Passvogel, T., {et~al.} 2010, Astronomy and
  Astrophysics, 518, L1, \dodoi{10.1051/0004-6361/201014759}

\bibitem[{Planck {et~al.}(2016)Planck, Ade, Aghanim, Arnaud, Ashdown, Aumont,
  Baccigalupi, Banday, Barreiro, Bartolo, Battaner, Benabed, Benoît,
  Benoit-Lévy, Bernard, Bersanelli, Bielewicz, Bonaldi, Bonavera, Bond,
  Borrill, Bouchet, Boulanger, Bucher, Burigana, Butler, Calabrese, Catalano,
  Chamballu, Chiang, Christensen, Clements, Colombi, Colombo, Combet, Couchot,
  Coulais, Crill, Curto, Cuttaia, Danese, Davies, Davis, de~Bernardis, de~Rosa,
  de~Zotti, Delabrouille, Désert, Dickinson, Diego, Dole, Donzelli, Doré,
  Douspis, Ducout, Dupac, Efstathiou, Elsner, Enßlin, Eriksen, Falgarone,
  Fergusson, Finelli, Forni, Frailis, Fraisse, Franceschi, Frejsel, Galeotta,
  Galli, Ganga, Giard, Giraud-Héraud, Gjerløw, González-Nuevo, Górski,
  Gratton, Gregorio, Gruppuso, Gudmundsson, Hansen, Hanson, Harrison, Helou,
  Henrot-Versillé, Hernández-Monteagudo, Herranz, Hildebrandt, Hivon, Hobson,
  Holmes, Hornstrup, Hovest, Huffenberger, Hurier, Jaffe, Jaffe, Jones, Juvela,
  Keihänen, {et~al.}}]{2016Planck_PGCC}
Planck, C., Ade, P. A.~R., Aghanim, N., {et~al.} 2016, Astronomy and
  Astrophysics, 594, A28, \dodoi{10.1051/0004-6361/201525819}

\bibitem[{Poglitsch {et~al.}(2010)Poglitsch, Waelkens, Geis, Feuchtgruber,
  Vandenbussche, Rodriguez, Krause, Renotte, van Hoof, Saraceno, Cepa,
  Kerschbaum, Agnèse, Ali, Altieri, Andreani, Augueres, Balog, Barl, Bauer,
  Belbachir, Benedettini, Billot, Boulade, Bischof, Blommaert, Callut, Cara,
  Cerulli, Cesarsky, Contursi, Creten, De~Meester, Doublier, Doumayrou, Duband,
  Exter, Genzel, Gillis, Grözinger, Henning, Herreros, Huygen, Inguscio,
  Jakob, Jamar, Jean, de~Jong, Katterloher, Kiss, Klaas, Lemke, Lutz, Madden,
  Marquet, Martignac, Mazy, Merken, Montfort, Morbidelli, Müller, Nielbock,
  Okumura, Orfei, Ottensamer, Pezzuto, Popesso, Putzeys, Regibo, Reveret,
  Royer, Sauvage, Schreiber, Stegmaier, Schmitt, Schubert, Sturm, Thiel,
  Tofani, Vavrek, Wetzstein, Wieprecht, \& Wiezorrek}]{2010Poglitsch_PACS}
Poglitsch, A., Waelkens, C., Geis, N., {et~al.} 2010, Astronomy and
  Astrophysics, 518, L2, \dodoi{10.1051/0004-6361/201014535}

\bibitem[{{Price-Whelan} {et~al.}(2018){Price-Whelan}, {Sip{\H{o}}cz},
  {G{\"u}nther}, {Lim}, {Crawford}, {Conseil}, {Shupe}, {Craig}, {Dencheva},
  {Ginsburg}, {VanderPlas}, {Bradley}, {P{\'e}rez-Su{\'a}rez}, {de Val-Borro},
  {Paper Contributors}, {Aldcroft}, {Cruz}, {Robitaille}, {Tollerud},
  {Coordination Committee}, {Ardelean}, {Babej}, {Bach}, {Bachetti}, {Bakanov},
  {Bamford}, {Barentsen}, {Barmby}, {Baumbach}, {Berry}, {Biscani}, {Boquien},
  {Bostroem}, {Bouma}, {Brammer}, {Bray}, {Breytenbach}, {Buddelmeijer},
  {Burke}, {Calderone}, {Cano Rodr{\'\i}guez}, {Cara}, {Cardoso}, {Cheedella},
  {Copin}, {Corrales}, {Crichton}, {D{\textquoteright}Avella}, {Deil},
  {Depagne}, {Dietrich}, {Donath}, {Droettboom}, {Earl}, {Erben}, {Fabbro},
  {Ferreira}, {Finethy}, {Fox}, {Garrison}, {Gibbons}, {Goldstein}, {Gommers},
  {Greco}, {Greenfield}, {Groener}, {Grollier}, {Hagen}, {Hirst}, {Homeier},
  {Horton}, {Hosseinzadeh}, {Hu}, {Hunkeler}, {Ivezi{\'c}}, {Jain}, {Jenness},
  {Kanarek}, {Kendrew}, {Kern}, {Kerzendorf}, {Khvalko}, {King}, {Kirkby},
  {Kulkarni}, {Kumar}, {Lee}, {Lenz}, {Littlefair}, {Ma}, {Macleod},
  {Mastropietro}, {McCully}, {Montagnac}, {Morris}, {Mueller}, {Mumford},
  {Muna}, {Murphy}, {Nelson}, {Nguyen}, {Ninan}, {N{\"o}the}, {Ogaz}, {Oh},
  {Parejko}, {Parley}, {Pascual}, {Patil}, {Patil}, {Plunkett}, {Prochaska},
  {Rastogi}, {Reddy Janga}, {Sabater}, {Sakurikar}, {Seifert}, {Sherbert},
  {Sherwood-Taylor}, {Shih}, {Sick}, {Silbiger}, {Singanamalla}, {Singer},
  {Sladen}, {Sooley}, {Sornarajah}, {Streicher}, {Teuben}, {Thomas},
  {Tremblay}, {Turner}, {Terr{\'o}n}, {van Kerkwijk}, {de la Vega}, {Watkins},
  {Weaver}, {Whitmore}, {Woillez}, {Zabalza}, \& {Contributors}}]{astropy:2018}
{Price-Whelan}, A.~M., {Sip{\H{o}}cz}, B.~M., {G{\"u}nther}, H.~M., {et~al.}
  2018, \aj, 156, 123, \dodoi{10.3847/1538-3881/aabc4f}

\bibitem[{Quénard {et~al.}(2018)Quénard, Jiménez-Serra, Viti, Holdship, \&
  Coutens}]{2018Quenard_NH2CHO_HNCO}
Quénard, D., Jiménez-Serra, I., Viti, S., Holdship, J., \& Coutens, A. 2018,
  Monthly Notices of the Royal Astronomical Society, 474, 2796,
  \dodoi{10.1093/mnras/stx2960}

\bibitem[{Rieke {et~al.}(2004)Rieke, Young, Engelbracht, Kelly, Low, Haller,
  Beeman, Gordon, Stansberry, Misselt, Cadien, Morrison, Rivlis, Latter,
  Noriega-Crespo, Padgett, Stapelfeldt, Hines, Egami, Muzerolle,
  Alonso-Herrero, Blaylock, Dole, Hinz, Le~Floc'h, Papovich, Pérez-González,
  Smith, Su, Bennett, Frayer, Henderson, Lu, Masci, Pesenson, Rebull, Rho,
  Keene, Stolovy, Wachter, Wheaton, Werner, \& Richards}]{2004Rieke_MIPS}
Rieke, G.~H., Young, E.~T., Engelbracht, C.~W., {et~al.} 2004, The
  Astrophysical Journal Supplement Series, 154, 25, \dodoi{10.1086/422717}

\bibitem[{Robitaille {et~al.}(2007)Robitaille, Whitney, Indebetouw, \&
  Wood}]{2007Robitaille_sedfitter}
Robitaille, T.~P., Whitney, B.~A., Indebetouw, R., \& Wood, K. 2007, The
  Astrophysical Journal Supplement Series, 169, 328, \dodoi{10.1086/512039}

\bibitem[{Robitaille {et~al.}(2006)Robitaille, Whitney, Indebetouw, Wood, \&
  Denzmore}]{2006Robitaille_sedfitter}
Robitaille, T.~P., Whitney, B.~A., Indebetouw, R., Wood, K., \& Denzmore, P.
  2006, The Astrophysical Journal Supplement Series, 167, 256,
  \dodoi{10.1086/508424}

\bibitem[{Sahu {et~al.}(2019)Sahu, Liu, Su, Li, Lee, Hirano, \&
  Takakuwa}]{2019Sahu_IRAS4A1}
Sahu, D., Liu, S.-Y., Su, Y.-N., {et~al.} 2019, The Astrophysical Journal, 872,
  196, \dodoi{10.3847/1538-4357/aaffda}

\bibitem[{Sahu {et~al.}(2021)Sahu, Liu, Liu, Evans, Hirano, Tatematsu, Lee,
  Kim, Dutta, Alina, Bronfman, Cunningham, Eden, Garay, Goldsmith, He, Hsu,
  Jhan, Johnstone, Juvela, Kim, Kuan, Kwon, Lee, Lee, Li, Li, Li, Luo,
  Montillaud, Moraghan, Pelkonen, Qin, Ristorcelli, Sanhueza, Shang, Shen,
  Soam, Wu, Zhang, \& Zhou}]{2021Sahu_ALMASOP}
Sahu, D., Liu, S.-Y., Liu, T., {et~al.} 2021, The Astrophysical Journal, 907,
  L15, \dodoi{10.3847/2041-8213/abd3aa}

\bibitem[{Saladino {et~al.}(2012)Saladino, Botta, Pino, Costanzo, \&
  Di~Mauro}]{2012Saladino_NH2CHO}
Saladino, R., Botta, G., Pino, S., Costanzo, G., \& Di~Mauro, E. 2012, Chem.
  Soc. Rev., 41, 5526, \dodoi{10.1039/C2CS35066A}

\bibitem[{Siringo {et~al.}(2009)Siringo, Kreysa, Kovács, Schuller, Weiß,
  Esch, Gemünd, Jethava, Lundershausen, Colin, Güsten, Menten, Beelen,
  Bertoldi, Beeman, \& Haller}]{2009Siringo_APEX870_LABOCA}
Siringo, G., Kreysa, E., Kovács, A., {et~al.} 2009, Astronomy and
  Astrophysics, 497, 945, \dodoi{10.1051/0004-6361/200811454}

\bibitem[{Siringo {et~al.}(2010)Siringo, Kreysa, De~Breuck, Kovacs, Lundgren,
  Schuller, Stanke, Weiss, Guesten, Jethava, May, Menten, Meyer, Starkloff, \&
  Zakosarenko}]{2010Siringo_APEX350_SABOCA}
Siringo, G., Kreysa, E., De~Breuck, C., {et~al.} 2010, The Messenger, 139, 20.
\newblock \url{https://ui.adsabs.harvard.edu/abs/2010Msngr.139...20S}

\bibitem[{Stutz {et~al.}(2013)Stutz, Tobin, Stanke, Megeath, Fischer,
  Robitaille, Henning, Ali, di~Francesco, Furlan, Hartmann, Osorio, Wilson,
  Allen, Krause, \& Manoj}]{2013Stutz_HOPS_APEX}
Stutz, A.~M., Tobin, J.~J., Stanke, T., {et~al.} 2013, The Astrophysical
  Journal, 767, 36, \dodoi{10.1088/0004-637X/767/1/36}

\bibitem[{Takahashi \& Ho(2012)}]{2012Takahashi_HOPS87_outflow}
Takahashi, S., \& Ho, P. T.~P. 2012, The Astrophysical Journal, 745, L10,
  \dodoi{10.1088/2041-8205/745/1/L10}

\bibitem[{Takahashi {et~al.}(2019)Takahashi, Machida, Tomisaka, Ho, Fomalont,
  Nakanishi, \& Girart}]{2019Takahashi_HOPS87_young}
Takahashi, S., Machida, M.~N., Tomisaka, K., {et~al.} 2019, The Astrophysical
  Journal, 872, 70, \dodoi{10.3847/1538-4357/aaf6ed}

\bibitem[{Taquet {et~al.}(2019)Taquet, Bianchi, Codella, Persson, Ceccarelli,
  Cabrit, Jørgensen, Kahane, López-Sepulcre, \& Neri}]{2019Taquet_DHratio}
Taquet, V., Bianchi, E., Codella, C., {et~al.} 2019, Astronomy and
  Astrophysics, 632, A19, \dodoi{10.1051/0004-6361/201936044}

\bibitem[{Tatematsu {et~al.}(2017)Tatematsu, Liu, Ohashi, Sanhueza,
  Nguyen~Lu'o'ng, Hirota, Liu, Hirano, Choi, Kang, Thompson, Fuller, Wu, Li,
  Di~Francesco, Kim, Wang, Ristorcelli, Juvela, Shinnaga, Cunningham, Saito,
  Lee, Tóth, He, Sakai, Kim, Collaboration, \&
  Collaboration}]{2016Tatematsu_PGCC_N2Dp}
Tatematsu, K., Liu, T., Ohashi, S., {et~al.} 2017, The Astrophysical Journal
  Supplement Series, 228, 12, \dodoi{10.3847/1538-4365/228/2/12}

\bibitem[{Tatematsu {et~al.}(2020)Tatematsu, Liu, Kim, Yi, Lee, Hirano, Liu,
  Ohashi, Sanhueza, Francesco, Evans~Ii, Fuller, Kandori, Choi, Kang, Feng,
  Hirota, Sakai, Lu, Lu’o’ng, Thompson, Wu, Li, Kim, Wang, Ristorcelli,
  Juvela, \& Tóth}]{2020Tatematsu_ALMASOP}
Tatematsu, K., Liu, T., Kim, G., {et~al.} 2020, The Astrophysical Journal, 895,
  119, \dodoi{10.3847/1538-4357/ab8d3e}

\bibitem[{Tobin {et~al.}(2015)Tobin, Stutz, Megeath, Fischer, Henning, Ragan,
  Ali, Stanke, Manoj, Calvet, \& Hartmann}]{2015Tobin_SED_HOPS}
Tobin, J.~J., Stutz, A.~M., Megeath, S.~T., {et~al.} 2015, The Astrophysical
  Journal, 798, 128, \dodoi{10.1088/0004-637X/798/2/128}

\bibitem[{Tobin {et~al.}(2020)Tobin, Sheehan, Megeath, Díaz-Rodríguez,
  Offner, Murillo, van~'t Hoff, van Dishoeck, Osorio, Anglada, Furlan, Stutz,
  Reynolds, Karnath, Fischer, Persson, Looney, Li, Stephens, Chandler, Cox,
  Dunham, Tychoniec, Kama, Kratter, Kounkel, Mazur, Maud, Patel, Perez,
  Sadavoy, Segura-Cox, Sharma, Stephenson, Watson, \&
  Wyrowski}]{2020Tobin_VANDAM-II}
Tobin, J.~J., Sheehan, P.~D., Megeath, S.~T., {et~al.} 2020, The Astrophysical
  Journal, 890, 130, \dodoi{10.3847/1538-4357/ab6f64}

\bibitem[{van Gelder {et~al.}(2020)van Gelder, Tabone, Tychoniec, van Dishoeck,
  Beuther, Boogert, Caratti~o Garatti, Klaassen, Linnartz, Müller, \&
  Taquet}]{2020vanGelder_COMs}
van Gelder, M.~L., Tabone, B., Tychoniec, {\L}., {et~al.} 2020, Astronomy and
  Astrophysics, 639, A87, \dodoi{10.1051/0004-6361/202037758}

\bibitem[{Watson(2020)}]{2020Watson_B335_distance}
Watson, D.~M. 2020, Research Notes of the American Astronomical Society, 4, 88,
  \dodoi{10.3847/2515-5172/ab9df4}

\bibitem[{Werner {et~al.}(2004)Werner, Roellig, Low, Rieke, Rieke, Hoffmann,
  Young, Houck, Brandl, Fazio, Hora, Gehrz, Helou, Soifer, Stauffer, Keene,
  Eisenhardt, Gallagher, Gautier, Irace, Lawrence, Simmons, Van~Cleve, Jura,
  Wright, \& Cruikshank}]{2004Werner_Spitzer}
Werner, M.~W., Roellig, T.~L., Low, F.~J., {et~al.} 2004, The Astrophysical
  Journal Supplement Series, 154, 1, \dodoi{10.1086/422992}

\bibitem[{Whitney {et~al.}(2003)Whitney, Wood, Bjorkman, \&
  Wolff}]{2003Whitney_SED}
Whitney, B.~A., Wood, K., Bjorkman, J.~E., \& Wolff, M.~J. 2003, The
  Astrophysical Journal, 591, 1049, \dodoi{10.1086/375415}

\bibitem[{Wilson \& Rood(1994)}]{1994Wilson_isotope}
Wilson, T.~L., \& Rood, R. 1994, Annual Review of Astronomy and Astrophysics,
  32, 191, \dodoi{10.1146/annurev.aa.32.090194.001203}

\bibitem[{Wirstr\"om {et~al.}(2011)Wirstr\"om, {Geppert, W. D.}, {Hjalmarson,
  \AA{}.}, {Persson, C. M.}, {Black, J. H.}, {Bergman, P.}, {Millar, T. J.},
  {Hamberg, M.}, \& {Vigren, E.}}]{2011Wirstrom_12C13C_70}
Wirstr\"om, E.~S., {Geppert, W. D.}, {Hjalmarson, \AA{}.}, {et~al.} 2011, \aap,
  533, A24, \dodoi{10.1051/0004-6361/201116525}

\bibitem[{Wood {et~al.}(2002)Wood, Wolff, Bjorkman, \& Whitney}]{2002Wood_dust}
Wood, K., Wolff, M.~J., Bjorkman, J.~E., \& Whitney, B. 2002, The Astrophysical
  Journal, 564, 887, \dodoi{10.1086/324285}

\bibitem[{Wright {et~al.}(2010)Wright, Eisenhardt, Mainzer, Ressler, Cutri,
  Jarrett, Kirkpatrick, Padgett, McMillan, Skrutskie, Stanford, Cohen, Walker,
  Mather, Leisawitz, Gautier, McLean, Benford, Lonsdale, Blain, Mendez, Irace,
  Duval, Liu, Royer, Heinrichsen, Howard, Shannon, Kendall, Walsh, Larsen,
  Cardon, Schick, Schwalm, Abid, Fabinsky, Naes, \& Tsai}]{2010Wright_SED_WISE}
Wright, E.~L., Eisenhardt, P. R.~M., Mainzer, A.~K., {et~al.} 2010, The
  Astronomical Journal, 140, 1868, \dodoi{10.1088/0004-6256/140/6/1868}

\bibitem[{Yamamura(2010)}]{2010Yamamura_SED_AKARI_FIS}
Yamamura, I. 2010, in 38th COSPAR Scientific Assembly, Vol.~38, 2.
\newblock \url{https://ui.adsabs.harvard.edu/abs/2010cosp...38.2496Y}

\bibitem[{Yang {et~al.}(2020)Yang, Evans, Smith, Lee, Tobin, Terebey, Calcutt,
  Jørgensen, Green, \& Bourke}]{2020Yang_BHR-71-IRS1}
Yang, Y.-L., Evans, Neal~J., I., Smith, A., {et~al.} 2020, The Astrophysical
  Journal, 891, 61, \dodoi{10.3847/1538-4357/ab7201}

\bibitem[{Yang {et~al.}(2021)Yang, Sakai, Zhang, Murillo, Zhang, Higuchi, Zeng,
  López-Sepulcre, Yamamoto, Lefloch, Bouvier, Ceccarelli, Hirota, Imai, Oya,
  Sakai, \& Watanabe}]{2021Yang_PEACHES}
Yang, Y.-L., Sakai, N., Zhang, Y., {et~al.} 2021, The Astrophysical Journal,
  910, 20, \dodoi{10.3847/1538-4357/abdfd6}

\bibitem[{Yi {et~al.}(2018)Yi, Lee, Liu, Kim, Choi, Eden, Evans, Di~Francesco,
  Fuller, Hirano, Juvela, Kang, Kim, Koch, Lee, Li, Liu, Liu, Liu, Rawlings,
  Ristorcelli, Sanhueza, Soam, Tatematsu, Thompson, Toth, Wang, White, Wu,
  Yang, Collaboration, \& Collaboration}]{2018Yi_PGCC_Orion}
Yi, H.-W., Lee, J.-E., Liu, T., {et~al.} 2018, The Astrophysical Journal
  Supplement Series, 236, 51, \dodoi{10.3847/1538-4365/aac2e0}

\bibitem[{Zapata {et~al.}(2013)Zapata, Loinard, Rodríguez,
  Hernández-Hernández, Takahashi, Trejo, \&
  Parise}]{2013Zapata_IRAS16293B_depth}
Zapata, L.~A., Loinard, L., Rodríguez, L.~F., {et~al.} 2013, The Astrophysical
  Journal, 764, L14, \dodoi{10.1088/2041-8205/764/1/L14}

\bibitem[{Zucker {et~al.}(2020)Zucker, Speagle, Schlafly, Green, Finkbeiner,
  Goodman, \& Alves}]{2020Zucker_Perseus_distance}
Zucker, C., Speagle, J.~S., Schlafly, E.~F., {et~al.} 2020, Astronomy and
  Astrophysics, 633, A51, \dodoi{10.1051/0004-6361/201936145}

\end{thebibliography}
